\documentclass[twocolumn]{aastex631}
\usepackage{amsmath} 
\usepackage{graphicx}

\newcommand*{\hi}{\rm{H}\,\rm{\textsc{i}}}
\newcommand*{\msun}{\ensuremath{\rm{M}_{\odot}}}

\newcommand*{\kms}{\text{km}\,\text{s}\ensuremath{^{-1}}}

\begin{document}

\title{Streams, Shells, and Substructures in the Accretion-Built Stellar Halo of NGC~300}

\correspondingauthor{Catherine E. Fielder}
\email{cfielder@arizona.edu}

\author[0000-0001-8245-779X]{Catherine E. Fielder}
\affiliation{Steward Observatory, University of Arizona, 933 North Cherry Avenue, Tucson, AZ 85721-0065, USA}

\author[0000-0003-4102-380X]{David J. Sand}
\affiliation{Steward Observatory, University of Arizona, 933 North Cherry Avenue, Tucson, AZ 85721-0065, USA}

\author[0000-0002-5434-4904]{Michael G. Jones}
\affiliation{Steward Observatory, University of Arizona, 933 North Cherry Avenue, Tucson, AZ 85721-0065, USA}

\author[0000-0002-1763-4128]{Denija Crnojevi\'{c}}
\affil{Department of Physics and Astronomy, University of Tampa, 401 West Kennedy Boulevard, Tampa, FL 33606, USA}

\author[0000-0001-8251-933X]{Alex Drlica-Wagner}
\affil{Astronomy \& Astrophysics, University of Chicago, Chicago, IL 60637 USA}
\affil{Fermi National Accelerator Laboratory, Batavia, IL, USA}

\author[0000-0001-8354-7279]{Paul Bennet}
\affiliation{Space Telescope Science Institute, 3700 San Martin Drive, Baltimore, MD 21218, USA}

\author[0000-0002-3936-9628]{Jeffrey L. Carlin}
\affil{AURA/Rubin Observatory, 950 North Cherry Avenue, Tucson, AZ 85719, USA} 

\author[0000-0003-1697-7062]{William Cerny}
\affiliation{Department of Astronomy, Yale University, New Haven, CT 06520, USA}

\author[0000-0001-9775-9029]{Amandine Doliva-Dolinsky}
\affil{Department of Physics and Astronomy, University of Tampa, 401 West Kennedy Boulevard, Tampa, FL 33606, USA}
\affil{Department of Physics and Astronomy, Dartmouth College, Hanover, NH 03755, USA}

\author[0000-0001-5368-3632]{Laura C. Hunter}
\affil{Department of Physics and Astronomy, Dartmouth College, Hanover, NH 03755, USA}

\author[0000-0001-8855-3635]{Ananthan Karunakaran}
\affiliation{Department of Astronomy \& Astrophysics, University of Toronto, Toronto, ON M5S 3H4, Canada}
\affiliation{Dunlap Institute for Astronomy and Astrophysics, University of Toronto, Toronto ON, M5S 3H4, Canada}

\author[0000-0002-9269-8287]{Guilherme Limberg}
\affil{Kavli Institute for Cosmological Physics, University of Chicago, Chicago, IL 60637, USA}

\author[0000-0001-9649-4815]{Bur\c{c}in Mutlu-Pakdil}
\affil{Department of Physics and Astronomy, Dartmouth College, Hanover, NH 03755, USA}

\author[0000-0002-6021-8760]{Andrew B. Pace}
\affil{Department of Astronomy, University of Virginia, 530 McCormick Rd, Charlottesville, VA 22904, USA}

\author[0000-0003-0256-5446]{Sarah Pearson}
\affil{Niels Bohr Institute, University of Copenhagen, Copenagen, Denmark}

\author[0000-0003-2599-7524]{Adam Smercina}
\affil{Space Telescope Science Institute, 3700 San Martin Drive, Baltimore, MD 21218, USA}

\author[0000-0002-0956-7949]{Kristine Spekkens}
\affiliation{Department of Physics and Space Science, Royal Military College of Canada P.O. Box 17000, Station Forces Kingston, ON K7K 7B4, Canada}
\affiliation{Department of Physics, Engineering Physics and Astronomy, Queen’s University, Kingston, ON K7L 3N6, Canada}

\author[0000-0003-2539-8206]{Tjitske Starkenburg}
\affil{Department of Physics \& Astronomy and CIERA, Northwestern University, Evanston, IL, USA}

\author[0000-0002-1468-9668]{Jay Strader}
\affil{Center for Data Intensive and Time Domain Astronomy, Department of Physics and Astronomy, Michigan State University, East Lansing, MI 48824, USA}

\author[0000-0003-1479-3059]{Guy S. Stringfellow}
\affil{Center for Astrophysics and Space Astronomy, University of Colorado, 389 UCB, Boulder, CO 80309-0389, USA}

\author[0000-0002-9599-310X]{Erik Tollerud}
\affil{Space Telescope Science Institute, 3700 San Martin Drive, Baltimore, MD 21218, USA}

\author[0000-0003-4383-2969]{Clecio R. Bom}
\affil{Centro Brasileiro de Pesquisas F\'isicas, Rua Dr. Xavier Sigaud 150, 22290-180 Rio de Janeiro, RJ, Brazil}

\author[0000-0002-3690-105X]{Julio A. Carballo-Bello}
\affil{Instituto de Alta Investigaci\'on, Universidad de Tarapac\'a, Casilla 7D, Arica, Chile}

\author[0000-0001-5143-1255]{Astha Chaturvedi}
\affil{Department of Physics, University of Surrey, Guildford GU2 7XH, UK}

\author[0000-0003-1680-1884]{Yumi Choi}
\affiliation{NSF NOIRLab, 950 N. Cherry Ave., Tucson, AZ 85719, USA}

\author[0000-0001-5160-4486]{David J. James}
\affil{ASTRAVEO LLC, PO Box 1668, Gloucester, MA 01931}
\affil{Applied Materials Inc., 35 Dory Road, Gloucester, MA 01930}

\author[0000-0002-9144-7726]{Clara E. Mart\'inez-V\'azquez}
\affil{International Gemini Observatory/NSF NOIRLab, 670 N. A'ohoku Place, Hilo, Hawai'i, 96720, USA}

\author[0000-0001-5805-5766]{Alex Riley}
\affil{Institute for Computational Cosmology, Department of Physics, Durham University, South Road, Durham DH1 3LE, UK}

\author[0000-0002-1594-1466]{Joanna Sakowska}
\affil{Department of Physics, University of Surrey, Guildford GU2 7XH, UK}

\author[0000-0003-4341-6172]{Kathy Vivas}
\affil{Cerro Tololo Inter-American Observatory/NSF NOIRLab, Casilla 603, La Serena, Chile}



\begin{abstract}

We present deep optical observations of the stellar halo of NGC~300, an LMC-mass galaxy, acquired with the DEEP sub-component of the DECam Local Volume Exploration survey (DELVE) using the 4~m Blanco Telescope. Our resolved star analysis reveals a large, low surface brightness stellar stream ($M_{V}\sim-8.5$;  [Fe/H] $= -1.4\pm0.15$) extending more than 40~kpc north from the galaxy's center. We also find other halo structures, including potentially an additional stream wrap to the south, which may be associated with the main stream. The morphology and derived low metallicities of the streams and shells discovered surrounding NGC~300 are highly suggestive of a past accretion event. Assuming a single progenitor, the accreted system is approximately Fornax-like in luminosity, with an inferred mass ratio to NGC~300 of approximately $1:15$. We also present the discovery of a metal-poor globular cluster ($R_{\rm{proj}}=23.3$~kpc; $M_{V}=-8.99\pm0.16$; [Fe/H] $\approx-1.6\pm0.6$) in the halo of NGC~300, the furthest identified globular cluster associated with NGC~300. The stellar structures around NGC~300 represent the richest features observed in a Magellanic Cloud analog to date, strongly supporting the idea that accretion and subsequent disruption is an important mechanism in the assembly of dwarf galaxy stellar halos.

\end{abstract}



\section{Introduction} 
\label{sec:intro}

In modern cosmological theory, structure forms hierarchically within the $\Lambda$CDM framework (cold dark matter with a cosmological constant), where galaxies grow within massive dark matter (DM) halos by accreting smaller DM halos \citep[e.g.,][]{white1991,kauffmann1993,springel2006}. These lower-mass halos often host lower-mass dwarf galaxies, which can leave distinct imprints in the stellar halo far from the main galaxy when accreted. The stellar halo itself is comprised of a smooth component formed from ancient, phase-mixed accretions, interspersed with visible substructures such as stellar streams, plumes, and shells that are remnants from past mergers or interactions. Stellar halos constitute a small portion of a galaxy's stellar mass ($\sim1$\% in the Milky Way; \citealt{deason2019,mackereth2020}), but the long dynamical timescales at play preserve the evolutionary history of the system. Stellar halo structures are predicted in $\Lambda$CDM simulations \citep[e.g.][]{bullock2005,cooper2010}, and extensive studies of M31 and the Milky Way's stellar halo have been performed \citep[e.g.,][and many others]{grillmair2016,shipp2018,ibata2019,helmi2020}. Similarly, the latest generation of low surface brightness (e.g., NGC~922, \citealt{martinezdelgado2023}), deep resolved imaging work (e.g., M31, \citealt{ibata2014}; Cen~A, \citealt{crnojevic2016}; M64, \citealt{smercina2023}; M81, \citealt{okamoto2019}; M94, \citealt{gozman2023}), and growing wide-field stellar stream searches \citep[e.g.,][]{martinezdelgado2023b,mirro-carretero2024} have revealed the prevalence of primarily accreted stellar halos in systems as massive or larger than the Milky Way \citep[see also][]{harmsen2017}. 


At dwarf galaxy mass scales (stellar mass $M_{*}\lesssim10^{9}$~\msun), however, stellar halo assembly has been less explored. While dwarf galaxies in the Local Group are well studied they are heavily influenced by massive companions and often experience tidal stripping \citep[e.g.,][]{dooley2017,higgs2021}, which complicates efforts to isolate and study their stellar halos. 
In contrast, evidence from beyond the Local Group suggests that dwarf galaxies may possess stellar halos, though their formation mechanisms remain uncertain. At smaller mass scales, the frequency of mergers is expected to be lower than in more massive galaxies, suggesting that `in situ' growth may dominate \citep[e.g.,][]{genel2010,gu2016,fitts2018}. Consequently, stellar halos of dwarf galaxies may originate from two potential mechanisms: accretion, or in situ processes. And just like more massive hosts there may be a combination of both components.



There is some evidence of accretion in isolated dwarf galaxies (in addition to the Large Magellanic Cloud, see e.g., \citealt{mucciarelli2021}). NGC~4449, a Large Magellanic Cloud (LMC) analog in luminosity but slightly lower mass ($M_{*}=7.46\times10^{8}$~\msun), has been found to have a stellar stream originating from a lower mass dwarf galaxy \citep{martinezdelgado2012}, with an accretion history further corroborated by the presence of metal-poor stellar halo GCs \citep{strader2012}. Similarly, the LMC-mass NGC~2403 \citep[$M_{*}=7.24\times10^{9}\msun$;][]{dooley2017b} is stripping stars from its massive dwarf companion DDO~44 \citep[$M_{*}\sim 2\times10^{7}\msun$;][]{carlin2019}. Serendipitous observations of Ark~226 ($M_{*}=5\times10^{9}$~\msun) reveal shelves indicative of accretion buildup \citep{conroy2023}. The systematic study performed in \citet{annibali2020} detected several streams suggestive of accretion in a sample of 45 isolated dwarfs ($D<10$~Mpc).

`In situ' stellar halos form when stars from the disk of the galaxy are displaced. This can occur due to star formation-driven radial migration via stellar or supernova feedback \citep{stinson2009,maxwell2012,elbadry2016}, dynamical heating from interactions with low mass DM halos bereft of stars \citep{starkenburg2016}, from interactions with giant molecular clouds \citep{sellwood2013}, or when stars are stripped from the host's disk by an interaction. Observational evidence supporting in situ formation includes the predominantly round halos observed in nearby dwarf galaxies \citep{kadofong2020,kadofong2022}, as well as the presence of both old and intermediate-age stellar populations in the halo -- contrasting with the exclusively old populations typically found in the halos of more massive galaxies \citep{minniti1996,vansevivcius2004,hidalgo2009}. Recent spectroscopic studies have found relatively metal-rich intermediate-age populations in the halo of M33 (about twice the mass of the LMC), which are unlikely to have originated from accreted satellites \citep{cullinane2023}. Some of the stream-like features coming from the LMC are likely caused by its interaction with the SMC \citep{belokurov2017}. These findings suggest that dwarf galaxies may have distinct halo formation histories compared to their more massive counterparts.



NGC~300 is an LMC-mass ($M_{*} = 2.6\times10^{9}~\msun$; derived from the $K$-band luminosity reported in \citealt{karachentsev2013}, see also \citealt{drlicawagner2021}) SA spiral galaxy that lies beyond the Local Group and along the edge of the Sculptor Group in relative isolation, making it an ideal candidate for studying its stellar halo without the complicated influence of a nearby massive companion. Its disk is nearly face-on and exhibits a near-perfect exponential profile \citep[e.g.,][]{devaucouleurs1962,carignan1985,blandhawthorn2005,vlajic2009} with a radial scale length of 1.3~kpc \citep{gogarten2010}, consistent with expectations for undisturbed disk galaxies. Stellar population studies of the disk reveal a standard inside-out growth pattern, with older stars dominating in the inner regions and younger populations increasingly prominent at larger radii \citep{butler2004,tikhonov2005,gogarten2010,hillis2016,kang2016}. NGC~300 has a measured tip of the red giant branch (TRGB) distance of $2.01\pm0.03$~Mpc \citep{dalcaton2009}, which we adopt for this work. Prior to the study presented here only a few narrow and deep fields from HST \citep{tikhonov2005,jang2020} and Gemini \citep{vlajic2009} existed beyond the apparent optical disk of NGC~300, with inconclusive results regarding the presence of a stellar halo. \citet{vlajic2009} identified an 8~Gyr, [Fe/H] $= -1$ population ($R_{proj} = 15'$ east along the major axis), while \citet{jang2020} found a 10~Gyr, [Fe/H] $= -1.6$ population ($R_{proj} = 32'$ west along the major axis) with kinks in the radial density profile. The systematic wide-field search presented in this paper aims to elucidate the presence and general properties of NGC~300's stellar halo.

NGC~300 is one of the targets selected by the DECam Local Volume Exploration survey (DELVE; \citealt{drlicawagner2021,drlicawagner2022}), a long-term imaging program conducted with the Dark Energy Camera (DECam) on the Blanco 4m telescope \citep{flaugher2015}. The DELVE-DEEP sub-survey was designed to image four relatively isolated LMC and SMC (Small Magellanic Cloud) analogs within the Local Volume, out to their estimated virial radii (complementary to the MADCASH survey; \citealt{carlin2016,carlin2024}). This approach aims to identify stellar substructures and smaller satellite companions with resolved stars, probing the hierarchical assembly of dwarf galaxies.

In this paper we present several previously undetected and metal poor stream and shell substructures around NGC~300, providing strong evidence for an accretion built stellar halo. \autoref{sec:data} describes the DELVE-DEEP data acquisition and reduction in addition to spectroscopic data of a newly discovered globular cluster (GC). In \autoref{sec:halo_features}, we present the identification process and photometric properties of the features surrounding NGC~300. \autoref{sec:stellar_pops} includes color-magnitude diagrams (CMDs) and derived metallicities of the streams and shells, and analyses of the GC. In \autoref{sec:discussion} we explore the possible origins of these streams and shells and compare to prior studies of NGC~300. Lastly, we summarize our results and conclude in \autoref{sec:conclusion}.

\section{Data and Reduction} 
\label{sec:data}

\subsection{DELVE-DEEP Imaging}
\label{subsec:delve}

The DELVE survey is a long term multi-component imaging program designed to acquire contiguous multi-band coverage of the southern sky (Galactic latitude $|b| > 10^{\circ}$). NGC~300 lies at a latitude of $b\sim 79.4^{\circ}$ \citep{jang2020}. DELVE data is acquired with DECam, which is mounted on the 4~m Blanco telescope at the Cerro Tololo Inter-American Observatory, with a field of view of 3 square degrees. The DELVE-DEEP component performed deep imaging in $g$ and $i$-bands within the virial radius of four nearby and isolated Magellanic Cloud (MC) analogs in the Local Volume. The program was designed to reach $\gtrsim1.5$ mag below the TRGB ($g<26, i<25$) to $5\sigma$. For additional details on the DELVE survey design, please see \citet{drlicawagner2021} and \citet{drlicawagner2022} (which includes data releases DR1 and DR2 respectively). DR1 and DR2 of DELVE does not include the DEEP portion and the data presented here will be available in the forthcoming third data release. In this work we present our findings for the search for stellar structures within the halo of NGC~300.

NGC~300 DELVE-DEEP observations were acquired over the course of 21 nights between July 2021 and July 2023. This includes a total of 14 distinct fields taken as $12\times300$~s $g$-band exposures and $7\times300$~s $i$-band exposures per field, which are also supplemented with existing Dark Energy Survey (DES) data to attain the desired depth. Dithering of the fields was designed for uniform depth within the DEEP area to cover the chip gaps. For NGC~300, the average $5\sigma$ magnitude depth is $g=26.1$ and $i=25.3$, and the average $10\sigma$ magnitude depth is $g=25.7$ and $i=24.4$. In this work we present the more conservative $10\sigma$ depth, but we arrive at very similar results going to the $5\sigma$ depth. 
The fields taken around NGC~300 are designed to approximately cover the extent of a dark matter halo of the mass of the LMC (virial radius $r_{vir}\approx110~\rm{kpc}$; \citealt{dooley2017}), where the approximate virial radius of NGC~300 is found to be 120~kpc \citep{mutlupakdil2021} using the \citet{bryan1998} virial radius definition (the enclosed halo density is 104 times
the critical density of the universe $\rho_{c}$). This corresponds to a projected $r_{vir}=3.4$~deg, with the final DEEP coverage extending radially to $\sim3.1$~deg from the center of NGC~300, covering $\sim30.2$ square deg on the sky 
(although DEEP coverage is hexagonal and not perfectly circular).

For a detailed discussion on the data reduction for the DELVE early data release 3 used in this work please refer to Section 2 of \citet{tan2024}, which we summarize here. Multi-epoch photometric fitting is performed, using the processes developed for DES, with the DES Data Management \citep{morganson2018} co-addition (coadd) pipeline \citep{hartley2022,everett2022}. Multi-epoch data structures (MEDS; \citealt{jarvis2016,zuntz2018}) are created, which are constructed from cutouts centered on each object detected in the coadd images. Using the \texttt{fitvd} tool developed for DES \citep{hartley2022}, built on the \texttt{ngmix} framework \citep{sheldon2014}, multi-band, multi-epoch fitting is performed. This includes point spread function (PSF) model fits and bulge + disc (BDF) model fits while masking neighboring sources, known as a ``single-object fit'' (SOF) in DES terminology. PSF fits involve varying the amplitude of individual-epoch PSF models, while BDF fits use Sérsic profiles with fixed indices of $n=4$ for bulges and $n=1$ for discs, with their relative effective radii fixed to unity to reduce parameter degeneracies. The magnitudes referenced in this paper come from the \texttt{fitvd} PSF fit, which provides better photometry for point-like sources in crowded fields \citep{abbott2021}.

Star-galaxy separation is then computed using the sizes and signal-to-noise (S/N) measured from the \texttt{fitvd} assuming the BDF model. This process is performed with interpolation functions in the 2D space of size and S/N, developed for DES Y6 (Bechtol et al. in prep), which act as a classifier. Sources are classified with numerical values spanning from $0-4$. A $0$ classification is a high likelihood star, while a $4$ classification is a high likelihood galaxy. As described in \citet{tan2024} a complete stellar sample selected with $0 \leq \rm{class} \leq3$ has a stellar efficiency (true positive) rate of $96\%$, which starts to diminish below $i\sim23$. We choose to select sources with classifications in the range $0 \leq \rm{class} \leq 2$ to ensure reliable star identification at fainter magnitudes.  
Note that our analysis yields near identical results for $0 \leq \rm{class} \leq 1$.

The photometry is then corrected for Galactic extinction using the \citet{SFD1998} and the DES extinction coefficients \citep{abbott2021} which are updated with the re-normalization of \citet{schlafly2011} (see \citealt{tan2024} for further details). Unless otherwise indicated, all magnitudes presented are Milky Way extinction-corrected AB magnitudes.

\subsection{Spectroscopy}
\label{subsec:spec_data}

During the preparation of this paper, we serendipitously identified a GC in the southwest halo region of NGC~300, for which radial velocity information was required to confirm membership. Optical spectroscopy of this GC was procured on Aug. 11, 2024 on the Southern Astrophysical Research (SOAR) Telescope with the Goodman High Throughput Spectrograph \citep{clemens2004}. The total exposure time was 1~hr ($2\times1800$~s exposures) with the 1.2'' long slit and 400 lines/mm grating.  The data was taken in good seeing conditions with a mid-exposure airmass of 1.04. The data were reduced (bias-subtraction, flat-fielding, wavelength calibration, and flux-calibration) using IRAF \citep{Tody1986}. The final spectrum covers a wavelength range of 3798 -- 7845 \r{A}, with a median signal to noise of 18 \r{A}$^{-1}$, and resolution (FWHM) of 7.1 \r{A}$^{-1}$. This GC is further discussed in \autoref{subsection:GC}.

\section{Substructures Within the Halo of NGC~300}
\label{sec:halo_features}

\begin{figure}
    \centering
    \includegraphics[width=0.7\linewidth]{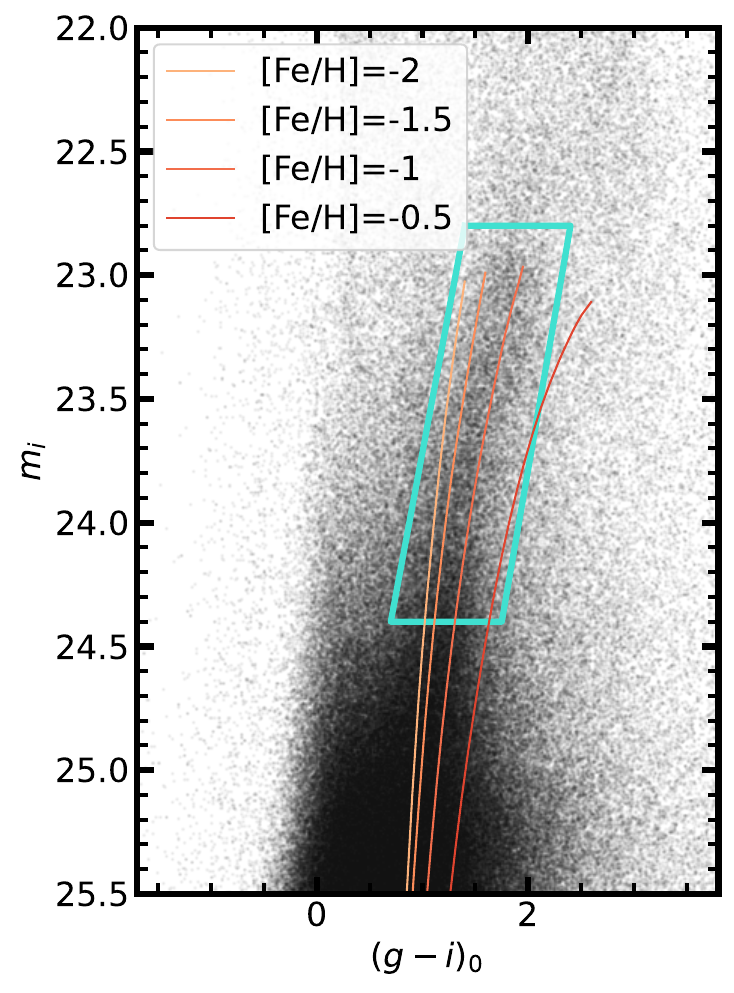}
    \caption{Color-magnitude diagram of a subset of DELVE-DEEP sources in the field of NGC~300 out to $0.5\times r_{vir}$. We over-plot our RGB selection box in cyan. For reference we also plot a few 10~Gyr Dartmouth isochrones \citep{dotter2008} of different metallicities (see legend), shifted to the distance of NGC~300.}
    \label{fig:rgb_selection}
\end{figure}

\begin{figure*}
    \centering
    \includegraphics[width=0.9\textwidth]{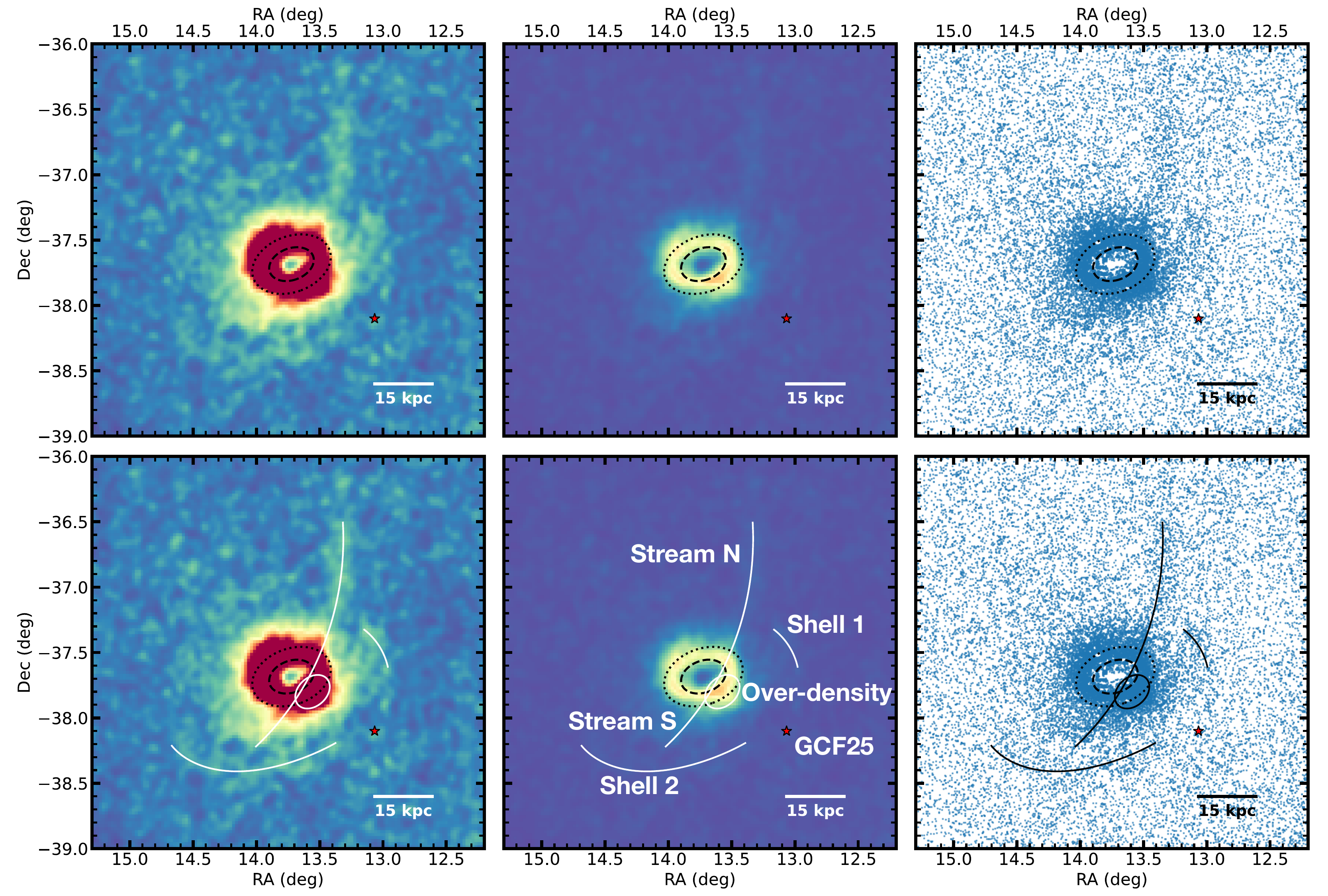}
    \caption{A red giant branch map of NGC~300 constructed from the DELVE-DEEP data. Crowding within the galaxy itself led to non-detections, hence the hole at the center of the density map. We draw a dashed ellipse to denote NGC~300 using a position angle of $\theta=109^{\circ}$ \citep{devaucouleurs1962} for the major axis (which spans southeast to northwest), axis ratio $b/a=0.67$ \citep{lauberts1989}, and $R_{25} = 10.9'$ \citep{blandhawthorn2005}. The dotted ellipse is $2\times R_{25}$. The left and central panels have the RGB stars placed in $1.6'$ bins, which are then convolved with a Gaussian kernel of $\sigma=1.2'$. The left panel is scaled to emphasize low surface brightness features and the middle to emphasize features near the disk. Blue is the lowest density of RGB stars while red is the highest. The right panel shows the RGB point sources. The bottom row is the same as the top, but we highlight and label the halo features. The red star shows our newly discovered metal poor GC, GCF25 (see \autoref{subsection:GC} for details).}
    \label{fig:rgb_map}
\end{figure*}

Faint substructures around dwarf galaxies like NGC~300 can be identified through resolved star searches \citep[e.g.,][Doliva-Dolinsky et al. in prep, Medoff et al. in prep]{crnojevic2016,carlin2019,smercina2023,gozman2023,carlin2024}. Our search for substructure was performed using red giant branch (RGB) stars. We select RGB stars with CMD positions consistent with relatively metal-poor isochrones ($-2\lesssim \rm{[Fe/H]}\lesssim -0.5$) at the distance of NGC~300 (2.01~Mpc). We select RGB stars down to a magnitude of $m_{i} = 24.4$ (the $10\sigma$ depth), where the RGB is clearly separable from other contaminants. \autoref{fig:rgb_selection} depicts our selection box on top of a scatter plot of point sources in DELVE-DEEP within $0.5\times r_{vir}$ of NGC~300. Note that in the CMDs of the various halo features there is no clear young stellar population (e.g., a main sequence or bright helium burning stars) or intermediate age stellar population (e.g., an asymptotic giant branch) in so our RGB selection is sufficient for the analyses in this work (see \autoref{subsection:CMD} for more details). Additionally, our results are similar at the $5\sigma$ depth $m_{i} = 25.3$. 

An RGB selected stellar density map centered on NGC~300 is shown in \autoref{fig:rgb_map}, using all DELVE-DEEP point sources that fall within our RGB selection box. Due to crowding a majority of the disk of the galaxy itself is not included in our figure, although some of the spiral structure is detectable in the point source panel. The optical disk/spiral features fall within $R_{25} = 10.9'$ -- the radius at which the surface brightness of the galaxy falls below 25 mag/arcsec$^{2}$ in the $B$-band \citep{blandhawthorn2005}. 

\begin{figure*}
    \centering
    \includegraphics[width=0.9\textwidth]{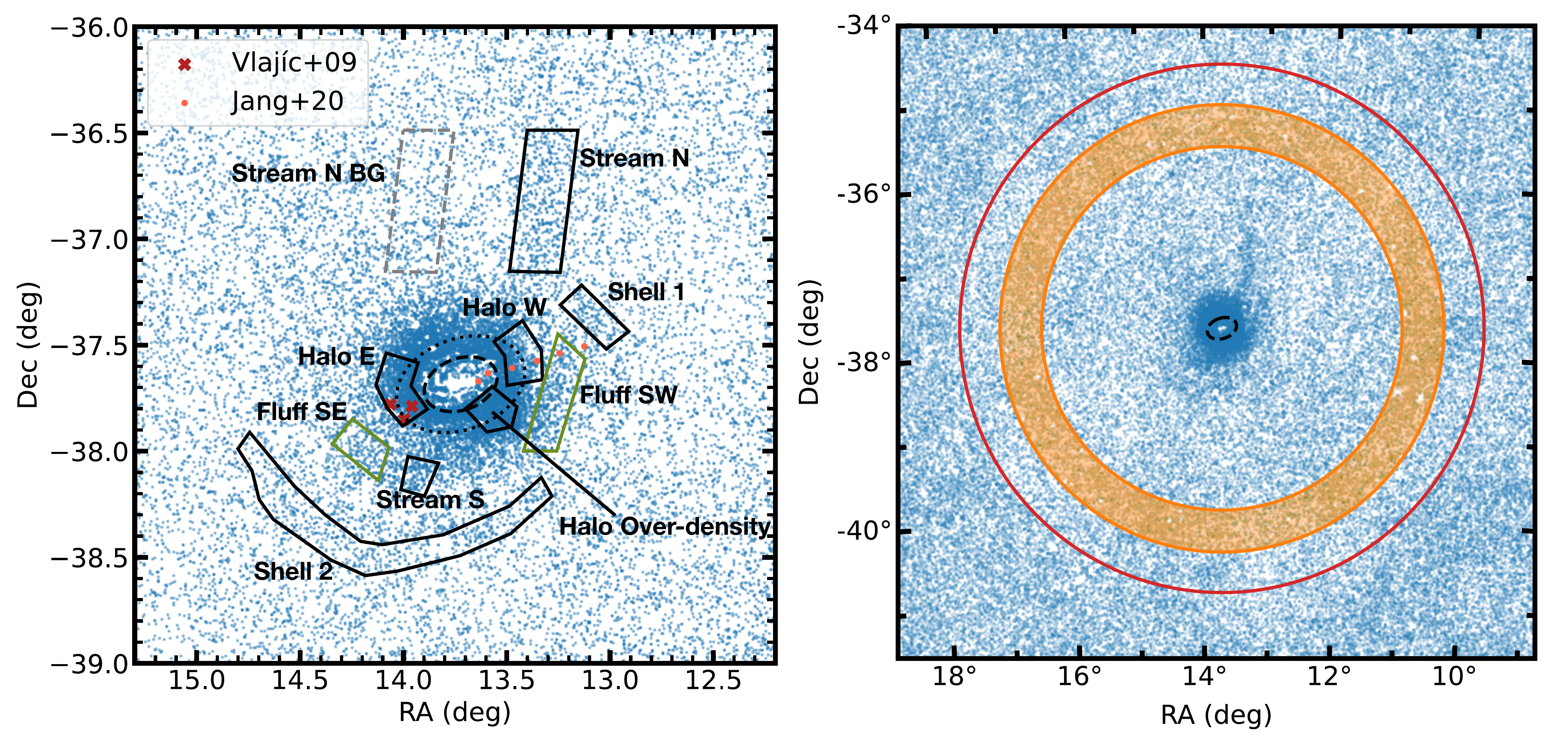}
    \caption{\textit{Left:} Same as the right column of \autoref{fig:rgb_map}, but we show selection boxes of the various features and regions around NGC~300 studied in depth in this work. Black outlined boxes are placed on the primary features. The gray dashed box corresponds to a region at the same distance from NGC~300 as the Stream N box and used for background subtraction of the Stream N CMD. The green boxes are additional regions explored as comparisons, placed where it appears that there is an over-density of stars but avoiding the southern stream. Lastly, central pointings utilized in studies by \citet{vlajic2009} and \citet{jang2020} are marked in red and orange, respectively, for reference. \textit{Right:} Same as left, but zoomed much further out. The red circle marks the circularized boundary between the DELVE-DEEP data and the shallower DES data ($R\sim3.15$~deg). The orange annulus shows the region selected for background subtraction ($130'<r<160'$; 7.59 square deg) in subsequent analyses.}
    \label{fig:selections}
\end{figure*}

The distribution of stars in the outer regions of our RGB density map (see the outskirts of \autoref{fig:rgb_map} and also the right panel of \autoref{fig:selections}) is low density, uniform, and unstructured, which can generally be attributed to background contaminants (e.g. compact galaxies) given the sub-dominance of Milky Way foreground stars in this portion of the sky ($b\sim 79.4^{\circ}$). Closer to the galaxy, however, a number of features emerge in addition to a possible smooth halo component at the outskirts of NGC~300. Due to the exceptionally low surface brightness of the features surrounding NGC~300 they are not visible in diffuse light, but are only detectable as individual resolved stars. For convenience we have highlighted and labeled the various features in \autoref{fig:rgb_map}, which we discuss below. The properties of these features are summarized in \autoref{tab:halo_features}. 

There is a long stellar stream that extends north of NGC~300 (Stream N in \autoref{fig:rgb_map}). This is the most distinct and farthest detectable feature near NGC~300, reaching a projected distance of 1.3~deg or 40.1~kpc from the center. Approximately $180$ degrees from Stream N lies another radial protrusion (Stream S). This feature is much closer to NGC~300, reaching $\sim$ 0.5~deg/17.1~kpc in projection from the center. Two separate shell-like features are also apparent in the RGB map. While there is no obvious connection between the two, it is possible that they could both be part of the same structure. The average distance of RGB stars within Shell 1 is 0.6~deg/19~kpc in projection from the center of NGC~300 and for Shell 2 is slightly further at 0.8~deg/25.4~kpc. The length of Shell 1 in projection is $\sim$ 0.3~deg/9.7~kpc and the length of Shell 2 is $\sim$ 1.3~deg/47.1~kpc. 

There are a few additional subtle features we note as well. Closer towards the galaxy on the southwestern side resides a notable over-density of RGB stars ($\sim$ 0.2~deg/5.4~kpc from the center) in addition to what appears to be excess RGB stars along the south, possibly even connecting to the shells. The smoothed RGB maps also hint that there may be excess RGB stars on the north and eastern sides of the galaxy. However, they are not clearly detected as individual and the stars here have similar properties to other stars in proximity to NGC~300. Thus while it is possible that extended or continuous substructure may be present around NGC~300 we focus on the most prominent features in this work.

We study each feature by drawing selection boxes, which we show in \autoref{fig:selections}. For the stream features our boxes are drawn to avoid the inner halo region of the galaxy to mitigate confusion between feature stars and halo stars. All other boxes are placed around the highest density of stars, but drawing precise boundaries around these features is non-trivial (and some may even overlap, such as the Stream N and Shell 1). The approximate area ($A$) covered by each feature is determined using the vertices of the feature selection boxes in (RA, Dec) and calculating the area as projected on the sky, making use of the $\tt{spherical\_geometry}$ Python library.

We estimate the luminosities and magnitudes of the features by using their CMDs, which we discuss in detail in \autoref{sec:stellar_pops}, and a luminosity function generated from a Kroupa initial mass function \citep{kroupa2001,kroupa2002}. 
We determine the cumulative luminosity function and derive the fraction of the total luminosity that lies below our RGB magnitude limits ($m_{i} = 24.4$, $m_{g} = 25.7$; below which the RGB becomes progressively difficult to separate from contaminants), and then use this to correct for the luminosity contribution below our imposed limit. 76\% of the luminosity is in our RGB selection in $i$-band and 84\% of the luminosity is in our RGB selection in $g$-band. This correction is applied to the luminosity of the background subtracted RGB populations of each feature before we convert to magnitude. Corrected $g$ and $i$-band magnitudes are presented in \autoref{tab:halo_features}. The conversion to $V$-band is derived using the transformations from DES DR2 \citep{abbott2021} adapted to $g$ and $i$-band; $M_{V} = 0.644M_{g} + 0.356M_{i} - 0.03$. Then we convert to luminosity using the relation $\log{L_{V} = \frac{4.80-M_{V}}{2.5}}$ where 4.80 is the absolute AB $V$-band magnitude of the sun \citep{willmer2018}. Given the faint nature of these features and the likelihood of missing luminosity or stellar mass due to the streams dipping toward the central galaxy, we emphasize that the brightness estimates derived here are only approximate, making accurate error estimation challenging. We also make an approximate estimate for the average surface brightness of each feature with the relation $\langle \mu_{V}\rangle = m_{V} + 2.5\log_{10}{A}$. Note that the derived surface brightnesses represent a faint limit, as the features are likely slightly brighter than calculated due to the fact that our selection boxes are slightly larger than the features themselves. 

Stellar masses are derived using the relation between stellar mass and magnitude and color from \citet{taylor2011} (their Eq. 8) and our derived $(g-i)$ colors and $M_{i}$ magnitudes. The colors of the stellar halo features generally fall within the regime where the \citet{taylor2011} relation aligns with \citet{zibetti2009}, with the exception of Shell 1. For this feature, a mass estimate based on \citet{zibetti2009} would yield a slightly higher value. Among the features Stream N has the highest derived mass, followed by Shell 2, and then Stream S and Shell 1. These values roughly follow the inferred luminosity of the features, where the interplay between color and brightness yields a similar result for Stream S and Shell 1.

\begin{table*}[t!]
     \centering
     \begin{tabular}{c|c|c|c|c|c|c|c|c|c|c|c}
         \hline \hline
          Halo Feature & $M_{g}$ & $M_{i}$ & $M_{V}$ & $\log{L_{V}/\rm{L}_{\odot}}$ & $A$ & $A$ & $\langle \mu_{V}\rangle$ & R$_{\rm{proj}}$  &R$_{\rm{proj}}$  & $M_{*}$ & $\langle\rm{[Fe/H]}\rangle$ \\ 
           & & & & & [deg$^{2}$] & [kpc$^{2}$] & [mag/arcsec$^{2}$] & [kpc] & [deg] & [$\times10^{5}$\msun] &  \\ \hline
          Stream N & $-8.2$ & $-9.0$ & $-8.5$ & $5.3$ & 0.13 & 161.8 & $< 33.6$ & 44.8 & 1.28 & 2.0 & $-1.4$ \\
          Stream S & $-7.5$ & $-8.5$ & $-7.9$ & $5.1$ & 0.02 & 21.4 & $<32.0$ & 19.1 & 0.55 & 1.8 & $-1.2$ \\
          Shell 1 & $-6.4$ & $-7.8$ & $-6.9$ & $4.7$ & 0.03 & 41.5 & $<33.7$ & 21.2 & 0.60 & 1.8 & $-1.2$ \\
          Shell 2 & $-8.1$ & $-8.9$ & $-8.4$ & $5.3$ & 0.19 & 230.8 & $<34.1$ & 28.4 & 0.81 & 1.9 & $-1.4$ \\ \hline
     \end{tabular}
     \caption{The notable features within the halo region of NGC~300 and their inferred properties. Refer to \autoref{fig:rgb_map} and \autoref{fig:selections} for labeling. Note that the low surface brightness nature of these features implies that these derived luminosities and masses are only approximate. The columns are: 1) $M_{g}$: $g$-band absolute magnitude. 2) $M_{i}$: $i$-band absolute magnitude. 3) $M_{V}$: $V$-band absolute magnitude, derived using the DES DR2 transformations. 4) $\log{L_{V}}$: $V$-band luminosity. 5) and 6) $A$: approximate area in square degrees and square kpc. 7) $\langle \mu_{V}\rangle$: average $V$-band surface brightness. 8) and 9) R$_{\rm{proj}}$: Projected distance from the center of NGC~300 in kpc and degrees. For the streams this is the maximum projected distance from the center of NGC~300, and for the shells this is the mean projected distance. 10) $M_{*}$: The stellar mass in units of $10^{5}$\msun\ derived using the color and magnitude relation from \citet{taylor2011}. 11) $\langle\rm{[Fe/H]}\rangle$: The photometrically derived mean metallicity and spread, see \autoref{subsection:MDF}. 
     }
     \label{tab:halo_features}
\end{table*}

\section{Stellar Populations}
\label{sec:stellar_pops}

In this section we discuss the properties of the RGB stars identified in the halo features of NGC~300. \autoref{subsection:stellar_overview} describes the RGB stars outside of the optical disk, \autoref{subsection:CMD} discusses the CMDs of each of the NGC~300 halo features and a few other additional regions, \autoref{subsection:MDF} derives metallicities for the RGB stars, and in \autoref{subsection:GC} we derive the properties of the new halo GC.

\subsection{Metal-rich and Metal-poor Substructures}
\label{subsection:stellar_overview}
\begin{figure*}
    \centering
    \includegraphics[width=0.9\textwidth]{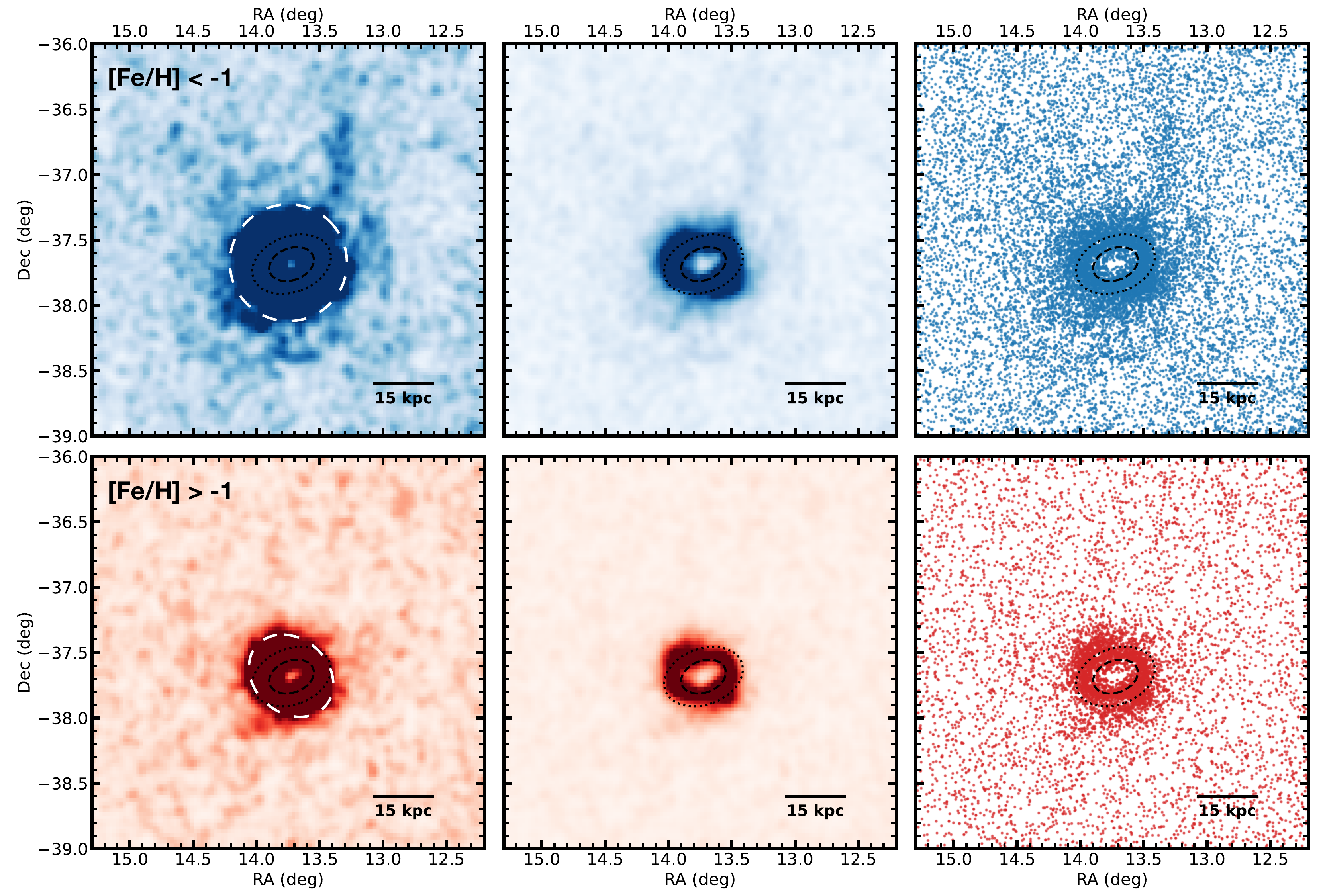}
    \caption{Same as \autoref{fig:rgb_map} but the stars that fall in our RGB selection box (see \autoref{fig:rgb_selection}) are split on a 10~Gyr [Fe/H] = $-1$ isochrone ([$\alpha$/Fe] $= 0$). RGB stars more metal-poor than [Fe/H] $= -1$ are colored blue and plotted in the top row, with more metal-rich stars plotted in the bottom row. The stellar halo features we identify are all more metal-poor than [Fe/H] $= -1$. Our identified smooth stellar halo is dominated by metal-poor stars further out, and it becomes more metal-rich closer to the edge of the disk. These components are highlighted by the white dashed circle (upper left) and ellipse (lower left; see text for details).}
    \label{fig:split_rgb_map}
\end{figure*}

Given the dominance of RGB stars in the features around NGC~300 we expect old stellar populations, with literature studies finding populations within the outskirts of NGC~300 approximately 10~Gyr old \citep[e.g.,][]{vlajic2009,jang2020}. For an assumed 10~Gyr stellar population we split our RGB map into metal-rich and metal-poor populations with an $\rm{[Fe/H]} = -1$ Dartmouth isochrone \citep{dotter2008} in \autoref{fig:split_rgb_map}. The blue top row shows stars more metal-poor than [Fe/H] $=-1$ and the red bottom row shows stars more metal-rich than [Fe/H] $= -1$. It is clear that the various features (i.e. the streams and shells) we detect have a metal-poor stellar population that is not apparent in the metal-rich map. There is a more metal-rich component beyond $2\times R_{25}$, falling approximately within an ellipse perpendicular to the major axis of the NGC~300 disk, in addition to a circular and more extended metal-poor component. Both features are approximately outlined in the left column of \autoref{fig:split_rgb_map} with a white long dashed circle (metal-poor) and ellipse (metal-rich). 

The appearance of the metal-poor stars is consistent with the detection of round old stellar populations identified around dwarfs \citep[e.g.,][]{strader2012,nidever2019,pucha2019,kadofong2020}
similar to the stellar halos of more massive galaxies. Therefore it is likely that we are seeing a similar type of structure in NGC~300, and refer to this region as the smooth inner halo. We discuss the implications of this in more detail in \autoref{subsection:interpretation}. The smoothed RGB maps also hint that there may be  RGB structures on the north and eastern sides of NGC~300. Again, these structures are not clearly detected as individual resolved stars and the stars here have similar properties (colors, brightnesses, and photometric metallicities) to other stars in the inner stellar halo region so we do not treat them separately from the smooth inner stellar halo.

\subsection{Color-magnitude Diagrams}
\label{subsection:CMD}

\begin{figure*}
    \centering
    \includegraphics[width=0.6\textwidth]{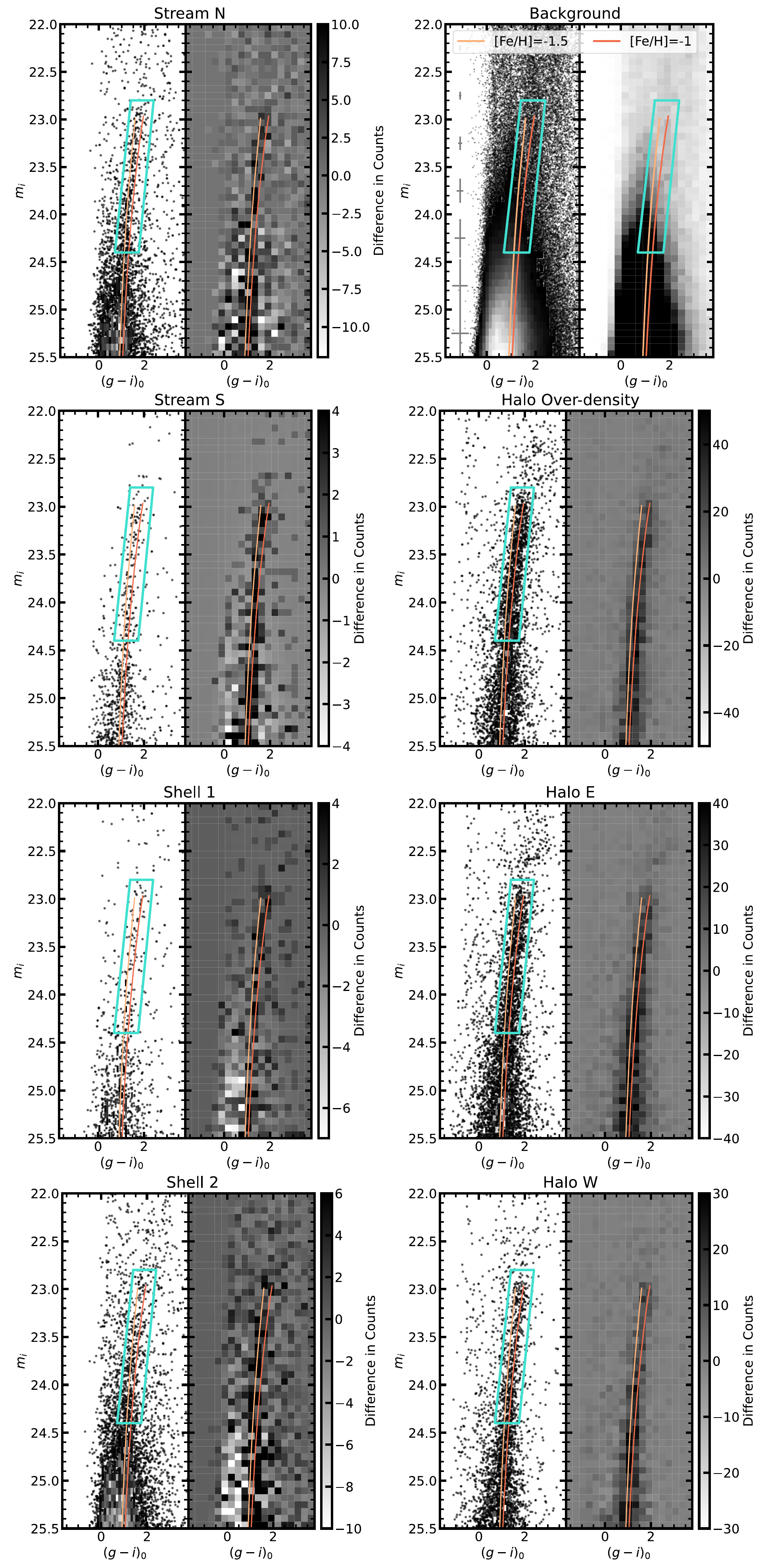}
    \caption{CMDs of the various features in proximity to NGC~300. The cyan box marks the RGB selection used to construct the RGB density maps (same as \autoref{fig:rgb_selection}). The background region (top right panel) is selected as a circular annulus as shown in \autoref{fig:selections} around NGC~300 ($130' < r < 160'$). Over-plotted are two metal-poor 10 Gyr Dartmouth isochrones that have been shifted to the distance of NGC~300. The background has no strong RGB component that aligns with the isochrones. The other panels depict CMDs of the various halo over-densities highlighted in \autoref{fig:selections}. In each CMD the right panel is background subtracted and binned (other than the background panel, which is only binned). Derived metallicites are presented in \autoref{fig:MDF}.}
    \label{fig:CMDs}
\end{figure*}

\autoref{fig:CMDs} shows the CMDs for point sources within the notable features and other areas of interest around NGC~300 (see the selection boxes in \autoref{fig:selections}). The cyan box depicts the RGB selection used in constructing the RGB maps. Over-plotted are 10~Gyr Dartmouth isochrones shifted to the distance of NGC~300 in orange. We only show two metal-poor isochrones for visual clarity, which approximately span the metallicity range of the red giant branches observed around NGC~300. The left panel of each CMD shows all stars that fall within our selection box, while the right panels are background subtracted (save for the background CMD, in the top left panel). For Stream N instead of using the large background annulus the background subtraction is performed with a box placed at the same distance from NGC~300 as Stream N (see the dashed gray selection in \autoref{fig:selections}). This is to check if there is a population of similar stars at the same projected distance from NGC~300. The background subtracted CMD for Stream N looks approximately the same if the background annulus is used. Nominally for each feature we would background subtract stars at the same approximate distance from the galaxy, but this exercise is non-trivial due to the abundances of features around NGC~300. 

Each feature shows up clearly in the background subtracted CMDs, emphasizing that the over-densities are real. Overall the Stream N has the most prominent and metal-poor RGB of all of the features, with a metallicity close to [Fe/H] $= -1.5$ and the smallest spread in color. The shells and halo exhibit a broader metallicity range, with the halo showing the widest distribution and a metallicity skewed towards higher values, centered around [Fe/H] $= -1$. While we have taken 3 different selection boxes in the inner smooth stellar halo region of the galaxy (see \autoref{fig:selections}), overall they contain very similar CMDs. With the current uncertainties and low star counts we are unable to determine differences in distance between the features. And, while there are likely a few asymptotic giant branch (AGB) stars in our CMDs, they are difficult to differentiate from other contaminants and we do not analyze them further.

\subsection{Metallicity Distribution Functions}
\label{subsection:MDF}

\begin{figure*}
    \centering
    \includegraphics[width=0.95\textwidth]{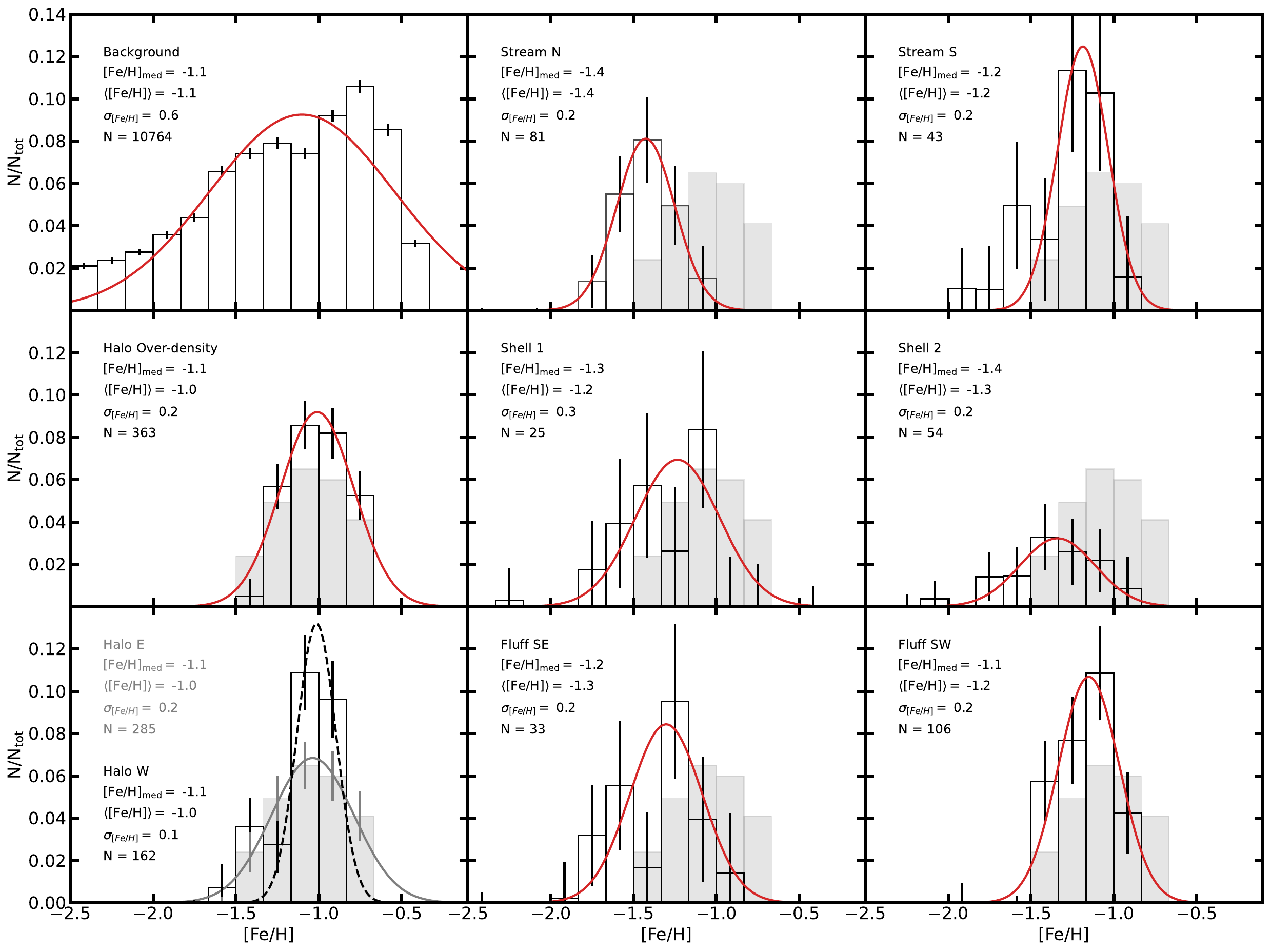}
    \caption{MDFs for each feature identified around NGC~300, derived by interpolating metallicity values for individual RGB stars between isochrones of fixed age and varied metallicity. The background MDF has been subtracted from each other MDF, which are then normalized to the total number of stars for which metallicites are derived. A best-fit Gaussian (red line) is plotted for each feature, from which we obtain the mean metallicity and metallicity dispersion. The gray MDF plotted in each panel shows the eastern inner halo region. The features identified around NGC~300 are more metal-poor than that of the smooth inner halo regions, indicative of an accretion origin.}
    \label{fig:MDF}
\end{figure*}

We construct metallicity distribution functions (MDFs) as a means to estimate the metallicity of each halo feature and a few additional regions around NGC~300. Our approach is similar to that of \citet{crnojevic2019} but our interpolation is different. To this end, we utilize a set of  10~Gyr Dartmouth isochrones in 0.25 dex metallicity intervals, ranging from $-2.5 < \rm{[Fe/H]} < 0.5$ and with no alpha-enhancement ([$\alpha$/Fe]=0). We construct a simple cubic interpolator mapping color and magnitude to metallicity with the $\tt{griddata}$ method in $\tt{scipy}$ in order to estimate the metallicity of every RGB star. To plot the MDFs we first construct histograms for each feature, normalized by the number of RGB stars found in the selection box. The MDFs are then background subtracted with the large annulus background region (see \autoref{fig:selections}, right panel) and normalized to the total number of stars for which metallicities have been derived. The MDFs are then fit to a Gaussian, weighted by Poisson errors (shown as errors on the bars), from which we derive a best fit mean metallicity and corresponding $1-\sigma$ spread. We also include the median background subtracted metallicity for each feature since some of these distributions are not well-fit by a Gaussian. The errors on the mean derived metallicities are small (0.01 dex order of magnitude) and do not fully represent the uncertainty in isochrone models. Instead we adopt an error for each derived mean metallicity that is equal to approximately half the isochrone separation, or 0.15 dex. These results are plotted in \autoref{fig:MDF}.

In each panel we show the eastern inner stellar halo box for comparison (Halo E, see also \autoref{fig:selections}), which has more stars than the western halo box (Halo W), with both showing the same metallicity distribution and best-fit mean metallicity $\langle\rm{[Fe/H]}\rangle=-1.0$. We want a direct comparison closer to the disk of NGC~300 to highlight any difference in metallicity we may find with the halo features. Despite the unexpected detection of an over-density within the smooth stellar halo, its metallicity is not clearly differentiable from less dense parts of the smooth stellar halo. The streams and shells are all more metal-poor on average than the inner halo regions of NGC~300.

Of the distinct halo features Stream N is the most metal-poor ($\langle\rm{[Fe/H]}\rangle=-1.4\pm0.15$) followed by Shell 2 ($\langle\rm{[Fe/H]}\rangle=-1.3\pm0.15$) then Shell 1 ($\langle\rm{[Fe/H]}\rangle=-1.2\pm0.15$) and Stream S ($\langle\rm{[Fe/H]}\rangle=-1.2\pm0.15$). That said, these four distinct halo features contain metallicities consistent with each other, with more metal poor peaks than Halo E (or Halo W). Shell 1 and Stream S exhibit somewhat bi-modal distributions, but the abundance of more metal-rich RGB stars is likely due to the proximity to NGC~300 where the more metal-rich inner smooth halo stars are falling within our selection. We also examine RGB stars in two boxes placed within the southern halo region (Fluff SE and Fluff SW; see \autoref{fig:selections} green boxes), where it appears there is an excess of RGB stars that may even connect to the shells. The southeastern box (Fluff SE), near Stream S and Shell 2, appears to have more metal-poor RGB stars than the southwestern box (Fluff SW) that extends near Shell 1. The shapes between the MDFs of the Shells and the Fluffs are different, so it is difficult to determine if these features are truly connected. While there are large uncertainties involved in the derivations of photometric metallicities, the relative comparisons of the MDFs is quite robust, where only bright AGB stars may slightly bias the MDFs with a slight metal-poor tail.
We compare these metallicites to those of previous studies of NGC~300 in \autoref{subsection:interpretation}.

\subsection{A Globular Cluster in the Halo of NGC~300}
\label{subsection:GC}

\begin{figure}
    \centering
    \includegraphics[width=0.95\linewidth]{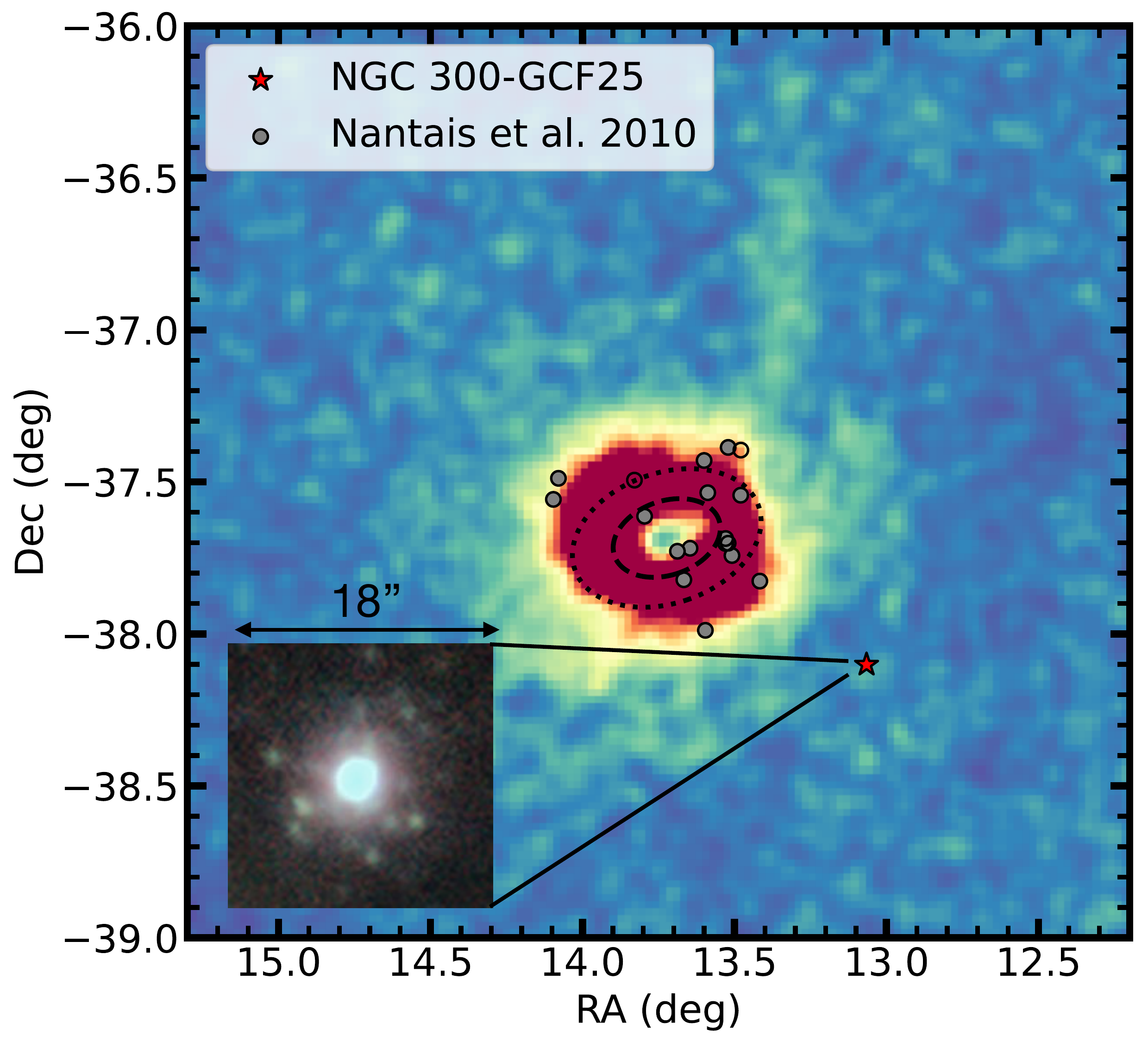}
    \caption{Distribution of GCs within NGC~300, first identified by \citet{kim2002} and \citet{olsen2004} and confirmed in \citet{nantais2010A}. High confidence GCs are marked by filled points (14), while lower confidence GCs are marked by open points (3). Our newly discovered NGC~300-GCF25 is marked by the red star, which falls in projection between the distinct shell 1 and shell 2 over-densities in the RGB map. In the lower left corner we include a full color $griz$ image of GCF25 obtained from the publicly available DESI Legacy imaging \citep{dey2019}. RGB stars in the outskirts of the GC are clearly resolved.}
    \label{fig:GC_distribution}
\end{figure}

We first identified NGC~300-GCF25 by eye in the DECaLS portion of the DESI Legacy Imaging Surveys \citep{dey2019}. RGB stars that lie in the outskirts of the GC-F25 are resolved, as seen in the full color $griz$ insert panel of \autoref{fig:GC_distribution}. A spectrum of this GC was obtained with SOAR as described in \autoref{subsec:spec_data}, while photometric properties are derived from the publicly available DR10 imaging from the DESI Legacy survey. The properties of GCF25 are summarized in \autoref{tab:gc_props}. While this is the 15th confirmed GC associated with NGC~300 \citep[see][]{nantais2010A}, in order to mitigate naming confusion we refer to this as GCF25 in reference to the first author and year of this paper. There will be a comprehensive search for and study of the halo GCs of NGC~300 presented in future work. 

\begin{figure}
    \centering
    \includegraphics[width=0.95\linewidth]{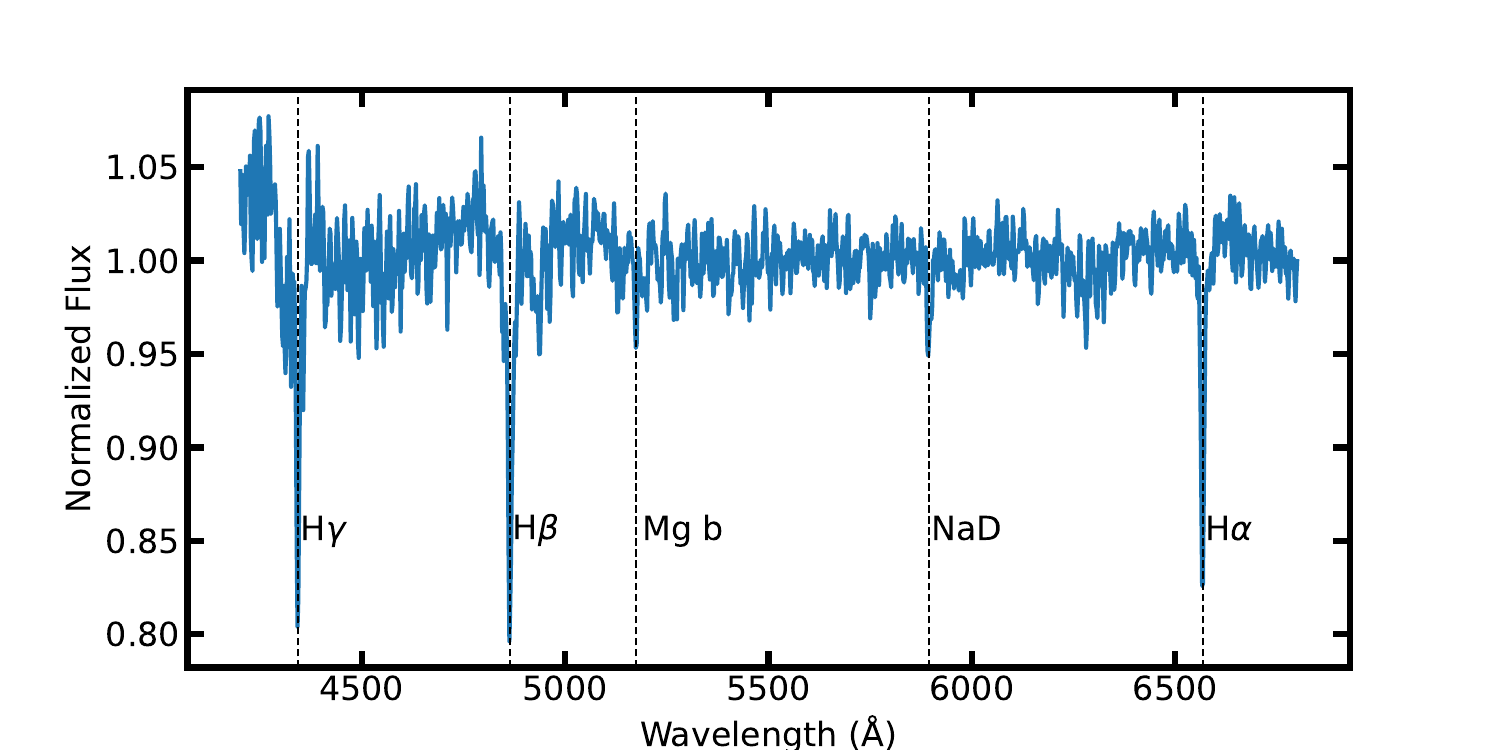}
    \caption{The normalized SOAR spectrum of NGC 300-GCF25. We highlight the strong Balmer lines as well as the Mg b and NaD absorption lines, which are common GC absorption features.}
    \label{fig:GC_spectrum}
\end{figure}

The barycentric radial velocity of NGC~300-GCF25 is $v_{\rm{helio}} = 190 \pm 8$ \kms, where a zero point correction using the telluric absorption is performed using the $\tt{TelFit}$ python package \citep{gullikson2014}, and the spectrum is then fit using the $\tt{RVSpecFit}$ python package \citep{koposov2011}. The spectrum over a portion of the wavelength range is plotted in \autoref{fig:GC_spectrum} where we highlight common absorption features in GC spectra. The \hi\ barycentric radial velocity for NGC~300 is $144\pm2$ km s$^{-1}$ \citep{westmeier2011}. 
Work by \citet{nantais2010A} spectroscopically followed up 42 GC candidates identified by \citet{kim2002} and \citet{olsen2004}, which resulted in the confirmation of 14 GCs with an additional 3 as possible GCs. \citet{nantais2010A} finds that positive radial velocities indicate that they are likely not associated with the Milky Way. For the confirmed GCs, \citet{nantais2010A} found a velocity dispersion of 68 \kms\, and the theoretical derivation in \citet{carignan1985} finds an expected velocity dispersion of $\sim$ 60 \kms.  In this case, a $46\pm8$ \kms\ offset is consistent, and we conclude that NGC~300-GCF25 is kinematically associated with NGC~300 and the furthest NGC~300 GC confirmed to date.

The photometric properties for NGC~300-GCF25 are calculated from the DECaLS images using the \texttt{Photutils v2.0.1} Python package for aperture photometry, with Milky Way extinction corrections derived from \citet{schlafly2011}. The photometry is converted to the Johnson-Cousins $V$ and $I$ bands (still AB magnitudes) using the DES DR2 conversions as before. In terms of magnitude and color, NGC~300-GCF25 is bright and relatively blue. The globular cluster luminosity function (GCLF) from \citet{peng2009} peaks at $\mu_{M_{I},\rm{AB}} = -8.12$ mag with $\sigma_{M_{I}} = 1.37$. NGC~300-GCF25 is slightly more than $1\sigma$ outside of the distribution at $M_{I} = -9.83$ but still well within expectations for a GC. GC colors are expected to range from $0.5 < (V-I) < 1.5$ \citep[e.g.,][]{brodie2006}, thus GCF25 is on the blue half of the distribution at $(V-I) = 0.86$ but not unusually blue. 

The half-light radius is determined by utilizing a series of finely spaced circular apertures of increasing radii and identifying the radius at which the flux is equal to half of the total flux. The PSF-corrected half-light radius derived for GCF25 is $r_{1/2} = 11.4\pm1.5$~pc assuming a distance of 2.01~Mpc. This is large for a GC, and only comparable to a small fraction of Milky Way GCs (about $10\%$ have similar or larger half-light radii; see e.g., \citealt{harris1996} and the Holger Baumgardt GC database\footnote{\url{https://people.smp.uq.edu.au/HolgerBaumgardt/globular/}}, \citealt{baumgardt2020,baumgardt2021}). We note that this GC is in close proximity to a prominent background galaxy, which may be contaminating the measurement. 
The \texttt{Photutils} derived ellipticity of GCF25 is fairly typical at $\epsilon = 0.071\pm0.005$, which is very close to the median measurement for the Milky Way ($\epsilon = 0.06, \sigma_{\epsilon}=0.07$; \citealt{beasley2019}). Errors are derived by a Monte Carlo approach, where the data is perturbed to simulate the impact of noise and then the spread in ellipticity values is determined.

While the spectrum lacks sufficient resolution to derive a robust metallicity, we make an estimate based on a single-burst stellar population model and the derived photometric colors. Using the PAdova and TRieste Stellar Evolution Code (\textsc{PARSEC v.1.2S} \citealt{bressan2012,chen2014,tang2014,chen2015,marigo2017,pastorelli2019,pastorelli2020}) with a Kroupa IMF, we generate 8–12 Gyr stellar populations with no $\alpha$-enhancement and a range of metallicities. Then we derive metallicities where the simple stellar population matches the derived colors of GCF25. Since the \textsc{Parsec} metallicities are in [M/H] and assume no $\alpha$-enhancement, they closely approximate [Fe/H]. Given that the best-fitting isochrones for the NGC~300 features suggest an age of 10~Gyr, we adopt the same age for the GC and thus the metallicity that corresponds to the 10~Gyr stellar population: [M/H] = $-1.6\pm0.6$ (reported in \autoref{tab:gc_props}). If the GC is slightly younger at 8~Gyr then it is more metal rich, with [M/H] = $-1.4\pm0.6$. If the GC is slightly older at 12~Gyr then it is slightly more metal poor (but close to $-1.6$). For comparison, the NGC~300 GCs analyzed by \citet{nantais2010A} have metallicities ranging from $-1.61 <$ [Fe/H] $< -0.07$, with a mean of [Fe/H] = $-0.94 \pm 0.15$. 


\begin{table}[]
    \centering
    \caption{Properties of NGC~300-GCF25}
    \begin{tabular}{r|c}
    \hline\hline
        R.A. (J2000) & 00:52:15.74 \\
        Dec. (J2000) & $-$38:06:03.96 \\
        $M_{g}$ & $-8.68\pm0.13$ \\
        $M_{r}$ & $-9.20\pm0.10$ \\
        $M_{i}$ & $-9.40\pm0.09$\\
        $M_{V}$ & $-8.97\pm0.16$ \\
        $M_{I}$ & $-9.83\pm0.19$ \\
        $(g-r)$ & $0.53\pm0.17$ \\
        $(g-i)$ & $0.71\pm0.16$ \\
        $(g-z)$ & $0.80\pm0.16$ \\
        $(V-I)$ & $0.86\pm0.11$ \\
        $r_{1/2}$ (arcsec) & $1.17\pm0.15$ \\
        $r_{1/2}$ (pc) & $11.43\pm1.46$ \\
        $\epsilon$ & $0.071\pm0.005$ \\
        $v_{\rm{radial}}$ [\kms] & $190\pm8$ \\
        $M_{*}$ [\msun] & $(2.6\pm0.8)\times10^{5}$ \\
        $[\mathrm{Fe}/\mathrm{H}]$ & $-1.6\pm0.6$ \\
        Age (Gyr) & $\sim 10$\\
        $\rm{R}_{\rm{proj}}$ [kpc] & 23.3 \\
        $\rm{R}_{\rm{proj}}$ [deg] & 0.67 \\
    \hline
    \end{tabular} \\ [4pt]
    Various properties of NGC~300-GCF25. Note that magnitudes and colors are in the AB photometric system. The half light radius ($r_{1/2}$) and ellipticity ($\epsilon$) are measured from the $r$-band. \\
    \label{tab:gc_props}
\end{table}

\section{Discussion}
\label{sec:discussion}

\subsection{Potential Origins of Halo Substructures}
\label{subsection:origins}

We have identified several stellar substructures in the halo of NGC~300, all of which are clear over-densities with respect to the background, and contain a generally metal-poor population of stars. We now comment on the potential relationships between these features and their origins.

In general, stellar streams are the results of tidal interaction between a primary galaxy and a lower mass system, typically a satellite galaxy or a GC. It is likely that this is also the case for Stream N. This stream is distinct and contains a stellar population with a metallicity lower than the [Fe/H] $>-1$ observed in the outskirts of NGC~300 \citep[see][and references therein]{vlajic2009}. It is unlikely to pull out a stellar population with metallicity [Fe/H]$=-1.4\pm0.15$ from the galaxy, and there is no clearly detected nearby companion to have tidally influenced NGC~300 in such a way to result in the formation of the stream. Therefore we infer it to be the remains of an accreted dwarf.

The Stream S may be a wrap or extension of Stream N, where the disrupting object turned around in its orbit, or a completely separate smaller stream originating from a different object. Additionally, given its proximity to NGC~300 it may also be an in situ feature, possibly arising from significant supernova feedback \citep{elbadry2016} or a merger with a dark satellite that induced star formation offset from the galaxy itself, which has been seen in simulations \citep[e.g.,][]{starkenburg2016}. Since the CMDs for both features are similar and consistent within uncertainties, we favor a scenario where the two streams are related. Slight differences between the two MDFs can be explained from two factors. First, stars from the more metal-rich smooth inner stellar halo overlap with the Stream S selection box (see also \autoref{fig:split_rgb_map}) and are likely contributing to the more metal-rich end, since our background estimate consists of a region far away from the galaxy. Second, wrapped streams like the Sagittarius stream have been observed to have metallicity gradients, so a metallicity difference is not unexpected \citep{hayes2020,johnson2020,limberg2023,cuningham2024}.

The shells exhibit broader and less defined CMDs, partly due to their extremely low surface brightness. 
The similarities in the CMDs and MDFs between the shells suggest a shared origin, consistent with models predicting shell-like structures from minor mergers on radial orbits \citep{hernquist1987}. It is also plausible that the streams and shells are related; all features are more metal-poor than the smooth halo component and exhibit broadly consistent metallicities. In this scenario, the progenitor may have initially followed a radial orbit, forming the shells, before wrapping around the galaxy to create the streams. The converse may also be true, where the stream forms on a tangential orbit and then radializes the rest of the debris \citep{vasiliev2022}.

Alternatively, the shells might result from in situ extended disk star formation that has since passively faded -- a phenomenon observed in NGC~404 through ultraviolet imaging and attributed to merger-induced processes \citep{thilker2010}. However, GALEX \citep[GAlaxy Evolution EXplorer;][]{galex2005} all-sky survey ultraviolet observations of NGC~300 show no detections beyond the spiral features \citep[see also][]{mondal2019}, indicating that any extended star formation ceased over 100~Myr ago. Likewise, in the CMDs we do not detect evidence of a young stellar population. The comparable and old stellar populations of the shells and streams favor a single origin -- an accreted dwarf galaxy.

\subsection{Potential Progenitor Properties}
\label{subsection:progenitor}

Using the luminosity lower limits derived for each distinct NGC~300 feature (see \autoref{tab:halo_features}), and assuming all of the features are from one or more accretion events, we estimate the approximate luminosity of the progenitors of these features by summing the corrected luminosity derived for each feature (presented in \autoref{tab:halo_features}). We consider three possible combinations: a single progenitor for all of the streams and shells ($M_{V} = -9.5$), a single progenitor for the shells ($M_{V} = -8.6$), and a single progenitor for the streams ($M_{V} = -8.8$). The estimated luminosity is a lower limit -- NGC~300 may be blocking portions of these features, and the features may have been disrupted or partially dissipated, especially given that these features are detected in RGB star counts and not diffuse light.

Inferring a metallicity for a progenitor is less straightforward. We opt to use the derived metallicity of the most metal-poor contributor for a more conservative estimate. This means the metallicity we use for a single progenitor is [Fe/H] $= -1.4\pm0.3$, for a single progenitor for the shells is [Fe/H] $= -1.3\pm0.2$, and for a single progenitor of the streams is [Fe/H] $= -1.4\pm0.2$. The error bars plotted for [Fe/H] correspond to the metallicity error ($\sigma$) of each individual contributor added in quadrature.

\begin{figure*}
    \centering
    \includegraphics[width=0.95\textwidth]{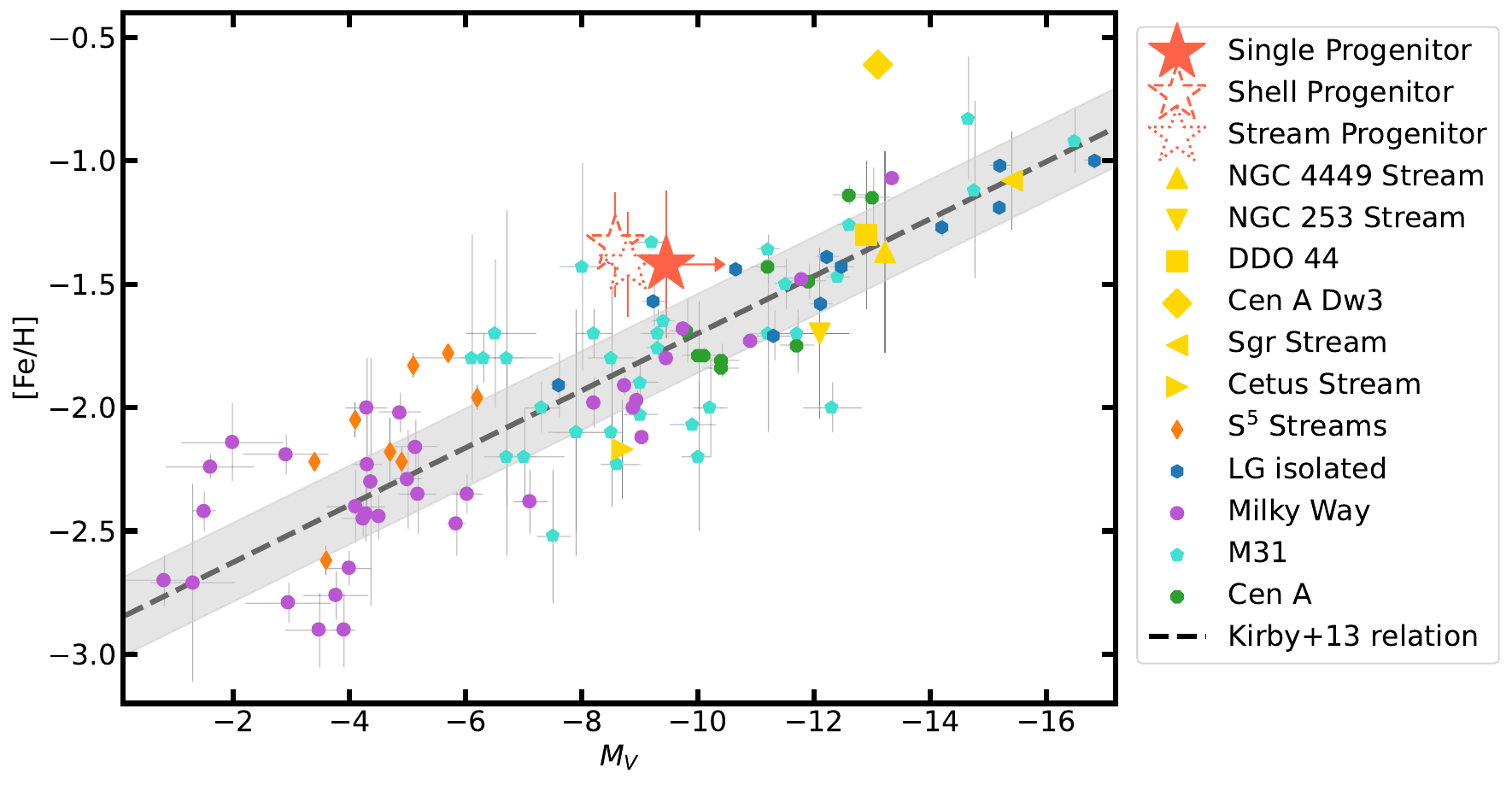}
    \caption{The stellar metallicity ([Fe/H]) - luminosity ($M_{V}$) relation for dwarf galaxies, adapted from \citet{collins2022}. Observations of isolated Local Group dwarfs and dwarfs around well-studied massive hosts are included, in addition to a few stream progenitor estimates (see text for references). The dashed line and shaded region corresponds to the derived best-fit relation and $1\sigma$ scatter from \citet{kirby2013}.  We plot potential progenitors of the NGC~300 features as orange stars. The solid star represents a single progenitor for all of the features, while the dashed star represents a progenitor for Shell 1 and Shell 2 and the dotted star represents a single progenitor for Stream N and Stream S. These are plotted at the mean metallicity of the most metal-poor feature contributing to the measurement, while the errors on [Fe/H] include the errors added in quadrature of the mean metallicities of each contributing feature. Note these magnitude estimates are a lower limit.
    }
    \label{fig:metal_lum}
\end{figure*}

We plot these three progenitor scenarios in \autoref{fig:metal_lum}. We compare the potential progenitors for the stellar halo features to isolated Local Group dwarfs \citep[blue hexagons;][]{geha2010,kirby2013,taibi2018,taibi2020}, Milky Way dwarfs \citep[purple circles;][]{simon2019,pace2024}, M31 dwarfs \citep[cyan pentagons;][]{collins2013,collins2020,wojno2020,collins2021,pace2024}, and Cen~A dwarfs \citep[green octagons;][]{muller2021}. We also include the several Milky Way streams from S$^{5}$ (orange diamonds; metallicites from \citealt{li2022} and magnitudes from \citealt{shipp2018}) in addition to the Milky Way Sagittarius (Sgr) stream progenitor \citep{niederste2012,ramos2022}, the Milky Way Cetus stream \citep{yuan2022}, the stream from Sc1-MM-Dw2 interacting with NGC~253 \citep{toloba2016a}, the disrupting Dw3 of Cen~A \citep{crnojevic2019}, DD0~44 which has tidal streams due to its host NGC~2403 \citep{carlin2019}, and stream progenitor estimates for the NGC~4449 stream \citep{martinezdelgado2012,toloba2016}.  

The established metallicity-luminosity relation and $1\sigma$ scatter from \citet{kirby2013} is plotted for reference, derived from bound satellites (e.g., not tidally disrupted). Given that our progenitor magnitude estimates are lower limits it is physically reasonable that they lie above the scaling relation (similar to expectations for tidally disrupted dwarfs). At a metallicity of [Fe/H]$=-1.4$ a single progenitor of the NGC~300 features would fall within the $1\sigma$ range of the \citet{kirby2013} metallicity-luminosity relation at a magnitude of $11.2\lesssim M_{V}\lesssim-13.8$. The means that the progenitor of the NGC~300 features could be comparable to some of the brighter Local Group dwarf spheroidals such as AndII ($M_{V}=-11.72$; \citealt{martin2016}) and AndVII ($M_{V}=-12.78$; \citealt{martin2016}) of M31, or LeoI ($M_{V}=-11.82$; \citealt{munoz2018}) and plausibly even Fornax ($M_{V} = -13.39$; \citealt{munoz2018}) of the Milky Way. Interestingly a plausible progenitor of the features around NGC~300 may be comparable to the progenitor of the stream close to NGC~4449, which is also considered an LMC analog. 

There is evidence that a single progenitor can result in all of the features that NGC~300 displays. For instance, the Gaia-Sausage/Enceladus merger with the Milky Way is thought to have produced additional over-densities within the stellar halo \citep{naidou2021} and predict a stream wrap. Similarly, \citet{fardal2013} revealed the connection between the southern stream of M31 and other tidal debris within the stellar halo, including two accretion ``shelves'' on the northeastern and western sides of the galaxy. Notably, the morphology seen in M31 bears a striking resemblance to the features identified in NGC~300 (and the straight nature of the northern stream could make it particularly useful for constraints on halo shape parameters, see \citealt[][]{nibauer2023}). At dwarf masses, gas modeling of a 1:8 merger by \citet{pearson2018} reproduces a morphology similar to NGC~300’s streams (see the 0.5~Gyr panel of their Fig. 5). The models of \citet{deason2022} explore smaller mass ratios, where the 1:10 merger of a low-mass dwarf with a $10^{10}$\msun\ dark matter halo also exhibits extended stream and shell-like features (see Fig. 8 of \citealt{deason2022}). Depending on the satellite binding energy and the amount of stellar material within, minor mergers can deposit stripped material at larger distances from the host galaxy’s center than major mergers - similar to the streams and shells detected around NGC~300 \citep[see also][]{pearson2018}.

Assuming a single progenitor, the estimated stellar mass we infer is $\log{({M_*}/\msun) = 5.9}$ (estimated by summing the derived masses in \autoref{tab:halo_features}). Using the low-mass stellar-to-halo mass relation derived in \citet{munshi2021}, 
this yields an estimated dark matter halo mass of $\log{({M_h}/\msun) \sim 9.6}
$. The estimate of the halo mass for NGC~300 is $\log{({M_h}/\msun) = 11.28}$ from \citet{mutlupakdil2021}. This suggests a dark matter mass ratio of $\approx1:50$ between the the dark matter mass of the progenitor of the features and dark matter mass of NGC~300, as the remnants we detect represent only a fraction of the progenitor's total mass. Assuming a progenitor with a metallicity of the northern stream ([Fe/H]$=-1.4$), and the mass-metallicity relation of \citet{kirby2013}, we may expect a progenitor mass of $\log{({M_*}/\msun) \sim 7
}$ or $\log{({M_h}/\msun) = 10
}$. This would yield a dark matter mass ratio of 1:15 between the progenitor and NGC~300. Although \citet{munshi2021} adopts the $200\times\rho_{c}$ virial definition rather than the $104\times\rho_{c}$ definition, the uncertainty in the stellar-to-halo mass relation means these mass ratios should be regarded as approximate estimates.

\subsection{Interpreting the Halo Substructures with Prior Studies of NGC~300}
\label{subsection:interpretation}

NGC~300 is a near face-on and near bulge-less disk galaxy \citep{vlajic2009,williams2013}, allowing for the stellar population within the disk to be well-studied. In summary the stellar population of the disk of NGC~300 appears to be relatively old. \citet{gogarten2010} finds 80\% of the stars are older than 6~Gyr, where in the very inner disk $>6$~Gyr stars dominate (90\%) and then decrease radially (40\% at 5.4~kpc). \citet{gogarten2010} and \citet{kang2016} both conclude that significant radial migration of the disk stars is unlikely and that NGC~300 has experienced rapid disk growth from the inside out (i.e., the outer parts of the disk formed a greater fraction of its stars more recently than the inner parts). \citet{hillis2016} comes to a similar conclusion based on the chemical abundance gradient. 

Historically, studies of the disk of NGC~300 ($R\lesssim 25'$) concluded this galaxy contains a nearly perfect exponential disk. This includes analysis of the luminosity profile \citep[$R<10'$;][]{devaucouleurs1962,carignan1985}, integrated light profiles \citep{kim2004, munoz2007, laine2016}, resolved star counts \citep{kim2004, munoz2007, laine2016}, and the young stellar population \citep[$R<18.8'$][]{hillis2016}. Given the uniformity in the optical disk within $10'$/5.8~kpc of the center, processes that affected the outer regions of the galaxy have left little imprint on the very inner region. For example, the disk of NGC~300 is comparable to that of NGC~2403 (another LMC-mass galaxy), with no evident breaks in the young stellar population \citep{hillis2016}. However, NGC~2403 is disrupting a nearby companion dwarf \citep{carlin2019}, DDO~44, so it is possible that these types of interactions may not leave a clear imprint on the inner stellar disk. Beyond the optical disk and expanding out to $R\sim32'$, with resolved star counts in archival HST data, \citet{jang2020} identified two breaks in the stellar density profile at $R\sim10'$ and again at $R\sim14.2'$. A similar break at $R\sim10'$ in UV profile studies is also detected \citep{roussel2005,GildePaz2007}. Breaks in the stellar density profile like this can be attributed to some sort of gravitational perturbation, in particular an accretion event, with similar profile breaks seen in simulations \citep{younger2007,amorisco2017,deason2022} and observations \citep{conroy2023}. This also implies that NGC~300 is not perfectly exponential in its outskirts.

\begin{figure}
    \centering
    \includegraphics[width=0.9\linewidth]{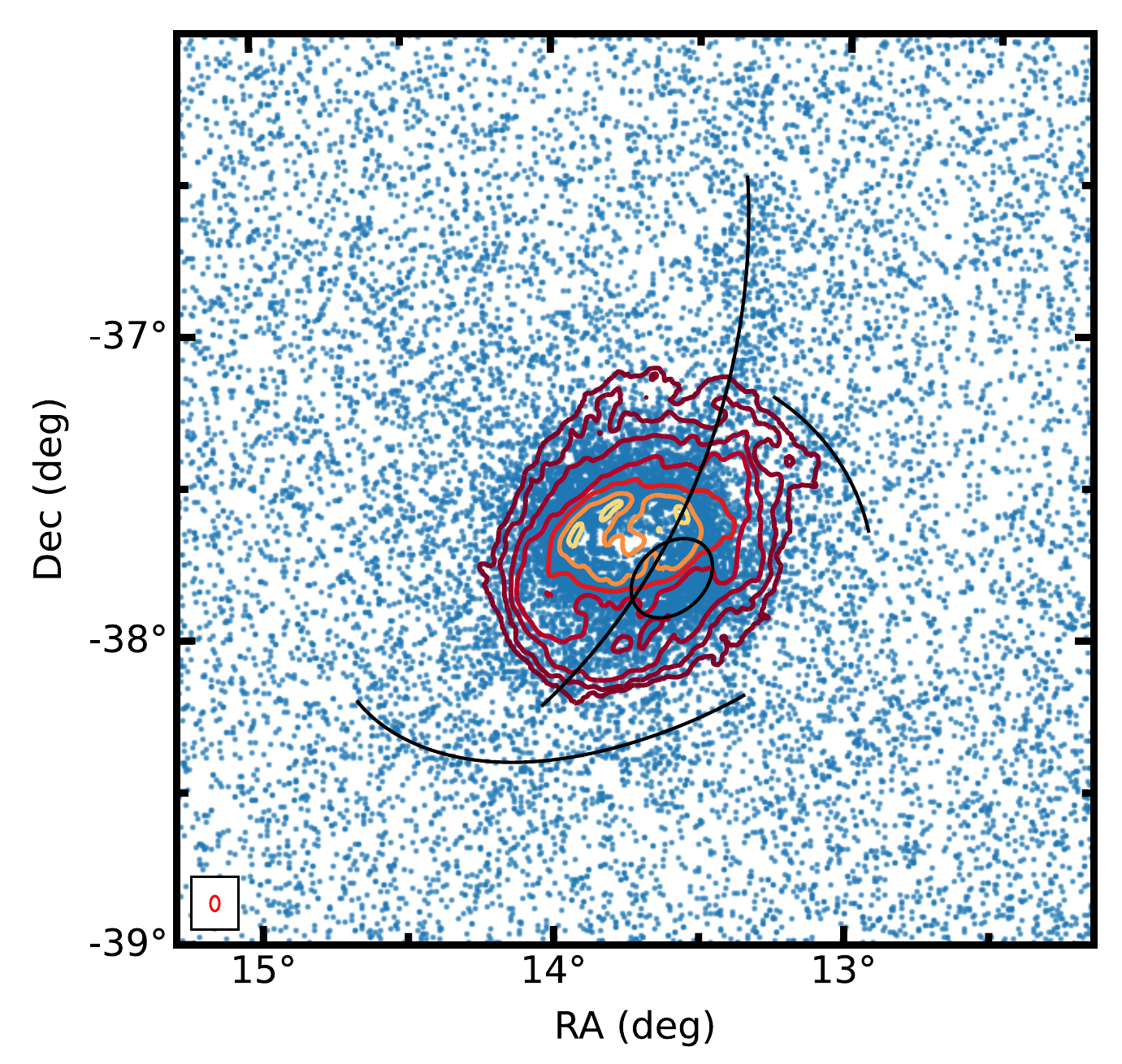}
    \caption{\hi\ column density contours from \citet{westmeier2011} taken with the ATCA overlaid on the same RGB point source map as \autoref{fig:rgb_map}. The beam size is plotted as the red circle in the lower left. The contour levels correspond to 0.1, 0.4, 1, 2, 4, 8 and 12 $\times10^{20}~\rm{cm}^{-2}$.}
    \label{fig:hicontour}
\end{figure}

In the \hi\ study performed by \citet{westmeier2011} with the Australia Telescope Compact Array it was revealed that NGC~300 contains a dense inner gas disk ($R\lesssim 10'$) and an extended outer disk ($10'\lesssim R\lesssim 30'$). The inner \hi\ disk is observed to be well aligned with the optical disk, but the identified outer \hi\ disk has an entirely different orientation angle ($\Delta\theta\approx41$~deg, as seen in \autoref{fig:hicontour}). In this work the authors speculate that the twist in the outer \hi\ disk is suggestive of a tidal interaction (possibly by a dwarf galaxy), which is also seen as a general trend in \hi\ studies of dwarf-dwarf interactions \citep{pearson2016}. This idea was initially inconsistent with the undisturbed stellar disk, but the findings of \citet{jang2020} demonstrate that the stellar density appears affected at a similar radius. We plot the \hi\ contours of \citet{westmeier2011} along with our RGB map in \autoref{fig:hicontour}. Part of the outer \hi\ disk abuts Shell 1, and both Stream N and Stream S follow approximately along the major axis of the outer \hi\ disk. It is possible that the source of the streams and shells around NGC~300 are connected to the distortion of the outer \hi\ disk. 
The LMC-mass NGC~2403 contains \hi\ anomalies attributed to its dwarf companion DDO~44 \citep[][]{veronese2023}, which may be a similar situation to NGC~300.

Metallicity studies of NGC~300 have also been generally limited to the central regions of the galaxy, with the younger ($<6$~Gyr) and more metal rich ([M/H]$\sim-0.5$) stars dominating at $\sim10'$ \citep{gogarten2010}. In the accretion based evolutionary scenario we generally expect lower metallicities than in the disk due to the dominance of accreted stars from metal-poor dwarfs and the lack of sustained star formation. At the outskirts, using an MDF technique similar to ours, but assuming an 8 Gyr population from \citet{vandenberg2006}, \citet{vlajic2009} identified the most metal-poor component at $R \sim 15.5'$ with [Fe/H] $= -1$, followed by an increase in metallicity extending out to $R \sim 25'$ ([Fe/H] $\approx -0.85$). \citet{jang2020} finds similar results within $R\sim10'$ but at larger radii instead finds a decrease in metallicity of [Fe/H] = $-1.6^{+0.2}_{-0.4}$ with a 10~Gyr [$\alpha$/Fe] $= +0.3$ Padova isochrone \citep{bressan2012} and a color to metallicity conversion factor. 

Our metallicity measurements, based on three different selections (Halo E, Halo W, and the Halo Over-density at $R \lesssim 30'$, with Halo E including the \citealt{vlajic2009} pointings; see \autoref{fig:selections}), are consistent with the findings of \citet{vlajic2009}, with each region yielding a mean metallicity of $\langle\rm{[Fe/H]}\rangle = -1.0\pm0.15$. However, we do not detect an increase in metallicity like \citet{vlajic2009}. To compare to \citet{jang2020} we place a selection box around the 3 most distant HST pointings, deriving a metallicity of $\langle\rm{[Fe/H]}\rangle = -1.2 \pm0.15$ (comparable to the Fluff SW box distribution seen in \autoref{fig:MDF} but with a few more metal-poor stars). Although our metallicities in this region show slight tension with \citet{jang2020}, their use of an alpha-enhanced [Fe/H] $= -1.6$ isochrone is equivalent to an un-enhanced [Fe/H] $= -1.4$ isochrone, bringing the results into agreement within the stated uncertainties. Differences in stellar models, combined with the greater depth of HST compared to DECam -- which probes several magnitudes fainter and offers more robust metallicity constraints -- may also contribute to these discrepancies. Regardless of these differences, both \citet{jang2020} and our results demonstrate that the RGB stars outside of $2\times R_{25}=21.8'=12.7$~kpc are more metal-poor and older (10~Gyr) compared to the stellar populations closer to the optical disk (a majority of the stars in the outermost $4.5 < R < 5.4$~kpc bin of \citealt{gogarten2010} are younger than 6~Gyr).

With accretion based evolution, the inner stellar halo typically consists of a mix of radially migrated outer-lying in situ disk stars and accretion remnants, while the outer stellar halo becomes increasingly dominated by accretion debris \citep[e.g.,][]{cooper2010}. Several lines of evidence suggest this is true for NGC~300. Most notably, we have identified a distinctly metal-poor stellar stream extending far from the galaxy, a clear signature of an accretion remnant. We have observed other over-densities within the stellar halo with metallicites consistent with accretion, although their origins are less clear. Additionally, the outer \hi\ disk of NGC~300 is twisted relative to the inner \hi\ disk, which aligns closely with the optical disk. This twist, along with a break in the optical and UV light profiles at similar radii, supports the idea of a dwarf galaxy accretion event. In \autoref{fig:split_rgb_map}, we identify a circular, extensive metal-poor component approximately $\sim15$~kpc from the center, alongside a more central and elliptical, metal-rich component, indicating a mix of both metal-poor and metal-rich stars. Simulations by \citet{kadofong2022} suggest that round stellar halo components in dwarfs can form in situ. However, given the evidence for past accretion in NGC~300 and the mixed metallicities in this region, it is plausible that the smoother inner stellar halo we are detecting may have a mix of displaced in situ stars in addition to accreted stars from either the same accretion event that left the stream wake, or plausibly an even older accretion event. 

\subsection{The metal-poor Globular Cluster in the Halo of NGC~300}
\label{subsec:gc_discussion}

NGC~300-GCF25 is an extended, old, and metal poor GC (see \autoref{subsection:GC}) lying at a relatively large distance from NGC~300 (23.3~kpc). As a result, it is likely that this is an accreted GC and not an in situ one. While the ellipticity of this GC is relatively spherical, its large size could be an indicator that this is perhaps a nuclear star cluster remnant from an accreted dwarf. Similarly, GCF25’s proximity to Shell 1 and 2 may be indicative of an association, but the shell over-densities do not clearly overlap with its position.

Metal-poor GCs are stellar halo tracers in terms of both chemical composition and spatial distribution \citep{brodie2006}. Other than the LMC itself (which has some evidence of past accretion, see e.g., \citealt{mucciarelli2021}), NGC~4449 is the best-studied isolated LMC-analog dwarf galaxy that contains signs of a stellar halo \citep{rys2011} and is clearly interacting with a dwarf galaxy \citep{martinezdelgado2012}. A study of the GC population of NGC~4449 performed by \citet{strader2012} found a sub-population of metal-poor GCs ($-1.6<$[Fe/H]$<-1.1$), comparable to NGC~300-GCF25 with [Fe/H] $\approx -1.6$. In \autoref{fig:metal_lum} we show that the progenitor of the features around NGC~300 may be similar to the progenitor of the NGC~4449 stream, so finding a GC in NGC~300 with a metallicity similar to those within the halo of NGC~4449 is perhaps unsurprising. 
Additionally, NGC~4449's stellar halo also appears to be accretion built, just like NGC~300. A forthcoming systematic study of GCs within the stellar halo region of NGC~300 is a sensible next step, and star clusters associated with any of the stellar substructures would allow for strong mass constraints of NGC~300 \citep[e.g.,][]{pearson2022}. 

\section{Conclusion}
\label{sec:conclusion}

NGC~300 is an isolated, LMC-mass galaxy situated 2.01~Mpc away with $\sim$two plausible satellites within its virial radius\footnote{The satellite populations will be further revealed by a forthcoming systematic search of the DEEP data.} \citep{dalcanton2009,sand2024}. Using deep photometric data (with a average $10\sigma$ magnitude depth of $g=25.7$ and $i=24.4$) out to the virial radius of NGC~300 obtained with DECam on the Blanco Telescope as part of the DELVE-DEEP survey, our resolved RGB star search has revealed the presence of several stellar features beyond the disk of the galaxy. We find:
\begin{enumerate}
    \item A stellar stream extending north of NGC~300 $\sim40$~kpc from the galactic center (Stream N), with RGB stars notably more metal-poor ($\langle\rm{[Fe/H]}\rangle = -1.4\pm0.15$) than stars residing in the disk and  our proposed inner stellar halo ([Fe/H] $> -1$). We conclude that this stream is likely originated from an accretion event.    
    \item A smaller radial protrusion $\sim180^{\circ}$ opposite of the northern stream (Stream S). This feature is also metal-poor ($\langle\rm{[Fe/H]}\rangle = -1.2\pm0.15$), but the proximity of this feature to the galaxy complicates interpretation. We link this to the northern stream as a plausible stream wrap, but additional follow-up is needed to verify this conclusion as a similar feature may be able to form in situ.
    \item Two shell-like structures along the western and southern sides of the galaxy (Shell 1 and Shell 2). These features are also more metal-poor than the disk of NGC~300 ($\langle\rm{[Fe/H]}\rangle = -1.2\pm0.15$ and $\langle\rm{[Fe/H]}\rangle = -1.3\pm0.15$ respectively), but the low surface brightness of these shell features complicates interpretation. We again favor that the shells are connected to an accretion event given their metallicities, but we cannot completely rule out in situ formation.
    \item A metal-poor ([Fe/H] $\approx-1.6\pm0.6$) and old ($\sim10$~Gyr) halo GC (NGC~300-GCF25), falling between the shell features radially in projection. Dwarf galaxies with old halo GCs also show evidence of accretion (including these outer-lying metal-poor GCs), meaning that this metal-poor halo GC may have also been accreted.  
\end{enumerate}
The currently known possible satellites of NGC~300 are much further out from the galaxy than these stellar halo features, with no current detected relation between the two.

The evolutionary history of NGC~300 is far more complex than initially thought. NGC~300 was once considered the quintessential example of a pure exponential disk with no stellar halo component at all \citep[e.g.,][]{devaucouleurs1962,carignan1985,blandhawthorn2005,vlajic2009}. It has an un-disturbed optical disk and it contains a stellar population consistent with classic inside-out growth \citep{gogarten2010,hillis2016,kang2016}. However, outside of the optical disk, a new picture is emerging. NGC~300 has been found to have breaks in its optical outer stellar density profile by \citet{jang2020} with comparable findings in the UV \citep{roussel2005,GildePaz2007}, and \citet{westmeier2011} finds that it contains a twisted outer \hi\ disk starting at a similar radius. Our discovery of streams and shells outside of the disk of NGC~300 explain these discrepancies with an accretion event, where the halo features align well with the extent of the disturbed \hi\ disk. 

Our data is most suggestive of a single accreted progenitor for the over-densities, plausibly Fornax-like but certainly comparable to a dwarf spheroidal galaxy of M31 or the Milky Way ($M_{V}<-9.5$), with a mass ratio of approximately 1:15 with NGC~300. However, presently we cannot definitively rule out multiple progenitors of the features, and it is possible that some of the features may have an in situ origin).
Future targeted high resolution studies of the features around NGC~300 can enable more discrete measurements of the stellar populations, providing better constraints on the origins of these features.

NGC~300 is a strong piece of the accumulating evidence of the importance of accretion in galactic assembly at dwarf mass scales, with our results suggesting that significant accretion can occur at the LMC mass scale. 
Yet the question remains on whether what we have observed around NGC~300 is rare, or expected. Theoretical models by \citet{deason2022} predict that galaxies in dwarf-mass dark matter halos ($\sim10^{10}$~\msun\ at $z=0$, about an order of magnitude lower than the mass of the LMC and NGC~300) experience $N=1.6$ major mergers and $N=7$ minor mergers over their lifetime in $\Lambda$CDM. However, depending on the galaxy formation model used, the contribution of accreted stars to stellar halo formation varies substantially in these models and intermediate mass mergers appear to dominate in contribution to the stellar halo. Answering whether dwarf galaxies host accreted or in situ stellar halos (or a mix of the two) likely depends strongly on both their formation histories and the types of mergers the galaxy experienced.

In this work we have demonstrated the power of wide and deep searches around dwarf galaxies in revealing the faint tracers of accretion. Narrow-field studies of NGC~300 led to a misleading conclusion of a galaxy with a classical exponential disk presumed to be reflective of a quiescent evolutionary history. To fully assess the assembly histories of dwarf galaxies observationally, wide and deep imaging is necessary, which can then be followed-up with narrow field imaging for precision measurements of interesting features. For example, now that we have identified the locations of features of note around NGC~300, follow-up pointings performed with HST or JWST will enable detection of more fine metallicity differences between the features and even be able to discern metallicity and/or distance gradients. A wide-field search combined with narrow-field follow-up is a promising strategy in the forthcoming era of Euclid \citep{euclid}, LSST \citep{lsst} and Roman \citep{roman}, where sufficiently deep, wide-field imaging will be obtained for a larger number of dwarf galaxies. It is this two-pronged approach that will enable us to understand the stellar halos of dwarf galaxies. 


\begin{acknowledgments}
DJS and the Arizona team acknowledges support from NSF grant AST-2205863.
WC gratefully acknowledges support from a Gruber Science Fellowship at Yale University. This material is based upon work supported by the National Science Foundation Graduate Research Fellowship Program under Grant No. DGE2139841. Any opinions, findings, and conclusions or recommendations expressed in this material are those of the author(s) and do not necessarily reflect the views of the National Science Foundation.
SP acknowledges support by a research grant (VIL53081) from VILLUM FONDEN.

This project used data obtained with the Dark Energy Camera (DECam), which was constructed by the Dark Energy Survey (DES) collaboration. Funding for the DES Projects has been provided by the DOE and NSF (USA), MISE (Spain), STFC (UK), HEFCE (UK), NCSA (UIUC), KICP (U. Chicago), CCAPP (Ohio State), MIFPA (Texas A\&M), CNPQ, FAPERJ, FINEP (Brazil), MINECO (Spain), DFG (Germany) and the Collaborating Institutions in the Dark Energy Survey, which are Argonne Lab, UC Santa Cruz, University of Cambridge, CIEMAT-Madrid, University of Chicago, University College London, DES-Brazil Consortium, University of Edinburgh, ETH Zürich, Fermilab, University of Illinois, ICE (IEEC-CSIC), IFAE Barcelona, Lawrence Berkeley Lab, LMU München and the associated Excellence Cluster Universe, University of Michigan, NOIRLab, University of Nottingham, Ohio State University, OzDES Membership Consortium, University of Pennsylvania, University of Portsmouth, SLAC National Lab, Stanford University, University of Sussex, and Texas A\&M University.

The DELVE project is partially supported by Fermilab LDRD project L2019-011, the NASA Fermi Guest Investigator Program Cycle 9 grant 91201, and the U.S. National Science Foundation (NSF) under grants AST2108168, AST-2307126, and AST-2108169.

This manuscript has been authored by Fermi Research Alliance, LLC under Contract No. DE-AC02-07CH11359 with the U.S. Department of Energy, Office of Science, Office of High Energy Physics. The United States Government retains and the publisher, by accepting the article for publication, acknowledges that the United States Government retains a non-exclusive, paid-up, irrevocable, world-wide license to publish or reproduce the published form of this manuscript, or allow others to do so, for United States Government purposes.

\end{acknowledgments}

%

\vspace{5mm}

\facilities{DECam, SOAR}


\software{Astropy \citep{astropy2013,astropy2018,astropy2022}, 
          Numpy \citep{numpy},
          Scipy \citep{2020SciPy-NMeth,scipy_14593523},
          Pandas \citep{pandas_13819579},
          Photutils \citep{bradley2022}, 
          Iraf \citep{iraf}
          }




\bibliography{refs}{}

\begin{thebibliography}{}
\expandafter\ifx\csname natexlab\endcsname\relax\def\natexlab#1{#1}\fi
\providecommand{\url}[1]{\href{#1}{#1}}
\providecommand{\dodoi}[1]{doi:~\href{http://doi.org/#1}{\nolinkurl{#1}}}
\providecommand{\doeprint}[1]{\href{http://ascl.net/#1}{\nolinkurl{http://ascl.net/#1}}}
\providecommand{\doarXiv}[1]{\href{https://arxiv.org/abs/#1}{\nolinkurl{https://arxiv.org/abs/#1}}}

\bibitem[{{Abbott} {et~al.}(2021){Abbott}, {Adam{\'o}w}, {Aguena}, {Allam}, {Amon}, {Annis}, {Avila}, {Bacon}, {Banerji}, {Bechtol}, {Becker}, {Bernstein}, {Bertin}, {Bhargava}, {Bridle}, {Brooks}, {Burke}, {Carnero Rosell}, {Carrasco Kind}, {Carretero}, {Castander}, {Cawthon}, {Chang}, {Choi}, {Conselice}, {Costanzi}, {Crocce}, {da Costa}, {Davis}, {De Vicente}, {DeRose}, {Desai}, {Diehl}, {Dietrich}, {Drlica-Wagner}, {Eckert}, {Elvin-Poole}, {Everett}, {Evrard}, {Ferrero}, {Fert{\'e}}, {Flaugher}, {Fosalba}, {Friedel}, {Frieman}, {Garc{\'\i}a-Bellido}, {Gaztanaga}, {Gelman}, {Gerdes}, {Giannantonio}, {Gill}, {Gruen}, {Gruendl}, {Gschwend}, {Gutierrez}, {Hartley}, {Hinton}, {Hollowood}, {Honscheid}, {Huterer}, {James}, {Jeltema}, {Johnson}, {Kent}, {Kron}, {Kuehn}, {Kuropatkin}, {Lahav}, {Li}, {Lidman}, {Lin}, {MacCrann}, {Maia}, {Manning}, {Maloney}, {March}, {Marshall}, {Martini}, {Melchior}, {Menanteau}, {Miquel}, {Morgan}, {Myles}, {Neilsen}, {Ogando}, {Palmese}, {Paz-Chinch{\'o}n}, {Petravick},
  {Pieres}, {Plazas}, {Pond}, {Rodriguez-Monroy}, {Romer}, {Roodman}, {Rykoff}, {Sako}, {Sanchez}, {Santiago}, {Scarpine}, {Serrano}, {Sevilla-Noarbe}, {Smith}, {Smith}, {Soares-Santos}, {Suchyta}, {Swanson}, {Tarle}, {Thomas}, {To}, {Tremblay}, {Troxel}, {Tucker}, {Turner}, {Varga}, {Walker}, {Wechsler}, {Weller}, {Wester}, {Wilkinson}, {Yanny}, {Zhang}, {Nikutta}, {Fitzpatrick}, {Jacques}, {Scott}, {Olsen}, {Huang}, {Herrera}, {Juneau}, {Nidever}, {Weaver}, {Adean}, {Correia}, {de Freitas}, {Freitas}, {Singulani}, {Vila-Verde}, \& {Linea Science Server}}]{abbott2021}
{Abbott}, T.~M.~C., {Adam{\'o}w}, M., {Aguena}, M., {et~al.} 2021, \apjs, 255, 20, \dodoi{10.3847/1538-4365/ac00b3}

\bibitem[{{Akeson} {et~al.}(2019){Akeson}, {Armus}, {Bachelet}, {Bailey}, {Bartusek}, {Bellini}, {Benford}, {Bennett}, {Bhattacharya}, {Bohlin}, {Boyer}, {Bozza}, {Bryden}, {Calchi Novati}, {Carpenter}, {Casertano}, {Choi}, {Content}, {Dayal}, {Dressler}, {Dor{\'e}}, {Fall}, {Fan}, {Fang}, {Filippenko}, {Finkelstein}, {Foley}, {Furlanetto}, {Kalirai}, {Gaudi}, {Gilbert}, {Girard}, {Grady}, {Greene}, {Guhathakurta}, {Heinrich}, {Hemmati}, {Hendel}, {Henderson}, {Henning}, {Hirata}, {Ho}, {Huff}, {Hutter}, {Jansen}, {Jha}, {Johnson}, {Jones}, {Kasdin}, {Kelly}, {Kirshner}, {Koekemoer}, {Kruk}, {Lewis}, {Macintosh}, {Madau}, {Malhotra}, {Mandel}, {Massara}, {Masters}, {McEnery}, {McQuinn}, {Melchior}, {Melton}, {Mennesson}, {Peeples}, {Penny}, {Perlmutter}, {Pisani}, {Plazas}, {Poleski}, {Postman}, {Ranc}, {Rauscher}, {Rest}, {Roberge}, {Robertson}, {Rodney}, {Rhoads}, {Rhodes}, {Ryan}, {Sahu}, {Sand}, {Scolnic}, {Seth}, {Shvartzvald}, {Siellez}, {Smith}, {Spergel}, {Stassun}, {Street}, {Strolger}, {Szalay},
  {Trauger}, {Troxel}, {Turnbull}, {van der Marel}, {von der Linden}, {Wang}, {Weinberg}, {Williams}, {Windhorst}, {Wollack}, {Wu}, {Yee}, \& {Zimmerman}}]{roman}
{Akeson}, R., {Armus}, L., {Bachelet}, E., {et~al.} 2019, arXiv e-prints, arXiv:1902.05569, \dodoi{10.48550/arXiv.1902.05569}

\bibitem[{{Amorisco}(2017)}]{amorisco2017}
{Amorisco}, N.~C. 2017, \mnras, 464, 2882, \dodoi{10.1093/mnras/stw2229}

\bibitem[{{Annibali} {et~al.}(2020){Annibali}, {Beccari}, {Bellazzini}, {Tosi}, {Cusano}, {Paris}, {Cignoni}, {Ciotti}, {Nipoti}, \& {Sacchi}}]{annibali2020}
{Annibali}, F., {Beccari}, G., {Bellazzini}, M., {et~al.} 2020, \mnras, 491, 5101, \dodoi{10.1093/mnras/stz3185}

\bibitem[{{Astropy Collaboration} {et~al.}(2013){Astropy Collaboration}, {Robitaille}, {Tollerud}, {Greenfield}, {Droettboom}, {Bray}, {Aldcroft}, {Davis}, {Ginsburg}, {Price-Whelan}, {Kerzendorf}, {Conley}, {Crighton}, {Barbary}, {Muna}, {Ferguson}, {Grollier}, {Parikh}, {Nair}, {Unther}, {Deil}, {Woillez}, {Conseil}, {Kramer}, {Turner}, {Singer}, {Fox}, {Weaver}, {Zabalza}, {Edwards}, {Azalee Bostroem}, {Burke}, {Casey}, {Crawford}, {Dencheva}, {Ely}, {Jenness}, {Labrie}, {Lim}, {Pierfederici}, {Pontzen}, {Ptak}, {Refsdal}, {Servillat}, \& {Streicher}}]{astropy2013}
{Astropy Collaboration}, {Robitaille}, T.~P., {Tollerud}, E.~J., {et~al.} 2013, \aap, 558, A33, \dodoi{10.1051/0004-6361/201322068}

\bibitem[{{Astropy Collaboration} {et~al.}(2018){Astropy Collaboration}, {Price-Whelan}, {Sip{\H{o}}cz}, {G{\"u}nther}, {Lim}, {Crawford}, {Conseil}, {Shupe}, {Craig}, {Dencheva}, {Ginsburg}, {Vand erPlas}, {Bradley}, {P{\'e}rez-Su{\'a}rez}, {de Val-Borro}, {Aldcroft}, {Cruz}, {Robitaille}, {Tollerud}, {Ardelean}, {Babej}, {Bach}, {Bachetti}, {Bakanov}, {Bamford}, {Barentsen}, {Barmby}, {Baumbach}, {Berry}, {Biscani}, {Boquien}, {Bostroem}, {Bouma}, {Brammer}, {Bray}, {Breytenbach}, {Buddelmeijer}, {Burke}, {Calderone}, {Cano Rodr{\'\i}guez}, {Cara}, {Cardoso}, {Cheedella}, {Copin}, {Corrales}, {Crichton}, {D'Avella}, {Deil}, {Depagne}, {Dietrich}, {Donath}, {Droettboom}, {Earl}, {Erben}, {Fabbro}, {Ferreira}, {Finethy}, {Fox}, {Garrison}, {Gibbons}, {Goldstein}, {Gommers}, {Greco}, {Greenfield}, {Groener}, {Grollier}, {Hagen}, {Hirst}, {Homeier}, {Horton}, {Hosseinzadeh}, {Hu}, {Hunkeler}, {Ivezi{\'c}}, {Jain}, {Jenness}, {Kanarek}, {Kendrew}, {Kern}, {Kerzendorf}, {Khvalko}, {King}, {Kirkby}, {Kulkarni},
  {Kumar}, {Lee}, {Lenz}, {Littlefair}, {Ma}, {Macleod}, {Mastropietro}, {McCully}, {Montagnac}, {Morris}, {Mueller}, {Mumford}, {Muna}, {Murphy}, {Nelson}, {Nguyen}, {Ninan}, {N{\"o}the}, {Ogaz}, {Oh}, {Parejko}, {Parley}, {Pascual}, {Patil}, {Patil}, {Plunkett}, {Prochaska}, {Rastogi}, {Reddy Janga}, {Sabater}, {Sakurikar}, {Seifert}, {Sherbert}, {Sherwood-Taylor}, {Shih}, {Sick}, {Silbiger}, {Singanamalla}, {Singer}, {Sladen}, {Sooley}, {Sornarajah}, {Streicher}, {Teuben}, {Thomas}, {Tremblay}, {Turner}, {Terr{\'o}n}, {van Kerkwijk}, {de la Vega}, {Watkins}, {Weaver}, {Whitmore}, {Woillez}, {Zabalza}, \& {Astropy Contributors}}]{astropy2018}
{Astropy Collaboration}, {Price-Whelan}, A.~M., {Sip{\H{o}}cz}, B.~M., {et~al.} 2018, \aj, 156, 123, \dodoi{10.3847/1538-3881/aabc4f}

\bibitem[{{Astropy Collaboration} {et~al.}(2022){Astropy Collaboration}, {Price-Whelan}, {Lim}, {Earl}, {Starkman}, {Bradley}, {Shupe}, {Patil}, {Corrales}, {Brasseur}, {N{"o}the}, {Donath}, {Tollerud}, {Morris}, {Ginsburg}, {Vaher}, {Weaver}, {Tocknell}, {Jamieson}, {van Kerkwijk}, {Robitaille}, {Merry}, {Bachetti}, {G{"u}nther}, {Aldcroft}, {Alvarado-Montes}, {Archibald}, {B{'o}di}, {Bapat}, {Barentsen}, {Baz{'a}n}, {Biswas}, {Boquien}, {Burke}, {Cara}, {Cara}, {Conroy}, {Conseil}, {Craig}, {Cross}, {Cruz}, {D'Eugenio}, {Dencheva}, {Devillepoix}, {Dietrich}, {Eigenbrot}, {Erben}, {Ferreira}, {Foreman-Mackey}, {Fox}, {Freij}, {Garg}, {Geda}, {Glattly}, {Gondhalekar}, {Gordon}, {Grant}, {Greenfield}, {Groener}, {Guest}, {Gurovich}, {Handberg}, {Hart}, {Hatfield-Dodds}, {Homeier}, {Hosseinzadeh}, {Jenness}, {Jones}, {Joseph}, {Kalmbach}, {Karamehmetoglu}, {Ka{l}uszy{'n}ski}, {Kelley}, {Kern}, {Kerzendorf}, {Koch}, {Kulumani}, {Lee}, {Ly}, {Ma}, {MacBride}, {Maljaars}, {Muna}, {Murphy}, {Norman}, {O'Steen},
  {Oman}, {Pacifici}, {Pascual}, {Pascual-Granado}, {Patil}, {Perren}, {Pickering}, {Rastogi}, {Roulston}, {Ryan}, {Rykoff}, {Sabater}, {Sakurikar}, {Salgado}, {Sanghi}, {Saunders}, {Savchenko}, {Schwardt}, {Seifert-Eckert}, {Shih}, {Jain}, {Shukla}, {Sick}, {Simpson}, {Singanamalla}, {Singer}, {Singhal}, {Sinha}, {Sip{H{o}}cz}, {Spitler}, {Stansby}, {Streicher}, {{{S}}umak}, {Swinbank}, {Taranu}, {Tewary}, {Tremblay}, {Val-Borro}, {Van Kooten}, {Vasovi{'c}}, {Verma}, {de Miranda Cardoso}, {Williams}, {Wilson}, {Winkel}, {Wood-Vasey}, {Xue}, {Yoachim}, {Zhang}, {Zonca}, \& {Astropy Project Contributors}}]{astropy2022}
{Astropy Collaboration}, {Price-Whelan}, A.~M., {Lim}, P.~L., {et~al.} 2022, apj, 935, 167, \dodoi{10.3847/1538-4357/ac7c74}

\bibitem[{{Baumgardt} {et~al.}(2020){Baumgardt}, {Sollima}, \& {Hilker}}]{baumgardt2020}
{Baumgardt}, H., {Sollima}, A., \& {Hilker}, M. 2020, \pasa, 37, e046, \dodoi{10.1017/pasa.2020.38}

\bibitem[{{Baumgardt} \& {Vasiliev}(2021)}]{baumgardt2021}
{Baumgardt}, H., \& {Vasiliev}, E. 2021, \mnras, 505, 5957, \dodoi{10.1093/mnras/stab1474}

\bibitem[{{Beasley} {et~al.}(2019){Beasley}, {Leaman}, {Gallart}, {Larsen}, {Battaglia}, {Monelli}, \& {Pedreros}}]{beasley2019}
{Beasley}, M.~A., {Leaman}, R., {Gallart}, C., {et~al.} 2019, \mnras, 487, 1986, \dodoi{10.1093/mnras/stz1349}

\bibitem[{{Belokurov} {et~al.}(2017){Belokurov}, {Erkal}, {Deason}, {Koposov}, {De Angeli}, {Evans}, {Fraternali}, \& {Mackey}}]{belokurov2017}
{Belokurov}, V., {Erkal}, D., {Deason}, A.~J., {et~al.} 2017, \mnras, 466, 4711, \dodoi{10.1093/mnras/stw3357}

\bibitem[{{Bland-Hawthorn} {et~al.}(2005){Bland-Hawthorn}, {Vlaji{\'c}}, {Freeman}, \& {Draine}}]{blandhawthorn2005}
{Bland-Hawthorn}, J., {Vlaji{\'c}}, M., {Freeman}, K.~C., \& {Draine}, B.~T. 2005, \apj, 629, 239, \dodoi{10.1086/430512}

\bibitem[{{Bradley} {et~al.}(2022){Bradley}, {Sip{\H{o}}cz}, {Robitaille}, {Tollerud}, {Vin{\'\i}cius}, {Deil}, {Barbary}, {Wilson}, {Busko}, {Donath}, {G{\"u}nther}, {Cara}, {Lim}, {Me{\ss}linger}, {Conseil}, {Bostroem}, {Droettboom}, {Bray}, {Andersen Bratholm}, {Barentsen}, {Craig}, {Rathi}, {Pascual}, {Perren}, {Georgiev}, {De Val-Borro}, {Kerzendorf}, {Bach}, {Quint}, \& {Souchereau}}]{bradley2022}
{Bradley}, L., {Sip{\H{o}}cz}, B., {Robitaille}, T., {et~al.} 2022, {astropy/photutils: 1.5.0}, 1.5.0,  Zenodo, \dodoi{10.5281/zenodo.6825092}

\bibitem[{{Bressan} {et~al.}(2012){Bressan}, {Marigo}, {Girardi}, {Salasnich}, {Dal Cero}, {Rubele}, \& {Nanni}}]{bressan2012}
{Bressan}, A., {Marigo}, P., {Girardi}, L., {et~al.} 2012, \mnras, 427, 127, \dodoi{10.1111/j.1365-2966.2012.21948.x}

\bibitem[{{Brodie} \& {Strader}(2006)}]{brodie2006}
{Brodie}, J.~P., \& {Strader}, J. 2006, \araa, 44, 193, \dodoi{10.1146/annurev.astro.44.051905.092441}

\bibitem[{{Bryan} \& {Norman}(1998)}]{bryan1998}
{Bryan}, G.~L., \& {Norman}, M.~L. 1998, \apj, 495, 80, \dodoi{10.1086/305262}

\bibitem[{{Bullock} \& {Johnston}(2005)}]{bullock2005}
{Bullock}, J.~S., \& {Johnston}, K.~V. 2005, \apj, 635, 931, \dodoi{10.1086/497422}

\bibitem[{{Butler} {et~al.}(2004){Butler}, {Mart{\'\i}nez-Delgado}, \& {Brandner}}]{butler2004}
{Butler}, D.~J., {Mart{\'\i}nez-Delgado}, D., \& {Brandner}, W. 2004, \aj, 127, 1472, \dodoi{10.1086/381922}

\bibitem[{{Carignan}(1985)}]{carignan1985}
{Carignan}, C. 1985, \apjs, 58, 107, \dodoi{10.1086/191031}

\bibitem[{{Carlin} {et~al.}(2016){Carlin}, {Sand}, {Price}, {Willman}, {Karunakaran}, {Spekkens}, {Bell}, {Brodie}, {Crnojevi{\'c}}, {Forbes}, {Hargis}, {Kirby}, {Lupton}, {Peter}, {Romanowsky}, \& {Strader}}]{carlin2016}
{Carlin}, J.~L., {Sand}, D.~J., {Price}, P., {et~al.} 2016, \apjl, 828, L5, \dodoi{10.3847/2041-8205/828/1/L5}

\bibitem[{{Carlin} {et~al.}(2019){Carlin}, {Garling}, {Peter}, {Crnojevi{\'c}}, {Forbes}, {Hargis}, {Mutlu-Pakdil}, {Pucha}, {Romanowsky}, {Sand}, {Spekkens}, {Strader}, \& {Willman}}]{carlin2019}
{Carlin}, J.~L., {Garling}, C.~T., {Peter}, A. H.~G., {et~al.} 2019, \apj, 886, 109, \dodoi{10.3847/1538-4357/ab4c32}

\bibitem[{{Carlin} {et~al.}(2024){Carlin}, {Sand}, {Mutlu-Pakdil}, {Crnojevi{\'c}}, {Doliva-Dolinsky}, {Garling}, {Peter}, {Brodie}, {Forbes}, {Hargis}, {Romanowsky}, {Spekkens}, {Strader}, \& {Willman}}]{carlin2024}
{Carlin}, J.~L., {Sand}, D.~J., {Mutlu-Pakdil}, B., {et~al.} 2024, \apj, 977, 112, \dodoi{10.3847/1538-4357/ad8dcd}

\bibitem[{Chen {et~al.}(2015)Chen, Bressan, Girardi, Marigo, Kong, \& Lanza}]{chen2015}
Chen, Y., Bressan, A., Girardi, L., {et~al.} 2015, Monthly Notices of the Royal Astronomical Society, 452, 1068, \dodoi{10.1093/mnras/stv1281}

\bibitem[{{Chen} {et~al.}(2014){Chen}, {Girardi}, {Bressan}, {Marigo}, {Barbieri}, \& {Kong}}]{chen2014}
{Chen}, Y., {Girardi}, L., {Bressan}, A., {et~al.} 2014, \mnras, 444, 2525, \dodoi{10.1093/mnras/stu1605}

\bibitem[{{Clemens} {et~al.}(2004){Clemens}, {Crain}, \& {Anderson}}]{clemens2004}
{Clemens}, J.~C., {Crain}, J.~A., \& {Anderson}, R. 2004, in Society of Photo-Optical Instrumentation Engineers (SPIE) Conference Series, Vol. 5492, Ground-based Instrumentation for Astronomy, ed. A.~F.~M. {Moorwood} \& M.~{Iye}, 331--340, \dodoi{10.1117/12.550069}

\bibitem[{{Collins} \& {Read}(2022)}]{collins2022}
{Collins}, M. L.~M., \& {Read}, J.~I. 2022, Nature Astronomy, 6, 647, \dodoi{10.1038/s41550-022-01657-4}

\bibitem[{{Collins} {et~al.}(2020){Collins}, {Tollerud}, {Rich}, {Ibata}, {Martin}, {Chapman}, {Gilbert}, \& {Preston}}]{collins2020}
{Collins}, M. L.~M., {Tollerud}, E.~J., {Rich}, R.~M., {et~al.} 2020, \mnras, 491, 3496, \dodoi{10.1093/mnras/stz3252}

\bibitem[{{Collins} {et~al.}(2013){Collins}, {Chapman}, {Rich}, {Ibata}, {Martin}, {Irwin}, {Bate}, {Lewis}, {Pe{\~n}arrubia}, {Arimoto}, {Casey}, {Ferguson}, {Koch}, {McConnachie}, \& {Tanvir}}]{collins2013}
{Collins}, M. L.~M., {Chapman}, S.~C., {Rich}, R.~M., {et~al.} 2013, \apj, 768, 172, \dodoi{10.1088/0004-637X/768/2/172}

\bibitem[{{Collins} {et~al.}(2021){Collins}, {Read}, {Ibata}, {Rich}, {Martin}, {Pe{\~n}arrubia}, {Chapman}, {Tollerud}, \& {Weisz}}]{collins2021}
{Collins}, M. L.~M., {Read}, J.~I., {Ibata}, R.~A., {et~al.} 2021, \mnras, 505, 5686, \dodoi{10.1093/mnras/stab1624}

\bibitem[{{Conroy} {et~al.}(2023){Conroy}, {Johnson}, {van Dokkum}, {Deason}, {Tacchella}, {Belli}, {Bowman}, {Naidu}, {Park}, {Abraham}, \& {Emami}}]{conroy2023}
{Conroy}, C., {Johnson}, B.~D., {van Dokkum}, P., {et~al.} 2023, arXiv e-prints, arXiv:2310.13048, \dodoi{10.48550/arXiv.2310.13048}

\bibitem[{Cooper {et~al.}(2010)Cooper, Cole, Frenk, White, Helly, Benson, De~Lucia, Helmi, Jenkins, Navarro, Springel, \& Wang}]{cooper2010}
Cooper, A.~P., Cole, S., Frenk, C.~S., {et~al.} 2010, Monthly Notices of the Royal Astronomical Society, 406, 744, \dodoi{10.1111/j.1365-2966.2010.16740.x}

\bibitem[{{Crnojevi{\'c}} {et~al.}(2016){Crnojevi{\'c}}, {Sand}, {Spekkens}, {Caldwell}, {Guhathakurta}, {McLeod}, {Seth}, {Simon}, {Strader}, \& {Toloba}}]{crnojevic2016}
{Crnojevi{\'c}}, D., {Sand}, D.~J., {Spekkens}, K., {et~al.} 2016, \apj, 823, 19, \dodoi{10.3847/0004-637X/823/1/19}

\bibitem[{{Crnojevi{\'c}} {et~al.}(2019){Crnojevi{\'c}}, {Sand}, {Bennet}, {Pasetto}, {Spekkens}, {Caldwell}, {Guhathakurta}, {McLeod}, {Seth}, {Simon}, {Strader}, \& {Toloba}}]{crnojevic2019}
{Crnojevi{\'c}}, D., {Sand}, D.~J., {Bennet}, P., {et~al.} 2019, \apj, 872, 80, \dodoi{10.3847/1538-4357/aafbe7}

\bibitem[{{Cullinane} {et~al.}(2023){Cullinane}, {Gilbert}, {Guhathakurta}, {Quirk}, {Escala}, {Smercina}, {Williams}, {Tollerud}, {Qu}, \& {McConnell}}]{cullinane2023}
{Cullinane}, L.~R., {Gilbert}, K.~M., {Guhathakurta}, P., {et~al.} 2023, \apj, 958, 157, \dodoi{10.3847/1538-4357/ad003b}

\bibitem[{{Cunningham} {et~al.}(2024){Cunningham}, {Hunt}, {Price-Whelan}, {Johnston}, {Ness}, {Lu}, {Escala}, \& {Stelea}}]{cuningham2024}
{Cunningham}, E.~C., {Hunt}, J. A.~S., {Price-Whelan}, A.~M., {et~al.} 2024, \apj, 963, 95, \dodoi{10.3847/1538-4357/ad187b}

\bibitem[{{Dalcanton} {et~al.}(2009{\natexlab{a}}){Dalcanton}, {Williams}, {Seth}, {Dolphin}, {Holtzman}, {Rosema}, {Skillman}, {Cole}, {Girardi}, {Gogarten}, {Karachentsev}, {Olsen}, {Weisz}, {Christensen}, {Freeman}, {Gilbert}, {Gallart}, {Harris}, {Hodge}, {de Jong}, {Karachentseva}, {Mateo}, {Stetson}, {Tavarez}, {Zaritsky}, {Governato}, \& {Quinn}}]{dalcaton2009}
{Dalcanton}, J.~J., {Williams}, B.~F., {Seth}, A.~C., {et~al.} 2009{\natexlab{a}}, \apjs, 183, 67, \dodoi{10.1088/0067-0049/183/1/67}

\bibitem[{{Dalcanton} {et~al.}(2009{\natexlab{b}}){Dalcanton}, {Williams}, {Seth}, {Dolphin}, {Holtzman}, {Rosema}, {Skillman}, {Cole}, {Girardi}, {Gogarten}, {Karachentsev}, {Olsen}, {Weisz}, {Christensen}, {Freeman}, {Gilbert}, {Gallart}, {Harris}, {Hodge}, {de Jong}, {Karachentseva}, {Mateo}, {Stetson}, {Tavarez}, {Zaritsky}, {Governato}, \& {Quinn}}]{dalcanton2009}
---. 2009{\natexlab{b}}, \apjs, 183, 67, \dodoi{10.1088/0067-0049/183/1/67}

\bibitem[{{de Vaucouleurs} \& {Page}(1962)}]{devaucouleurs1962}
{de Vaucouleurs}, G., \& {Page}, J. 1962, \apj, 136, 107, \dodoi{10.1086/147355}

\bibitem[{{Deason} {et~al.}(2019){Deason}, {Belokurov}, \& {Sanders}}]{deason2019}
{Deason}, A.~J., {Belokurov}, V., \& {Sanders}, J.~L. 2019, \mnras, 490, 3426, \dodoi{10.1093/mnras/stz2793}

\bibitem[{{Deason} {et~al.}(2022){Deason}, {Bose}, {Fattahi}, {Amorisco}, {Hellwing}, \& {Frenk}}]{deason2022}
{Deason}, A.~J., {Bose}, S., {Fattahi}, A., {et~al.} 2022, \mnras, 511, 4044, \dodoi{10.1093/mnras/stab3524}

\bibitem[{{Dey} {et~al.}(2019){Dey}, {Schlegel}, {Lang}, {Blum}, {Burleigh}, {Fan}, {Findlay}, {Finkbeiner}, {Herrera}, {Juneau}, {Landriau}, {Levi}, {McGreer}, {Meisner}, {Myers}, {Moustakas}, {Nugent}, {Patej}, {Schlafly}, {Walker}, {Valdes}, {Weaver}, {Y{\`e}che}, {Zou}, {Zhou}, {Abareshi}, {Abbott}, {Abolfathi}, {Aguilera}, {Alam}, {Allen}, {Alvarez}, {Annis}, {Ansarinejad}, {Aubert}, {Beechert}, {Bell}, {BenZvi}, {Beutler}, {Bielby}, {Bolton}, {Brice{\~n}o}, {Buckley-Geer}, {Butler}, {Calamida}, {Carlberg}, {Carter}, {Casas}, {Castander}, {Choi}, {Comparat}, {Cukanovaite}, {Delubac}, {DeVries}, {Dey}, {Dhungana}, {Dickinson}, {Ding}, {Donaldson}, {Duan}, {Duckworth}, {Eftekharzadeh}, {Eisenstein}, {Etourneau}, {Fagrelius}, {Farihi}, {Fitzpatrick}, {Font-Ribera}, {Fulmer}, {G{\"a}nsicke}, {Gaztanaga}, {George}, {Gerdes}, {Gontcho}, {Gorgoni}, {Green}, {Guy}, {Harmer}, {Hernandez}, {Honscheid}, {Huang}, {James}, {Jannuzi}, {Jiang}, {Joyce}, {Karcher}, {Karkar}, {Kehoe}, {Kneib}, {Kueter-Young}, {Lan},
  {Lauer}, {Le Guillou}, {Le Van Suu}, {Lee}, {Lesser}, {Perreault Levasseur}, {Li}, {Mann}, {Marshall}, {Mart{\'\i}nez-V{\'a}zquez}, {Martini}, {du Mas des Bourboux}, {McManus}, {Meier}, {M{\'e}nard}, {Metcalfe}, {Mu{\~n}oz-Guti{\'e}rrez}, {Najita}, {Napier}, {Narayan}, {Newman}, {Nie}, {Nord}, {Norman}, {Olsen}, {Paat}, {Palanque-Delabrouille}, {Peng}, {Poppett}, {Poremba}, {Prakash}, {Rabinowitz}, {Raichoor}, {Rezaie}, {Robertson}, {Roe}, {Ross}, {Ross}, {Rudnick}, {Safonova}, {Saha}, {S{\'a}nchez}, {Savary}, {Schweiker}, {Scott}, {Seo}, {Shan}, {Silva}, {Slepian}, {Soto}, {Sprayberry}, {Staten}, {Stillman}, {Stupak}, {Summers}, {Sien Tie}, {Tirado}, {Vargas-Maga{\~n}a}, {Vivas}, {Wechsler}, {Williams}, {Yang}, {Yang}, {Yapici}, {Zaritsky}, {Zenteno}, {Zhang}, {Zhang}, {Zhou}, \& {Zhou}}]{dey2019}
{Dey}, A., {Schlegel}, D.~J., {Lang}, D., {et~al.} 2019, \aj, 157, 168, \dodoi{10.3847/1538-3881/ab089d}

\bibitem[{{Dooley} {et~al.}(2017{\natexlab{a}}){Dooley}, {Peter}, {Carlin}, {Frebel}, {Bechtol}, \& {Willman}}]{dooley2017b}
{Dooley}, G.~A., {Peter}, A. H.~G., {Carlin}, J.~L., {et~al.} 2017{\natexlab{a}}, \mnras, 472, 1060, \dodoi{10.1093/mnras/stx2001}

\bibitem[{{Dooley} {et~al.}(2017{\natexlab{b}}){Dooley}, {Peter}, {Yang}, {Willman}, {Griffen}, \& {Frebel}}]{dooley2017}
{Dooley}, G.~A., {Peter}, A. H.~G., {Yang}, T., {et~al.} 2017{\natexlab{b}}, \mnras, 471, 4894, \dodoi{10.1093/mnras/stx1900}

\bibitem[{{Dotter} {et~al.}(2008){Dotter}, {Chaboyer}, {Jevremovi{\'c}}, {Kostov}, {Baron}, \& {Ferguson}}]{dotter2008}
{Dotter}, A., {Chaboyer}, B., {Jevremovi{\'c}}, D., {et~al.} 2008, \apjs, 178, 89, \dodoi{10.1086/589654}

\bibitem[{{Drlica-Wagner} {et~al.}(2021){Drlica-Wagner}, {Carlin}, {Nidever}, {Ferguson}, {Kuropatkin}, {Adam{\'o}w}, {Cerny}, {Choi}, {Esteves}, {Mart{\'\i}nez-V{\'a}zquez}, {Mau}, {Miller}, {Mutlu-Pakdil}, {Neilsen}, {Olsen}, {Pace}, {Riley}, {Sakowska}, {Sand}, {Santana-Silva}, {Tollerud}, {Tucker}, {Vivas}, {Zaborowski}, {Zenteno}, {Abbott}, {Allam}, {Bechtol}, {Bell}, {Bell}, {Bilaji}, {Bom}, {Carballo-Bello}, {Crnojevi{\'c}}, {Cioni}, {Diaz-Ocampo}, {de Boer}, {Erkal}, {Gruendl}, {Hernandez-Lang}, {Hughes}, {James}, {Johnson}, {Li}, {Mao}, {Mart{\'\i}nez-Delgado}, {Massana}, {McNanna}, {Morgan}, {Nadler}, {No{\"e}l}, {Palmese}, {Peter}, {Rykoff}, {S{\'a}nchez}, {Shipp}, {Simon}, {Smercina}, {Soares-Santos}, {Stringfellow}, {Tavangar}, {van der Marel}, {Walker}, {Wechsler}, {Wu}, {Yanny}, {Fitzpatrick}, {Huang}, {Jacques}, {Nikutta}, {Scott}, \& {Astro Data Lab}}]{drlicawagner2021}
{Drlica-Wagner}, A., {Carlin}, J.~L., {Nidever}, D.~L., {et~al.} 2021, \apjs, 256, 2, \dodoi{10.3847/1538-4365/ac079d}

\bibitem[{{Drlica-Wagner} {et~al.}(2022){Drlica-Wagner}, {Ferguson}, {Adam{\'o}w}, {Aguena}, {Allam}, {Andrade-Oliveira}, {Bacon}, {Bechtol}, {Bell}, {Bertin}, {Bilaji}, {Bocquet}, {Bom}, {Brooks}, {Burke}, {Carballo-Bello}, {Carlin}, {Carnero Rosell}, {Carrasco Kind}, {Carretero}, {Castander}, {Cerny}, {Chang}, {Choi}, {Conselice}, {Costanzi}, {Crnojevi{\'c}}, {da Costa}, {de Vicente}, {Desai}, {Esteves}, {Everett}, {Ferrero}, {Fitzpatrick}, {Flaugher}, {Friedel}, {Frieman}, {Garc{\'\i}a-Bellido}, {Gatti}, {Gaztanaga}, {Gerdes}, {Gruen}, {Gruendl}, {Gschwend}, {Hartley}, {Hernandez-Lang}, {Hinton}, {Hollowood}, {Honscheid}, {Hughes}, {Jacques}, {James}, {Johnson}, {Kuehn}, {Kuropatkin}, {Lahav}, {Li}, {Lidman}, {Lin}, {March}, {Marshall}, {Mart{\'\i}nez-Delgado}, {Mart{\'\i}nez-V{\'a}zquez}, {Massana}, {Mau}, {McNanna}, {Melchior}, {Menanteau}, {Miller}, {Miquel}, {Mohr}, {Morgan}, {Mutlu-Pakdil}, {Mu{\~n}oz}, {Neilsen}, {Nidever}, {Nikutta}, {Nilo Castellon}, {No{\"e}l}, {Ogando}, {Olsen}, {Pace},
  {Palmese}, {Paz-Chinch{\'o}n}, {Pereira}, {Pieres}, {Plazas Malag{\'o}n}, {Prat}, {Riley}, {Rodriguez-Monroy}, {Romer}, {Roodman}, {Sako}, {Sakowska}, {Sanchez}, {S{\'a}nchez}, {Sand}, {Santana-Silva}, {Santiago}, {Schubnell}, {Serrano}, {Sevilla-Noarbe}, {Simon}, {Smith}, {Soares-Santos}, {Stringfellow}, {Suchyta}, {Suson}, {Tan}, {Tarle}, {Tavangar}, {Thomas}, {To}, {Tollerud}, {Troxel}, {Tucker}, {Varga}, {Vivas}, {Walker}, {Weller}, {Wilkinson}, {Wu}, {Yanny}, {Zaborowski}, {Zenteno}, {Delve Collaboration}, {Des Collaboration}, \& {Astro Data Lab}}]{drlicawagner2022}
{Drlica-Wagner}, A., {Ferguson}, P.~S., {Adam{\'o}w}, M., {et~al.} 2022, \apjs, 261, 38, \dodoi{10.3847/1538-4365/ac78eb}

\bibitem[{{El-Badry} {et~al.}(2016){El-Badry}, {Wetzel}, {Geha}, {Hopkins}, {Kere{\v{s}}}, {Chan}, \& {Faucher-Gigu{\`e}re}}]{elbadry2016}
{El-Badry}, K., {Wetzel}, A., {Geha}, M., {et~al.} 2016, \apj, 820, 131, \dodoi{10.3847/0004-637X/820/2/131}

\bibitem[{{Euclid Collaboration} {et~al.}(2024){Euclid Collaboration}, {Mellier}, {Abdurro'uf}, {Acevedo Barroso}, {Ach{\'u}carro}, {Adamek}, {Adam}, {Addison}, {Aghanim}, {Aguena}, {Ajani}, {Akrami}, {Al-Bahlawan}, {Alavi}, {Albuquerque}, {Alestas}, {Alguero}, {Allaoui}, {Allen}, {Allevato}, {Alonso-Tetilla}, {Altieri}, {Alvarez-Candal}, {Amara}, {Amendola}, {Amiaux}, {Andika}, {Andreon}, {Andrews}, {Angora}, {Angulo}, {Annibali}, {Anselmi}, {Anselmi}, {Arcari}, {Archidiacono}, {Aric{\`o}}, {Arnaud}, {Arnouts}, {Asgari}, {Asorey}, {Atayde}, {Atek}, {Atrio-Barandela}, {Aubert}, {Aubourg}, {Auphan}, {Auricchio}, {Aussel}, {Aussel}, {Avelino}, {Avgoustidis}, {Avila}, {Awan}, {Azzollini}, {Baccigalupi}, {Bachelet}, {Bacon}, {Baes}, {Bagley}, {Bahr-Kalus}, {Balaguera-Antolinez}, {Balbinot}, {Balcells}, {Baldi}, {Baldry}, {Balestra}, {Ballardini}, {Ballester}, {Balogh}, {Ba{\~n}ados}, {Barbier}, {Bardelli}, {Barreiro}, {Barriere}, {Barros}, {Barthelemy}, {Bartolo}, {Basset}, {Battaglia}, {Battisti}, {Baugh},
  {Baumont}, {Bazzanini}, {Beaulieu}, {Beckmann}, {Belikov}, {Bel}, {Bellagamba}, {Bella}, {Bellini}, {Benabed}, {Bender}, {Benevento}, {Bennett}, {Benson}, {Bergamini}, {Bermejo-Climent}, {Bernardeau}, {Bertacca}, {Berthe}, {Berthier}, {Bethermin}, {Beutler}, {Bevillon}, {Bhargava}, {Bhatawdekar}, {Bisigello}, {Biviano}, {Blake}, {Blanchard}, {Blazek}, {Blot}, {Bosco}, {Bodendorf}, {Boenke}, {B{\"o}hringer}, {Bolzonella}, {Bonchi}, {Bonici}, {Bonino}, {Bonino}, {Bonvin}, {Bon}, {Booth}, {Borgani}, {Borlaff}, {Borsato}, {Bosco}, {Bose}, {Botticella}, {Boucaud}, {Bouche}, {Boucher}, {Boutigny}, {Bouvard}, {Bouy}, {Bowler}, {Bozza}, {Bozzo}, {Branchini}, {Brau-Nogue}, {Brekke}, {Bremer}, {Brescia}, {Breton}, {Brinchmann}, {Brinckmann}, {Brockley-Blatt}, {Brodwin}, {Brouard}, {Brown}, {Bruton}, {Bucko}, {Buddelmeijer}, {Buenadicha}, {Buitrago}, {Burger}, {Burigana}, {Busillo}, {Busonero}, {Cabanac}, {Cabayol-Garcia}, {Cagliari}, {Caillat}, {Caillat}, {Calabrese}, {Calabro}, {Calderone}, {Calura}, {Camacho
  Quevedo}, {Camera}, {Campos}, {Canas-Herrera}, {Candini}, {Cantiello}, {Capobianco}, {Cappellaro}, {Cappelluti}, {Cappi}, {Caputi}, {Cara}, {Carbone}, {Cardone}, {Carella}, {Carlberg}, {Carle}, {Carminati}, {Caro}, {Carrasco}, {Carretero}, {Carrilho}, {Carron Duque}, {Carry}, {Carvalho}, {Carvalho}, {Casas}, {Casas}, {Casenove}, {Casey}, {Cassata}, {Castander}, {Castelao}, {Castellano}, {Castiblanco}, {Castignani}, {Castro}, {Cavet}, {Cavuoti}, {Chabaud}, {Chambers}, {Charles}, {Charlot}, {Chartab}, {Chary}, {Chaumeil}, {Cho}, {Chon}, {Ciancetta}, {Ciliegi}, {Cimatti}, {Cimino}, {Cioni}, {Claydon}, {Cleland}, {Cl{\'e}ment}, {Clements}, {Clerc}, {Clesse}, {Codis}, {Cogato}, {Colbert}, {Cole}, {Coles}, {Collett}, {Collins}, {Colodro-Conde}, {Colombo}, {Combes}, {Conforti}, {Congedo}, {Conseil}, {Conselice}, {Contarini}, {Contini}, {Conversi}, {Cooray}, {Copin}, {Corasaniti}, {Corcho-Caballero}, {Corcione}, {Cordes}, {Corpace}, {Correnti}, {Costanzi}, {Costille}, {Courbin}, {Courcoult Mifsud}, {Courtois},
  {Cousinou}, {Covone}, {Cowell}, {Cragg}, {Cresci}, {Cristiani}, {Crocce}, {Cropper}, {E Crouzet}, {Csizi}, {Cuby}, {Cucchetti}, {Cucciati}, {Cuillandre}, {Cunha}, {Cuozzo}, {Daddi}, {D'Addona}, {Dafonte}, {Dagoneau}, {Dalessandro}, {Dalton}, {D'Amico}, {Dannerbauer}, {Danto}, {Das}, {Da Silva}, {da Silva}, {Daste}, {Davies}, {Davini}, {de Boer}, {Decarli}, {De Caro}, {Degaudenzi}, {Degni}, {de Jong}, {de la Bella}, {de la Torre}, {Delhaise}, {Delley}, {Delucchi}, {De Lucia}, {Denniston}, {De Paolis}, {De Petris}, {Derosa}, {Desai}, {Desjacques}, {Despali}, {Desprez}, {De Vicente-Albendea}, {Deville}, {Dias}, {D{\'\i}az-S{\'a}nchez}, {Diaz}, {Di Domizio}, {Diego}, {Di Ferdinando}, {Di Giorgio}, {Dimauro}, {Dinis}, {Dolag}, {Dolding}, {Dole}, {Dom{\'\i}nguez S{\'a}nchez}, {Dor{\'e}}, {Dournac}, {Douspis}, {Dreihahn}, {Droge}, {Dryer}, {Dubath}, {Duc}, {Ducret}, {Duffy}, {Dufresne}, {Duncan}, {Dupac}, {Duret}, {Durrer}, {Durret}, {Dusini}, {Ealet}, {Eggemeier}, {Eisenhardt}, {Elbaz}, {Elkhashab}, {Ellien},
  {Endicott}, {Enia}, {Erben}, {Escartin Vigo}, {Escoffier}, {Escudero Sanz}, {Essert}, {Ettori}, {Ezziati}, {Fabbian}, {Fabricius}, {Fang}, {Farina}, {Farina}, {Farinelli}, {Farrens}, {Faustini}, {Feltre}, {Ferguson}, {Ferrando}, {Ferrari}, {Ferr{\'e}-Mateu}, {Ferreira}, {Ferreras}, {Ferrero}, {Ferriol}, {Ferruit}, {Filleul}, {Finelli}, {Finkelstein}, {Finoguenov}, {Fiorini}, {Flentge}, {Focardi}, {Fonseca}, {Fontana}, {Fontanot}, {Fornari}, {Fosalba}, {Fossati}, {Fotopoulou}, {Fouchez}, {Fourmanoit}, {Frailis}, {Fraix-Burnet}, {Franceschi}, {Franco}, {Franzetti}, {Freihoefer}, {Frittoli}, {Frugier}, {Frusciante}, {Fumagalli}, {Fumagalli}, {Fumana}, {Fu}, {Gabarra}, {Galeotta}, {Galluccio}, {Ganga}, {Gao}, {Garc{\'\i}a-Bellido}, {Garcia}, {Gardner}, {Garilli}, {Gaspar-Venancio}, {Gasparetto}, {Gautard}, {Gavazzi}, {Gaztanaga}, {Genolet}, {Genova Santos}, {Gentile}, {George}, {Ghaffari}, {Giacomini}, {Gianotti}, {Gibb}, {Gillard}, {Gillis}, {Ginolfi}, {Giocoli}, {Girardi}, {Giri}, {Goh}, {G{\'o}mez-Alvarez},
  {Gonzalez}, {Gonzalez}, {Gonzalez}, {Gouyou Beauchamps}, {Gozaliasl}, {Gracia-Carpio}, {Grandis}, {Granett}, {Granvik}, {Grazian}, {Gregorio}, {Grenet}, {Grillo}, {Grupp}, {Gruppioni}, {Gruppuso}, {Guerbuez}, {Guerrini}, {Guidi}, {Guillard}, {Gutierrez}, {Guttridge}, {Guzzo}, {Gwyn}, {Haapala}, {Haase}, {Haddow}, {Hailey}, {Hall}, {Hall}, {Hamaus}, {Haridasu}, {Harnois-D{\'e}raps}, {Harper}, {Hartley}, {Hasinger}, {Hassani}, {Hatch}, {Haugan}, {H{\"a}u{\ss}ler}, {Heavens}, {Heisenberg}, {Helmi}, {Helou}, {Hemmati}, {Henares}, {Herent}, {Hern{\'a}ndez-Monteagudo}, {Heuberger}, {Hewett}, {Heydenreich}, {Hildebrandt}, {Hirschmann}, {Hjorth}, {Hoar}, {Hoekstra}, {Holland}, {Holliman}, {Holmes}, {Hook}, {Horeau}, {Hormuth}, {Hornstrup}, {Hosseini}, {Hu}, {Hudelot}, {Hudson}, {Huertas-Company}, {Huff}, {Hughes}, {Humphrey}, {Hunt}, {Huynh}, {Ibata}, {Ichikawa}, {Iglesias-Groth}, {Ilbert}, {Ili{\'c}}, {Ingoglia}, {Iodice}, {Israel}, {Israelsson}, {Izzo}, {Jablonka}, {Jackson}, {Jacobson}, {Jafariyazani}, {Jahnke},
  {Jansen}, {Jarvis}, {Jasche}, {Jauzac}, {Jeffrey}, {Jhabvala}, {Jimenez-Teja}, {Jimenez Mu{\~n}oz}, {Joachimi}, {Johansson}, {Joudaki}, {Jullo}, {Kajava}, {Kang}, {Kannawadi}, {Kansal}, {Karagiannis}, {K{\"a}rcher}, {Kashlinsky}, {Kazandjian}, {Keck}, {Keih{\"a}nen}, {Kerins}, {Kermiche}, {Khalil}, {Kiessling}, {Kiiveri}, {Kilbinger}, {Kim}, {King}, {Kirkpatrick}, {Kitching}, {Kluge}, {Knabenhans}, {Knapen}, {Knebe}, {Kneib}, {Kohley}, {Koopmans}, {Koskinen}, {Koulouridis}, {Kou}, {Kov{\'a}cs}, {Kova\{{\v{c}}\}i{\'c}}, {Kowalczyk}, {Koyama}, {Kraljic}, {Krause}, {Kruk}, {Kubik}, {Kuchner}, {Kuijken}, {K{\"u}mmel}, {Kunz}, {Kurki-Suonio}, {Lacasa}, {Lacey}, {La Franca}, {Lagarde}, {Lahav}, {Laigle}, {La Marca}, {La Marle}, {Lamine}, {Lam}, {Lan{\c{c}}on}, {Landt}, {Langer}, {Lapi}, {Larcheveque}, {Larsen}, {Lattanzi}, {Laudisio}, {Laugier}, {Laureijs}, {Lavaux}, {Lawrenson}, {Lazanu}, {Lazeyras}, {Le Boulc'h}, {Le Brun}, {Le Brun}, {Leclercq}, {Lee}, {Le Graet}, {Legrand}, {Leirvik}, {Le Jeune}, {Lembo}, {Le
  Mignant}, {Lepinzan}, {Lepori}, {Lesci}, {Lesgourgues}, {Leuzzi}, {Levi}, {Liaudat}, {Libet}, {Liebing}, {Ligori}, {Lilje}, {Lin}, {Linde}, {Linder}, {Lindholm}, {Linke}, {Li}, {Liu}, {Lloro}, {Lobo}, {Lodieu}, {Lombardi}, {Lombriser}, {Lonare}, {Longo}, {L{\'o}pez-Caniego}, {Lopez Lopez}, {Alvarez}, {Loureiro}, {Loveday}, {Lusso}, {Macias-Perez}, {Maciaszek}, {Magliocchetti}, {Magnard}, {Magnier}, {Magro}, {Mahler}, {Mainetti}, {Maino}, {Maiorano}, {Maiorano}, {Malavasi}, {Mamon}, {Mancini}, {Mandelbaum}, {Manera}, {Manj{\'o}n-Garc{\'\i}a}, {Mannucci}, {Mansutti}, {Manteiga Outeiro}, {Maoli}, {Maraston}, {Marcin}, {Marcos-Arenal}, {Margalef-Bentabol}, {Marggraf}, {Marinucci}, {Marinucci}, {Markovic}, {Marleau}, {Marpaud}, {Martignac}, {Mart{\'\i}n-Fleitas}, {Martin-Moruno}, {Martin}, {Martinelli}, {Martinet}, {Martin}, {Martins}, {Marulli}, {Massari}, {Massey}, {Masters}, {Matarrese}, {Matsuoka}, {Matthew}, {Maughan}, {Mauri}, {Maurin}, {Maurogordato}, {McCarthy}, {McConnachie}, {McCracken}, {McDonald},
  {McEwen}, {McPartland}, {Medinaceli}, {Mehta}, {Mei}, {Melchior}, {Melin}, {M{\'e}nard}, {Mendes}, {Mendez-Abreu}, {Meneghetti}, {Mercurio}, {Merlin}, {Metcalf}, {Meylan}, {Migliaccio}, {Mignoli}, {Miller}, {Miluzio}, {Milvang-Jensen}, {Mimoso}, {Miquel}, {Miyatake}, {Mobasher}, {Mohr}, {Monaco}, {Mongui{\'o}}, {Montoro}, {Mora}, {Moradinezhad Dizgah}, {Moresco}, {Moretti}, {Morgante}, {Morisset}, {Moriya}, {Morris}, {Mortlock}, {Moscardini}, {Mota}, {Moustakas}, {Moutard}, {M{\"u}ller}, {Munari}, {Murphree}, {Murray}, {Murray}, {Musi}, {Nadathur}, {Nagam}, {Nagao}, {Naidoo}, {Nakajima}, {Nally}, {Natoli}, {Navarro-Alsina}, {Navarro Girones}, {Neissner}, {Nersesian}, {Nesseris}, {Nguyen-Kim}, {Nicastro}, {Nichol}, {Nielbock}, {Niemi}, {Nieto}, {Nilsson}, {Noller}, {Norberg}, {Nourizonoz}, {Ntelis}, {Nucita}, {Nugent}, {Nunes}, {Nutma}, {Ocampo}, {Odier}, {Oesch}, {Oguri}, {Magalhaes Oliveira}, {Onoue}, {Oosterbroek}, {Oppizzi}, {Ordenovic}, {Osato}, {Pacaud}, {Pace}, {Padilla}, {Paech}, {Pagano}, {Page},
  {Palazzi}, {Paltani}, {Pamuk}, {Pandolfi}, {Paoletti}, {Paolillo}, {Papaderos}, {Pardede}, {Parimbelli}, {Parmar}, {Partmann}, {Pasian}, {Passalacqua}, {Paterson}, {Patrizii}, {Pattison}, {Paulino-Afonso}, {Paviot}, {Peacock}, {Pearce}, {Pedersen}, {Peel}, {Peletier}, {Pellejero Ibanez}, {Pello}, {Penny}, {Percival}, {Perez-Garrido}, {Perotto}, {Pettorino}, {Pezzotta}, {Pezzuto}, {Philippon}, {Piersanti}, {Pietroni}, {Piga}, {Pilo}, {Pires}, {Pisani}, {Pizzella}, {Pizzuti}, {Plana}, {Polenta}, {Pollack}, {Poncet}, {P{\"o}ntinen}, {Pool}, {Popa}, {Popa}, {Popp}, {Porciani}, {Porth}, {Potter}, {Poulain}, {Pourtsidou}, {Pozzetti}, {Prandoni}, {Pratt}, {Prezelus}, {Prieto}, {Pugno}, {Quai}, {Quilley}, {Racca}, {Raccanelli}, {R{\'a}cz}, {Radinovi{\'c}}, {Radovich}, {Ragagnin}, {Ragnit}, {Raison}, {Ramos-Chernenko}, {Ranc}, {Raylet}, {Rebolo}, {Refregier}, {Reimberg}, {Reiprich}, {Renk}, {Renzi}, {Retre}, {Revaz}, {Reyl{\'e}}, {Reynolds}, {Rhodes}, {Ricci}, {Ricci}, {Riccio}, {Ricken}, {Rissanen}, {Risso}, {Rix},
  {Robin}, {Rocca-Volmerange}, {Rocci}, {Rodenhuis}, {Rodighiero}, {Rodriguez Monroy}, {Rollins}, {Romanello}, {Roman}, {Romelli}, {Romero-Gomez}, {Roncarelli}, {Rosati}, {Rosset}, {Rossetti}, {Roster}, {Rottgering}, {Rozas-Fern{\'a}ndez}, {Ruane}, {Rubino-Martin}, {Rudolph}, {Ruppin}, {Rusholme}, {Sacquegna}, {S{\'a}ez-Casares}, {Saga}, {Saglia}, {Sahl{\'e}n}, {Saifollahi}, {Sakr}, {Salvalaggio}, {Salvaterra}, {Salvati}, {Salvato}, {Salvignol}, {S{\'a}nchez}, {Sanchez}, {Sanders}, {Sapone}, {Saponara}, {Sarpa}, {Sarron}, {Sartori}, {Sassolas}, {Sauniere}, {Sauvage}, {Sawicki}, {Scaramella}, {Scarlata}, {Scharr{\'e}}, {Schaye}, {Schewtschenko}, {Schindler}, {Schinnerer}, {Schirmer}, {Schmidt}, {Schmidt}, {Schmidt}, {Schneider}, {Schneider}, {Schneider}, {Sch{\"o}neberg}, {Schrabback}, {Schultheis}, {Schulz}, {Schwartz}, {Sciotti}, {Scodeggio}, {Scognamiglio}, {Scott}, {Scottez}, {Secroun}, {Sefusatti}, {Seidel}, {Seiffert}, {Sellentin}, {Selwood}, {Semboloni}, {Sereno}, {Serjeant}, {Serrano}, {Shankar},
  {Sharples}, {Short}, {Shulevski}, {Shuntov}, {Sias}, {Sikkema}, {Silvestri}, {Simon}, {Sirignano}, {Sirri}, {Skottfelt}, {Slezak}, {Sluse}, {Smith}, {Smith}, {Smith}, {Smit}, {Soldano}, {Solheim}, {Sorce}, {Sorrenti}, {Soubrie}, {Spinoglio}, {Spurio Mancini}, {Stadel}, {Stagnaro}, {Stanco}, {Stanford}, {Starck}, {Stassi}, {Steinwagner}, {Stern}, {Stone}, {Strada}, {Strafella}, {Stramaccioni}, {Surace}, {Sureau}, {Suyu}, {Swindells}, {Szafraniec}, {Szapudi}, {Taamoli}, {Talia}, {Tallada-Cresp{\'\i}}, {Tanidis}, {Tao}, {Tarr{\'\i}o}, {Tavagnacco}, {Taylor}, {Taylor}, {Taylor}, {Teixeira}, {Tenti}, {Teodoro Idiago}, {Teplitz}, {Tereno}, {Tessore}, {Testa}, {Testera}, {Tewes}, {Teyssier}, {Theret}, {Thizy}, {Thomas}, {Toba}, {Toft}, {Toledo-Moreo}, {Tolstoy}, {Tommasi}, {Torbaniuk}, {Torradeflot}, {Tortora}, {Tosi}, {Tosti}, {Trifoglio}, {Troja}, {Trombetti}, {Tronconi}, {Tsedrik}, {Tsyganov}, {Tucci}, {Tutusaus}, {Uhlemann}, {Ulivi}, {Urbano}, {Vacher}, {Vaillon}, {Valdes}, {Valentijn}, {Valenziano},
  {Valieri}, {Valiviita}, {Van den Broeck}, {Vassallo}, {Vavrek}, {Venemans}, {Venhola}, {Ventura}, {Verdoes Kleijn}, {Vergani}, {Verma}, {Vernizzi}, {Veropalumbo}, {Verza}, {Vescovi}, {Vibert}, {Viel}, {Vielzeuf}, {Viglione}, {Viitanen}, {Villaescusa-Navarro}, {Vinciguerra}, {Visticot}, {Voggel}, {von Wietersheim-Kramsta}, {Vriend}, {Wachter}, {Walmsley}, {Walth}, {Walton}, {Walton}, {Wander}, {Wang}, {Wang}, {Weaver}, {Weller}, {Whalen}, {Wiesmann}, {Wilde}, {Williams}, {Winther}, {Wittje}, {Wong}, {Wright}, {Yankelevich}, {Yeung}, {Youles}, {Yung}, {Zacchei}, {Zalesky}, {Zamorani}, {Zamorano Vitorelli}, {Zanoni Marc}, {Zennaro}, {Zerbi}, {Zinchenko}, {Zoubian}, {Zucca}, \& {Zumalacarregui}}]{euclid}
{Euclid Collaboration}, {Mellier}, Y., {Abdurro'uf}, {et~al.} 2024, arXiv e-prints, arXiv:2405.13491, \dodoi{10.48550/arXiv.2405.13491}

\bibitem[{{Everett} {et~al.}(2022){Everett}, {Yanny}, {Kuropatkin}, {Huff}, {Zhang}, {Myles}, {Masegian}, {Elvin-Poole}, {Allam}, {Bernstein}, {Sevilla-Noarbe}, {Splettstoesser}, {Sheldon}, {Jarvis}, {Amon}, {Harrison}, {Choi}, {Hartley}, {Alarcon}, {S{\'a}nchez}, {Gruen}, {Eckert}, {Prat}, {Tabbutt}, {Busti}, {Becker}, {MacCrann}, {Diehl}, {Tucker}, {Bertin}, {Jeltema}, {Drlica-Wagner}, {Gruendl}, {Bechtol}, {Carnero Rosell}, {Abbott}, {Aguena}, {Annis}, {Bacon}, {Bhargava}, {Brooks}, {Burke}, {Carrasco Kind}, {Carretero}, {Castander}, {Conselice}, {Costanzi}, {da Costa}, {Pereira}, {De Vicente}, {DeRose}, {Desai}, {Eifler}, {Evrard}, {Ferrero}, {Fosalba}, {Frieman}, {Garc{\'\i}a-Bellido}, {Gaztanaga}, {Gerdes}, {Gutierrez}, {Hinton}, {Hollowood}, {Honscheid}, {Huterer}, {James}, {Kent}, {Krause}, {Kuehn}, {Lahav}, {Lima}, {Lin}, {Maia}, {Marshall}, {Melchior}, {Menanteau}, {Miquel}, {Mohr}, {Morgan}, {Muir}, {Ogando}, {Palmese}, {Paz-Chinch{\'o}n}, {Plazas}, {Rodriguez-Monroy}, {Romer}, {Roodman},
  {Sanchez}, {Scarpine}, {Serrano}, {Smith}, {Soares-Santos}, {Suchyta}, {Swanson}, {Tarle}, {To}, {Troxel}, {Varga}, {Weller}, {Wilkinson}, \& {Wilkinson}}]{everett2022}
{Everett}, S., {Yanny}, B., {Kuropatkin}, N., {et~al.} 2022, \apjs, 258, 15, \dodoi{10.3847/1538-4365/ac26c1}

\bibitem[{{Fardal} {et~al.}(2013){Fardal}, {Weinberg}, {Babul}, {Irwin}, {Guhathakurta}, {Gilbert}, {Ferguson}, {Ibata}, {Lewis}, {Tanvir}, \& {Huxor}}]{fardal2013}
{Fardal}, M.~A., {Weinberg}, M.~D., {Babul}, A., {et~al.} 2013, \mnras, 434, 2779, \dodoi{10.1093/mnras/stt1121}

\bibitem[{{Fitts} {et~al.}(2018){Fitts}, {Boylan-Kolchin}, {Bullock}, {Weisz}, {El-Badry}, {Wheeler}, {Faucher-Gigu{\`e}re}, {Quataert}, {Hopkins}, {Kere{\v{s}}}, {Wetzel}, \& {Hayward}}]{fitts2018}
{Fitts}, A., {Boylan-Kolchin}, M., {Bullock}, J.~S., {et~al.} 2018, \mnras, 479, 319, \dodoi{10.1093/mnras/sty1488}

\bibitem[{{Flaugher} {et~al.}(2015){Flaugher}, {Diehl}, {Honscheid}, {Abbott}, {Alvarez}, {Angstadt}, {Annis}, {Antonik}, {Ballester}, {Beaufore}, {Bernstein}, {Bernstein}, {Bigelow}, {Bonati}, {Boprie}, {Brooks}, {Buckley-Geer}, {Campa}, {Cardiel-Sas}, {Castander}, {Castilla}, {Cease}, {Cela-Ruiz}, {Chappa}, {Chi}, {Cooper}, {da Costa}, {Dede}, {Derylo}, {DePoy}, {de Vicente}, {Doel}, {Drlica-Wagner}, {Eiting}, {Elliott}, {Emes}, {Estrada}, {Fausti Neto}, {Finley}, {Flores}, {Frieman}, {Gerdes}, {Gladders}, {Gregory}, {Gutierrez}, {Hao}, {Holland}, {Holm}, {Huffman}, {Jackson}, {James}, {Jonas}, {Karcher}, {Karliner}, {Kent}, {Kessler}, {Kozlovsky}, {Kron}, {Kubik}, {Kuehn}, {Kuhlmann}, {Kuk}, {Lahav}, {Lathrop}, {Lee}, {Levi}, {Lewis}, {Li}, {Mandrichenko}, {Marshall}, {Martinez}, {Merritt}, {Miquel}, {Mu{\~n}oz}, {Neilsen}, {Nichol}, {Nord}, {Ogando}, {Olsen}, {Palaio}, {Patton}, {Peoples}, {Plazas}, {Rauch}, {Reil}, {Rheault}, {Roe}, {Rogers}, {Roodman}, {Sanchez}, {Scarpine}, {Schindler}, {Schmidt},
  {Schmitt}, {Schubnell}, {Schultz}, {Schurter}, {Scott}, {Serrano}, {Shaw}, {Smith}, {Soares-Santos}, {Stefanik}, {Stuermer}, {Suchyta}, {Sypniewski}, {Tarle}, {Thaler}, {Tighe}, {Tran}, {Tucker}, {Walker}, {Wang}, {Watson}, {Weaverdyck}, {Wester}, {Woods}, {Yanny}, \& {DES Collaboration}}]{flaugher2015}
{Flaugher}, B., {Diehl}, H.~T., {Honscheid}, K., {et~al.} 2015, \aj, 150, 150, \dodoi{10.1088/0004-6256/150/5/150}

\bibitem[{{Geha} {et~al.}(2010){Geha}, {van der Marel}, {Guhathakurta}, {Gilbert}, {Kalirai}, \& {Kirby}}]{geha2010}
{Geha}, M., {van der Marel}, R.~P., {Guhathakurta}, P., {et~al.} 2010, \apj, 711, 361, \dodoi{10.1088/0004-637X/711/1/361}

\bibitem[{{Genel} {et~al.}(2010){Genel}, {Bouch{\'e}}, {Naab}, {Sternberg}, \& {Genzel}}]{genel2010}
{Genel}, S., {Bouch{\'e}}, N., {Naab}, T., {Sternberg}, A., \& {Genzel}, R. 2010, \apj, 719, 229, \dodoi{10.1088/0004-637X/719/1/229}

\bibitem[{{Gil de Paz} {et~al.}(2007){Gil de Paz}, {Boissier}, {Madore}, {Seibert}, {Joe}, {Boselli}, {Wyder}, {Thilker}, {Bianchi}, {Rey}, {Rich}, {Barlow}, {Conrow}, {Forster}, {Friedman}, {Martin}, {Morrissey}, {Neff}, {Schiminovich}, {Small}, {Donas}, {Heckman}, {Lee}, {Milliard}, {Szalay}, \& {Yi}}]{GildePaz2007}
{Gil de Paz}, A., {Boissier}, S., {Madore}, B.~F., {et~al.} 2007, \apjs, 173, 185, \dodoi{10.1086/516636}

\bibitem[{{Gogarten} {et~al.}(2010){Gogarten}, {Dalcanton}, {Williams}, {Ro{\v{s}}kar}, {Holtzman}, {Seth}, {Dolphin}, {Weisz}, {Cole}, {Debattista}, {Gilbert}, {Olsen}, {Skillman}, {de Jong}, {Karachentsev}, \& {Quinn}}]{gogarten2010}
{Gogarten}, S.~M., {Dalcanton}, J.~J., {Williams}, B.~F., {et~al.} 2010, \apj, 712, 858, \dodoi{10.1088/0004-637X/712/2/858}

\bibitem[{Gommers {et~al.}(2025)Gommers, Virtanen, Haberland, Burovski, Reddy, Weckesser, Oliphant, Cournapeau, Nelson, alexbrc, Roy, Peterson, Polat, Wilson, endolith, Mayorov, van~der Walt, Colley, Brett, Laxalde, Larson, Sakai, Millman, Bowhay, Lars, peterbell10, Carey, van Mulbregt, eric jones, \& Striega}]{scipy_14593523}
Gommers, R., Virtanen, P., Haberland, M., {et~al.} 2025, scipy/scipy: SciPy 1.15.0, v1.15.0,  Zenodo, \dodoi{10.5281/zenodo.14593523}

\bibitem[{{Gozman} {et~al.}(2023){Gozman}, {Bell}, {Smercina}, {Price}, {Bailin}, {de Jong}, {D'Souza}, {Jang}, {Monachesi}, \& {Slater}}]{gozman2023}
{Gozman}, K., {Bell}, E.~F., {Smercina}, A., {et~al.} 2023, \apj, 947, 21, \dodoi{10.3847/1538-4357/acbe3a}

\bibitem[{{Grillmair} \& {Carlin}(2016)}]{grillmair2016}
{Grillmair}, C.~J., \& {Carlin}, J.~L. 2016, in Astrophysics and Space Science Library, Vol. 420, Tidal Streams in the Local Group and Beyond, ed. H.~J. {Newberg} \& J.~L. {Carlin}, 87, \dodoi{10.1007/978-3-319-19336-6_4}

\bibitem[{{Gu} {et~al.}(2016){Gu}, {Conroy}, \& {Behroozi}}]{gu2016}
{Gu}, M., {Conroy}, C., \& {Behroozi}, P. 2016, \apj, 833, 2, \dodoi{10.3847/0004-637X/833/1/2}

\bibitem[{{Gullikson} {et~al.}(2014){Gullikson}, {Dodson-Robinson}, \& {Kraus}}]{gullikson2014}
{Gullikson}, K., {Dodson-Robinson}, S., \& {Kraus}, A. 2014, \aj, 148, 53, \dodoi{10.1088/0004-6256/148/3/53}

\bibitem[{{Harmsen} {et~al.}(2017){Harmsen}, {Monachesi}, {Bell}, {de Jong}, {Bailin}, {Radburn-Smith}, \& {Holwerda}}]{harmsen2017}
{Harmsen}, B., {Monachesi}, A., {Bell}, E.~F., {et~al.} 2017, \mnras, 466, 1491, \dodoi{10.1093/mnras/stw2992}

\bibitem[{Harris {et~al.}(2020)Harris, Millman, van~der Walt, Gommers, Virtanen, Cournapeau, Wieser, Taylor, Berg, Smith, Kern, Picus, Hoyer, van Kerkwijk, Brett, Haldane, del R{\'{i}}o, Wiebe, Peterson, G{\'{e}}rard-Marchant, Sheppard, Reddy, Weckesser, Abbasi, Gohlke, \& Oliphant}]{numpy}
Harris, C.~R., Millman, K.~J., van~der Walt, S.~J., {et~al.} 2020, Nature, 585, 357, \dodoi{10.1038/s41586-020-2649-2}

\bibitem[{{Harris}(1996)}]{harris1996}
{Harris}, W.~E. 1996, \aj, 112, 1487, \dodoi{10.1086/118116}

\bibitem[{{Hartley} {et~al.}(2022){Hartley}, {Choi}, {Amon}, {Gruendl}, {Sheldon}, {Harrison}, {Bernstein}, {Sevilla-Noarbe}, {Yanny}, {Eckert}, {Diehl}, {Alarcon}, {Banerji}, {Bechtol}, {Buchs}, {Cantu}, {Conselice}, {Cordero}, {Davis}, {Davis}, {Dodelson}, {Drlica-Wagner}, {Everett}, {Fert{\'e}}, {Gruen}, {Honscheid}, {Jarvis}, {Johnson}, {Kokron}, {MacCrann}, {Myles}, {Pace}, {Palmese}, {Paz-Chinch{\'o}n}, {Pereira}, {Plazas}, {Prat}, {Rodriguez-Monroy}, {Rykoff}, {Samuroff}, {S{\'a}nchez}, {Secco}, {Tarsitano}, {Tong}, {Troxel}, {Vasquez}, {Wang}, {Zhou}, {Abbott}, {Aguena}, {Allam}, {Annis}, {Bacon}, {Bertin}, {Bhargava}, {Brooks}, {Burke}, {Carnero Rosell}, {Carrasco Kind}, {Carretero}, {Castander}, {Costanzi}, {Crocce}, {da Costa}, {De Vicente}, {DeRose}, {Desai}, {Dietrich}, {Eifler}, {Elvin-Poole}, {Ferrero}, {Flaugher}, {Fosalba}, {Garc{\'\i}a-Bellido}, {Gaztanaga}, {Gerdes}, {Gschwend}, {Gutierrez}, {Hinton}, {Hollowood}, {Huterer}, {James}, {Kent}, {Krause}, {Kuehn}, {Kuropatkin}, {Lahav}, {Lin},
  {Maia}, {March}, {Marshall}, {Martini}, {Melchior}, {Menanteau}, {Miquel}, {Mohr}, {Morgan}, {Neilsen}, {Ogando}, {Pandey}, {Romer}, {Roodman}, {Sako}, {Sanchez}, {Scarpine}, {Serrano}, {Smith}, {Soares-Santos}, {Suchyta}, {Swanson}, {Tarle}, {Thomas}, {To}, {Varga}, {Walker}, {Wester}, {Wilkinson}, {Zuntz}, {Zuntz}, \& {DES Collaboration}}]{hartley2022}
{Hartley}, W.~G., {Choi}, A., {Amon}, A., {et~al.} 2022, \mnras, 509, 3547, \dodoi{10.1093/mnras/stab3055}

\bibitem[{{Hayes} {et~al.}(2020){Hayes}, {Majewski}, {Hasselquist}, {Anguiano}, {Shetrone}, {Law}, {Schiavon}, {Cunha}, {Smith}, {Beaton}, {Price-Whelan}, {Allende Prieto}, {Battaglia}, {Bizyaev}, {Brownstein}, {Cohen}, {Frinchaboy}, {Garc{\'\i}a-Hern{\'a}ndez}, {Lacerna}, {Lane}, {M{\'e}sz{\'a}ros}, {Bidin}, {M{\~{u}}noz}, {Nidever}, {Oravetz}, {Oravetz}, {Pan}, {Roman-Lopes}, {Sobeck}, \& {Stringfellow}}]{hayes2020}
{Hayes}, C.~R., {Majewski}, S.~R., {Hasselquist}, S., {et~al.} 2020, \apj, 889, 63, \dodoi{10.3847/1538-4357/ab62ad}

\bibitem[{{Helmi}(2020)}]{helmi2020}
{Helmi}, A. 2020, \araa, 58, 205, \dodoi{10.1146/annurev-astro-032620-021917}

\bibitem[{{Hernquist} \& {Quinn}(1987)}]{hernquist1987}
{Hernquist}, L., \& {Quinn}, P.~J. 1987, \apj, 312, 1, \dodoi{10.1086/164844}

\bibitem[{{Hidalgo} {et~al.}(2009){Hidalgo}, {Aparicio}, {Mart{\'\i}nez-Delgado}, \& {Gallart}}]{hidalgo2009}
{Hidalgo}, S.~L., {Aparicio}, A., {Mart{\'\i}nez-Delgado}, D., \& {Gallart}, C. 2009, \apj, 705, 704, \dodoi{10.1088/0004-637X/705/1/704}

\bibitem[{{Higgs} \& {McConnachie}(2021)}]{higgs2021}
{Higgs}, C.~R., \& {McConnachie}, A.~W. 2021, \mnras, 506, 2766, \dodoi{10.1093/mnras/stab1754}

\bibitem[{{Hillis} {et~al.}(2016){Hillis}, {Williams}, {Dolphin}, {Dalcanton}, \& {Skillman}}]{hillis2016}
{Hillis}, T.~J., {Williams}, B.~F., {Dolphin}, A.~E., {Dalcanton}, J.~J., \& {Skillman}, E.~D. 2016, \apj, 831, 191, \dodoi{10.3847/0004-637X/831/2/191}

\bibitem[{{Ibata} {et~al.}(2019){Ibata}, {Malhan}, \& {Martin}}]{ibata2019}
{Ibata}, R.~A., {Malhan}, K., \& {Martin}, N.~F. 2019, \apj, 872, 152, \dodoi{10.3847/1538-4357/ab0080}

\bibitem[{{Ibata} {et~al.}(2014){Ibata}, {Lewis}, {McConnachie}, {Martin}, {Irwin}, {Ferguson}, {Babul}, {Bernard}, {Chapman}, {Collins}, {Fardal}, {Mackey}, {Navarro}, {Pe{\~n}arrubia}, {Rich}, {Tanvir}, \& {Widrow}}]{ibata2014}
{Ibata}, R.~A., {Lewis}, G.~F., {McConnachie}, A.~W., {et~al.} 2014, \apj, 780, 128, \dodoi{10.1088/0004-637X/780/2/128}

\bibitem[{{Ivezi{\'c}} {et~al.}(2019){Ivezi{\'c}}, {Kahn}, {Tyson}, {Abel}, {Acosta}, {Allsman}, {Alonso}, {AlSayyad}, {Anderson}, {Andrew}, {Angel}, {Angeli}, {Ansari}, {Antilogus}, {Araujo}, {Armstrong}, {Arndt}, {Astier}, {Aubourg}, {Auza}, {Axelrod}, {Bard}, {Barr}, {Barrau}, {Bartlett}, {Bauer}, {Bauman}, {Baumont}, {Bechtol}, {Bechtol}, {Becker}, {Becla}, {Beldica}, {Bellavia}, {Bianco}, {Biswas}, {Blanc}, {Blazek}, {Blandford}, {Bloom}, {Bogart}, {Bond}, {Booth}, {Borgland}, {Borne}, {Bosch}, {Boutigny}, {Brackett}, {Bradshaw}, {Brandt}, {Brown}, {Bullock}, {Burchat}, {Burke}, {Cagnoli}, {Calabrese}, {Callahan}, {Callen}, {Carlin}, {Carlson}, {Chandrasekharan}, {Charles-Emerson}, {Chesley}, {Cheu}, {Chiang}, {Chiang}, {Chirino}, {Chow}, {Ciardi}, {Claver}, {Cohen-Tanugi}, {Cockrum}, {Coles}, {Connolly}, {Cook}, {Cooray}, {Covey}, {Cribbs}, {Cui}, {Cutri}, {Daly}, {Daniel}, {Daruich}, {Daubard}, {Daues}, {Dawson}, {Delgado}, {Dellapenna}, {de Peyster}, {de Val-Borro}, {Digel}, {Doherty}, {Dubois},
  {Dubois-Felsmann}, {Durech}, {Economou}, {Eifler}, {Eracleous}, {Emmons}, {Fausti Neto}, {Ferguson}, {Figueroa}, {Fisher-Levine}, {Focke}, {Foss}, {Frank}, {Freemon}, {Gangler}, {Gawiser}, {Geary}, {Gee}, {Geha}, {Gessner}, {Gibson}, {Gilmore}, {Glanzman}, {Glick}, {Goldina}, {Goldstein}, {Goodenow}, {Graham}, {Gressler}, {Gris}, {Guy}, {Guyonnet}, {Haller}, {Harris}, {Hascall}, {Haupt}, {Hernandez}, {Herrmann}, {Hileman}, {Hoblitt}, {Hodgson}, {Hogan}, {Howard}, {Huang}, {Huffer}, {Ingraham}, {Innes}, {Jacoby}, {Jain}, {Jammes}, {Jee}, {Jenness}, {Jernigan}, {Jevremovi{\'c}}, {Johns}, {Johnson}, {Johnson}, {Jones}, {Juramy-Gilles}, {Juri{\'c}}, {Kalirai}, {Kallivayalil}, {Kalmbach}, {Kantor}, {Karst}, {Kasliwal}, {Kelly}, {Kessler}, {Kinnison}, {Kirkby}, {Knox}, {Kotov}, {Krabbendam}, {Krughoff}, {Kub{\'a}nek}, {Kuczewski}, {Kulkarni}, {Ku}, {Kurita}, {Lage}, {Lambert}, {Lange}, {Langton}, {Le Guillou}, {Levine}, {Liang}, {Lim}, {Lintott}, {Long}, {Lopez}, {Lotz}, {Lupton}, {Lust}, {MacArthur}, {Mahabal},
  {Mandelbaum}, {Markiewicz}, {Marsh}, {Marshall}, {Marshall}, {May}, {McKercher}, {McQueen}, {Meyers}, {Migliore}, {Miller}, {Mills}, {Miraval}, {Moeyens}, {Moolekamp}, {Monet}, {Moniez}, {Monkewitz}, {Montgomery}, {Morrison}, {Mueller}, {Muller}, {Mu{\~n}oz Arancibia}, {Neill}, {Newbry}, {Nief}, {Nomerotski}, {Nordby}, {O'Connor}, {Oliver}, {Olivier}, {Olsen}, {O'Mullane}, {Ortiz}, {Osier}, {Owen}, {Pain}, {Palecek}, {Parejko}, {Parsons}, {Pease}, {Peterson}, {Peterson}, {Petravick}, {Libby Petrick}, {Petry}, {Pierfederici}, {Pietrowicz}, {Pike}, {Pinto}, {Plante}, {Plate}, {Plutchak}, {Price}, {Prouza}, {Radeka}, {Rajagopal}, {Rasmussen}, {Regnault}, {Reil}, {Reiss}, {Reuter}, {Ridgway}, {Riot}, {Ritz}, {Robinson}, {Roby}, {Roodman}, {Rosing}, {Roucelle}, {Rumore}, {Russo}, {Saha}, {Sassolas}, {Schalk}, {Schellart}, {Schindler}, {Schmidt}, {Schneider}, {Schneider}, {Schoening}, {Schumacher}, {Schwamb}, {Sebag}, {Selvy}, {Sembroski}, {Seppala}, {Serio}, {Serrano}, {Shaw}, {Shipsey}, {Sick}, {Silvestri},
  {Slater}, {Smith}, {Smith}, {Sobhani}, {Soldahl}, {Storrie-Lombardi}, {Stover}, {Strauss}, {Street}, {Stubbs}, {Sullivan}, {Sweeney}, {Swinbank}, {Szalay}, {Takacs}, {Tether}, {Thaler}, {Thayer}, {Thomas}, {Thornton}, {Thukral}, {Tice}, {Trilling}, {Turri}, {Van Berg}, {Vanden Berk}, {Vetter}, {Virieux}, {Vucina}, {Wahl}, {Walkowicz}, {Walsh}, {Walter}, {Wang}, {Wang}, {Warner}, {Wiecha}, {Willman}, {Winters}, {Wittman}, {Wolff}, {Wood-Vasey}, {Wu}, {Xin}, {Yoachim}, \& {Zhan}}]{lsst}
{Ivezi{\'c}}, {\v{Z}}., {Kahn}, S.~M., {Tyson}, J.~A., {et~al.} 2019, \apj, 873, 111, \dodoi{10.3847/1538-4357/ab042c}

\bibitem[{{Jang} {et~al.}(2020){Jang}, {de Jong}, {Minchev}, {Bell}, {Monachesi}, {Holwerda}, {Bailin}, {Smercina}, \& {D'Souza}}]{jang2020}
{Jang}, I.~S., {de Jong}, R.~S., {Minchev}, I., {et~al.} 2020, \aap, 640, L19, \dodoi{10.1051/0004-6361/202038651}

\bibitem[{{Jarvis} {et~al.}(2016){Jarvis}, {Sheldon}, {Zuntz}, {Kacprzak}, {Bridle}, {Amara}, {Armstrong}, {Becker}, {Bernstein}, {Bonnett}, {Chang}, {Das}, {Dietrich}, {Drlica-Wagner}, {Eifler}, {Gangkofner}, {Gruen}, {Hirsch}, {Huff}, {Jain}, {Kent}, {Kirk}, {MacCrann}, {Melchior}, {Plazas}, {Refregier}, {Rowe}, {Rykoff}, {Samuroff}, {S{\'a}nchez}, {Suchyta}, {Troxel}, {Vikram}, {Abbott}, {Abdalla}, {Allam}, {Annis}, {Benoit-L{\'e}vy}, {Bertin}, {Brooks}, {Buckley-Geer}, {Burke}, {Capozzi}, {Carnero Rosell}, {Carrasco Kind}, {Carretero}, {Castander}, {Clampitt}, {Crocce}, {Cunha}, {D'Andrea}, {da Costa}, {DePoy}, {Desai}, {Diehl}, {Doel}, {Fausti Neto}, {Flaugher}, {Fosalba}, {Frieman}, {Gaztanaga}, {Gerdes}, {Gruendl}, {Gutierrez}, {Honscheid}, {James}, {Kuehn}, {Kuropatkin}, {Lahav}, {Li}, {Lima}, {March}, {Martini}, {Miquel}, {Mohr}, {Neilsen}, {Nord}, {Ogando}, {Reil}, {Romer}, {Roodman}, {Sako}, {Sanchez}, {Scarpine}, {Schubnell}, {Sevilla-Noarbe}, {Smith}, {Soares-Santos}, {Sobreira}, {Swanson},
  {Tarle}, {Thaler}, {Thomas}, {Walker}, \& {Wechsler}}]{jarvis2016}
{Jarvis}, M., {Sheldon}, E., {Zuntz}, J., {et~al.} 2016, \mnras, 460, 2245, \dodoi{10.1093/mnras/stw990}

\bibitem[{{Johnson} {et~al.}(2020){Johnson}, {Conroy}, {Naidu}, {Bonaca}, {Zaritsky}, {Ting}, {Cargile}, {Han}, \& {Speagle}}]{johnson2020}
{Johnson}, B.~D., {Conroy}, C., {Naidu}, R.~P., {et~al.} 2020, \apj, 900, 103, \dodoi{10.3847/1538-4357/abab08}

\bibitem[{{Kado-Fong} {et~al.}(2020){Kado-Fong}, {Greene}, {Huang}, {Beaton}, {Goulding}, \& {Komiyama}}]{kadofong2020}
{Kado-Fong}, E., {Greene}, J.~E., {Huang}, S., {et~al.} 2020, \apj, 900, 163, \dodoi{10.3847/1538-4357/abacc2}

\bibitem[{{Kado-Fong} {et~al.}(2022){Kado-Fong}, {Sanderson}, {Greene}, {Cunningham}, {Wheeler}, {Chan}, {El-Badry}, {Hopkins}, {Wetzel}, {Boylan-Kolchin}, {Faucher-Gigu{\`e}re}, {Huang}, {Quataert}, \& {Starkenburg}}]{kadofong2022}
{Kado-Fong}, E., {Sanderson}, R.~E., {Greene}, J.~E., {et~al.} 2022, \apj, 931, 152, \dodoi{10.3847/1538-4357/ac6c88}

\bibitem[{{Kang} {et~al.}(2016){Kang}, {Zhang}, {Chang}, {Wang}, \& {Cheng}}]{kang2016}
{Kang}, X., {Zhang}, F., {Chang}, R., {Wang}, L., \& {Cheng}, L. 2016, \aap, 585, A20, \dodoi{10.1051/0004-6361/201527041}

\bibitem[{{Karachentsev} {et~al.}(2013){Karachentsev}, {Makarov}, \& {Kaisina}}]{karachentsev2013}
{Karachentsev}, I.~D., {Makarov}, D.~I., \& {Kaisina}, E.~I. 2013, \aj, 145, 101, \dodoi{10.1088/0004-6256/145/4/101}

\bibitem[{{Kauffmann} {et~al.}(1993){Kauffmann}, {White}, \& {Guiderdoni}}]{kauffmann1993}
{Kauffmann}, G., {White}, S.~D.~M., \& {Guiderdoni}, B. 1993, \mnras, 264, 201, \dodoi{10.1093/mnras/264.1.201}

\bibitem[{{Kim} {et~al.}(2002){Kim}, {Sung}, \& {Lee}}]{kim2002}
{Kim}, S.~C., {Sung}, H., \& {Lee}, M.~G. 2002, Journal of Korean Astronomical Society, 35, 9, \dodoi{10.5303/JKAS.2002.35.1.009}

\bibitem[{{Kim} {et~al.}(2004){Kim}, {Sung}, {Park}, \& {Sung}}]{kim2004}
{Kim}, S.~C., {Sung}, H., {Park}, H.~S., \& {Sung}, E.-C. 2004, \cjaa, 4, 299, \dodoi{10.1088/1009-9271/4/4/299}

\bibitem[{{Kirby} {et~al.}(2013){Kirby}, {Cohen}, {Guhathakurta}, {Cheng}, {Bullock}, \& {Gallazzi}}]{kirby2013}
{Kirby}, E.~N., {Cohen}, J.~G., {Guhathakurta}, P., {et~al.} 2013, \apj, 779, 102, \dodoi{10.1088/0004-637X/779/2/102}

\bibitem[{{Koposov} {et~al.}(2011){Koposov}, {Gilmore}, {Walker}, {Belokurov}, {Evans}, {Fellhauer}, {Gieren}, {Geisler}, {Monaco}, {Norris}, {Okamoto}, {Pe{\~n}arrubia}, {Wilkinson}, {Wyse}, \& {Zucker}}]{koposov2011}
{Koposov}, S.~E., {Gilmore}, G., {Walker}, M.~G., {et~al.} 2011, \apj, 736, 146, \dodoi{10.1088/0004-637X/736/2/146}

\bibitem[{{Kroupa}(2001)}]{kroupa2001}
{Kroupa}, P. 2001, \mnras, 322, 231, \dodoi{10.1046/j.1365-8711.2001.04022.x}

\bibitem[{{Kroupa}(2002)}]{kroupa2002}
---. 2002, Science, 295, 82, \dodoi{10.1126/science.1067524}

\bibitem[{{Laine} {et~al.}(2016){Laine}, {Laurikainen}, \& {Salo}}]{laine2016}
{Laine}, J., {Laurikainen}, E., \& {Salo}, H. 2016, \aap, 596, A25, \dodoi{10.1051/0004-6361/201628397}

\bibitem[{{Lauberts} \& {Valentijn}(1989)}]{lauberts1989}
{Lauberts}, A., \& {Valentijn}, E.~A. 1989, {The surface photometry catalogue of the ESO-Uppsala galaxies} (Simbad)

\bibitem[{{Li} {et~al.}(2022){Li}, {Ji}, {Pace}, {Erkal}, {Koposov}, {Shipp}, {Da Costa}, {Cullinane}, {Kuehn}, {Lewis}, {Mackey}, {Simpson}, {Zucker}, {Ferguson}, {Martell}, {Bland-Hawthorn}, {Balbinot}, {Tavangar}, {Drlica-Wagner}, {De Silva}, \& {Simon}}]{li2022}
{Li}, T.~S., {Ji}, A.~P., {Pace}, A.~B., {et~al.} 2022, \apj, 928, 30, \dodoi{10.3847/1538-4357/ac46d3}

\bibitem[{{Limberg} {et~al.}(2023){Limberg}, {Queiroz}, {Perottoni}, {Rossi}, {Amarante}, {Santucci}, {Chiappini}, {P{\'e}rez-Villegas}, \& {Lee}}]{limberg2023}
{Limberg}, G., {Queiroz}, A. B.~A., {Perottoni}, H.~D., {et~al.} 2023, \apj, 946, 66, \dodoi{10.3847/1538-4357/acb694}

\bibitem[{{Mackereth} \& {Bovy}(2020)}]{mackereth2020}
{Mackereth}, J.~T., \& {Bovy}, J. 2020, \mnras, 492, 3631, \dodoi{10.1093/mnras/staa047}

\bibitem[{{Marigo} {et~al.}(2017){Marigo}, {Girardi}, {Bressan}, {Rosenfield}, {Aringer}, {Chen}, {Dussin}, {Nanni}, {Pastorelli}, {Rodrigues}, {Trabucchi}, {Bladh}, {Dalcanton}, {Groenewegen}, {Montalb{\'a}n}, \& {Wood}}]{marigo2017}
{Marigo}, P., {Girardi}, L., {Bressan}, A., {et~al.} 2017, \apj, 835, 77, \dodoi{10.3847/1538-4357/835/1/77}

\bibitem[{{Martin} \& {GALEX Team}(2005)}]{galex2005}
{Martin}, C., \& {GALEX Team}. 2005, in Maps of the Cosmos, ed. M.~{Colless}, L.~{Staveley-Smith}, \& R.~A. {Stathakis}, Vol. 216, 221, \dodoi{10.1017/S0074180900196664}

\bibitem[{{Martin} {et~al.}(2016){Martin}, {Ibata}, {Lewis}, {McConnachie}, {Babul}, {Bate}, {Bernard}, {Chapman}, {Collins}, {Conn}, {Crnojevi{\'c}}, {Fardal}, {Ferguson}, {Irwin}, {Mackey}, {McMonigal}, {Navarro}, \& {Rich}}]{martin2016}
{Martin}, N.~F., {Ibata}, R.~A., {Lewis}, G.~F., {et~al.} 2016, \apj, 833, 167, \dodoi{10.3847/1538-4357/833/2/167}

\bibitem[{{Mart{\'\i}nez-Delgado} {et~al.}(2012){Mart{\'\i}nez-Delgado}, {Romanowsky}, {Gabany}, {Annibali}, {Arnold}, {Fliri}, {Zibetti}, {van der Marel}, {Rix}, {Chonis}, {Carballo-Bello}, {Aloisi}, {Macci{\`o}}, {Gallego-Laborda}, {Brodie}, \& {Merrifield}}]{martinezdelgado2012}
{Mart{\'\i}nez-Delgado}, D., {Romanowsky}, A.~J., {Gabany}, R.~J., {et~al.} 2012, \apjl, 748, L24, \dodoi{10.1088/2041-8205/748/2/L24}

\bibitem[{{Mart{\'\i}nez-Delgado} {et~al.}(2023{\natexlab{a}}){Mart{\'\i}nez-Delgado}, {Roca-F{\`a}brega}, {Mir{\'o}-Carretero}, {G{\'o}mez-Flechoso}, {Rom{\`a}n}, {Donatiello}, {Schmidt}, {Lang}, {Akhlaghi}, \& {Hanson}}]{martinezdelgado2023}
{Mart{\'\i}nez-Delgado}, D., {Roca-F{\`a}brega}, S., {Mir{\'o}-Carretero}, J., {et~al.} 2023{\natexlab{a}}, \aap, 669, A103, \dodoi{10.1051/0004-6361/202244832}

\bibitem[{{Mart{\'\i}nez-Delgado} {et~al.}(2023{\natexlab{b}}){Mart{\'\i}nez-Delgado}, {Cooper}, {Rom{\'a}n}, {Pillepich}, {Erkal}, {Pearson}, {Moustakas}, {Laporte}, {Laine}, {Akhlaghi}, {Lang}, {Makarov}, {Borlaff}, {Donatiello}, {Pearson}, {Mir{\'o}-Carretero}, {Cuillandre}, {Dom{\'\i}nguez}, {Roca-F{\`a}brega}, {Frenk}, {Schmidt}, {G{\'o}mez-Flechoso}, {Guzman}, {Libeskind}, {Dey}, {Weaver}, {Schlegel}, {Myers}, \& {Valdes}}]{martinezdelgado2023b}
{Mart{\'\i}nez-Delgado}, D., {Cooper}, A.~P., {Rom{\'a}n}, J., {et~al.} 2023{\natexlab{b}}, \aap, 671, A141, \dodoi{10.1051/0004-6361/202245011}

\bibitem[{{Maxwell} {et~al.}(2012){Maxwell}, {Wadsley}, {Couchman}, \& {Mashchenko}}]{maxwell2012}
{Maxwell}, A.~J., {Wadsley}, J., {Couchman}, H.~M.~P., \& {Mashchenko}, S. 2012, \apjl, 755, L35, \dodoi{10.1088/2041-8205/755/2/L35}

\bibitem[{{Minniti} \& {Zijlstra}(1996)}]{minniti1996}
{Minniti}, D., \& {Zijlstra}, A.~A. 1996, \apjl, 467, L13, \dodoi{10.1086/310189}

\bibitem[{{Mir{\'o}-Carretero} {et~al.}(2024){Mir{\'o}-Carretero}, {Mart{\'\i}nez-Delgado}, {G{\'o}mez-Flechoso}, {Cooper}, {Akhlaghi}, {Donatiello}, {Kuijken}, {Lang}, {Makarov}, {Laine}, \& {Roca-F{\`a}brega}}]{mirro-carretero2024}
{Mir{\'o}-Carretero}, J., {Mart{\'\i}nez-Delgado}, D., {G{\'o}mez-Flechoso}, M.~A., {et~al.} 2024, \aap, 691, A196, \dodoi{10.1051/0004-6361/202451685}

\bibitem[{{Mondal} {et~al.}(2019){Mondal}, {Subramaniam}, \& {George}}]{mondal2019}
{Mondal}, C., {Subramaniam}, A., \& {George}, K. 2019, Journal of Astrophysics and Astronomy, 40, 35, \dodoi{10.1007/s12036-019-9603-4}

\bibitem[{{Morganson} {et~al.}(2018){Morganson}, {Gruendl}, {Menanteau}, {Carrasco Kind}, {Chen}, {Daues}, {Drlica-Wagner}, {Friedel}, {Gower}, {Johnson}, {Johnson}, {Kessler}, {Paz-Chinch{\'o}n}, {Petravick}, {Pond}, {Yanny}, {Allam}, {Armstrong}, {Barkhouse}, {Bechtol}, {Benoit-L{\'e}vy}, {Bernstein}, {Bertin}, {Buckley-Geer}, {Covarrubias}, {Desai}, {Diehl}, {Goldstein}, {Gruen}, {Li}, {Lin}, {Marriner}, {Mohr}, {Neilsen}, {Ngeow}, {Paech}, {Rykoff}, {Sako}, {Sevilla-Noarbe}, {Sheldon}, {Sobreira}, {Tucker}, {Wester}, \& {DES Collaboration}}]{morganson2018}
{Morganson}, E., {Gruendl}, R.~A., {Menanteau}, F., {et~al.} 2018, \pasp, 130, 074501, \dodoi{10.1088/1538-3873/aab4ef}

\bibitem[{{Mu{\~n}oz} {et~al.}(2018){Mu{\~n}oz}, {C{\^o}t{\'e}}, {Santana}, {Geha}, {Simon}, {Oyarz{\'u}n}, {Stetson}, \& {Djorgovski}}]{munoz2018}
{Mu{\~n}oz}, R.~R., {C{\^o}t{\'e}}, P., {Santana}, F.~A., {et~al.} 2018, \apj, 860, 66, \dodoi{10.3847/1538-4357/aac16b}

\bibitem[{{Mu{\~n}oz-Mateos} {et~al.}(2007){Mu{\~n}oz-Mateos}, {Gil de Paz}, {Boissier}, {Zamorano}, {Jarrett}, {Gallego}, \& {Madore}}]{munoz2007}
{Mu{\~n}oz-Mateos}, J.~C., {Gil de Paz}, A., {Boissier}, S., {et~al.} 2007, \apj, 658, 1006, \dodoi{10.1086/511812}

\bibitem[{{Mucciarelli} {et~al.}(2021){Mucciarelli}, {Massari}, {Minelli}, {Romano}, {Bellazzini}, {Ferraro}, {Matteucci}, \& {Origlia}}]{mucciarelli2021}
{Mucciarelli}, A., {Massari}, D., {Minelli}, A., {et~al.} 2021, Nature Astronomy, 5, 1247, \dodoi{10.1038/s41550-021-01493-y}

\bibitem[{{M{\"u}ller} {et~al.}(2021){M{\"u}ller}, {Fahrion}, {Rejkuba}, {Hilker}, {Lelli}, {Lutz}, {Pawlowski}, {Coccato}, {Anand}, \& {Jerjen}}]{muller2021}
{M{\"u}ller}, O., {Fahrion}, K., {Rejkuba}, M., {et~al.} 2021, \aap, 645, A92, \dodoi{10.1051/0004-6361/202039359}

\bibitem[{{Munshi} {et~al.}(2021){Munshi}, {Brooks}, {Applebaum}, {Christensen}, {Quinn}, \& {Sligh}}]{munshi2021}
{Munshi}, F., {Brooks}, A.~M., {Applebaum}, E., {et~al.} 2021, \apj, 923, 35, \dodoi{10.3847/1538-4357/ac0db6}

\bibitem[{{Mutlu-Pakdil} {et~al.}(2021){Mutlu-Pakdil}, {Sand}, {Crnojevi{\'c}}, {Drlica-Wagner}, {Caldwell}, {Guhathakurta}, {Seth}, {Simon}, {Strader}, \& {Toloba}}]{mutlupakdil2021}
{Mutlu-Pakdil}, B., {Sand}, D.~J., {Crnojevi{\'c}}, D., {et~al.} 2021, \apj, 918, 88, \dodoi{10.3847/1538-4357/ac0db8}

\bibitem[{{Naidu} {et~al.}(2021){Naidu}, {Conroy}, {Bonaca}, {Zaritsky}, {Weinberger}, {Ting}, {Caldwell}, {Tacchella}, {Han}, {Speagle}, \& {Cargile}}]{naidou2021}
{Naidu}, R.~P., {Conroy}, C., {Bonaca}, A., {et~al.} 2021, \apj, 923, 92, \dodoi{10.3847/1538-4357/ac2d2d}

\bibitem[{{Nantais} {et~al.}(2010){Nantais}, {Huchra}, {Barmby}, \& {Olsen}}]{nantais2010A}
{Nantais}, J.~B., {Huchra}, J.~P., {Barmby}, P., \& {Olsen}, K. A.~G. 2010, \aj, 139, 1178, \dodoi{10.1088/0004-6256/139/3/1178}

\bibitem[{{Nibauer} {et~al.}(2023){Nibauer}, {Bonaca}, \& {Johnston}}]{nibauer2023}
{Nibauer}, J., {Bonaca}, A., \& {Johnston}, K.~V. 2023, \apj, 954, 195, \dodoi{10.3847/1538-4357/ace9bc}

\bibitem[{{Nidever} {et~al.}(2019){Nidever}, {Olsen}, {Choi}, {de Boer}, {Blum}, {Bell}, {Zaritsky}, {Martin}, {Saha}, {Conn}, {Besla}, {van der Marel}, {No{\"e}l}, {Monachesi}, {Stringfellow}, {Massana}, {Cioni}, {Gallart}, {Monelli}, {Martinez-Delgado}, {Mu{\~n}oz}, {Majewski}, {Vivas}, {Walker}, {Kaleida}, \& {Chu}}]{nidever2019}
{Nidever}, D.~L., {Olsen}, K., {Choi}, Y., {et~al.} 2019, \apj, 874, 118, \dodoi{10.3847/1538-4357/aafaf7}

\bibitem[{{Niederste-Ostholt} {et~al.}(2012){Niederste-Ostholt}, {Belokurov}, \& {Evans}}]{niederste2012}
{Niederste-Ostholt}, M., {Belokurov}, V., \& {Evans}, N.~W. 2012, \mnras, 422, 207, \dodoi{10.1111/j.1365-2966.2012.20602.x}

\bibitem[{{Okamoto} {et~al.}(2019){Okamoto}, {Arimoto}, {Ferguson}, {Irwin}, {Bernard}, \& {Utsumi}}]{okamoto2019}
{Okamoto}, S., {Arimoto}, N., {Ferguson}, A. M.~N., {et~al.} 2019, \apj, 884, 128, \dodoi{10.3847/1538-4357/ab44a7}

\bibitem[{{Olsen} {et~al.}(2004){Olsen}, {Miller}, {Suntzeff}, {Schommer}, \& {Bright}}]{olsen2004}
{Olsen}, K. A.~G., {Miller}, B.~W., {Suntzeff}, N.~B., {Schommer}, R.~A., \& {Bright}, J. 2004, \aj, 127, 2674, \dodoi{10.1086/383297}

\bibitem[{{Pace}(2024)}]{pace2024}
{Pace}, A.~B. 2024, arXiv e-prints, arXiv:2411.07424, \dodoi{10.48550/arXiv.2411.07424}

\bibitem[{pandas~development team(2024)}]{pandas_13819579}
pandas~development team, T. 2024, pandas-dev/pandas: Pandas, v2.2.3,  Zenodo, \dodoi{10.5281/zenodo.13819579}

\bibitem[{{Pastorelli} {et~al.}(2019){Pastorelli}, {Marigo}, {Girardi}, {Chen}, {Rubele}, {Trabucchi}, {Aringer}, {Bladh}, {Bressan}, {Montalb{\'a}n}, {Boyer}, {Dalcanton}, {Eriksson}, {Groenewegen}, {H{\"o}fner}, {Lebzelter}, {Nanni}, {Rosenfield}, {Wood}, \& {Cioni}}]{pastorelli2019}
{Pastorelli}, G., {Marigo}, P., {Girardi}, L., {et~al.} 2019, \mnras, 485, 5666, \dodoi{10.1093/mnras/stz725}

\bibitem[{{Pastorelli} {et~al.}(2020){Pastorelli}, {Marigo}, {Girardi}, {Aringer}, {Chen}, {Rubele}, {Trabucchi}, {Bladh}, {Boyer}, {Bressan}, {Dalcanton}, {Groenewegen}, {Lebzelter}, {Mowlavi}, {Chubb}, {Cioni}, {de Grijs}, {Ivanov}, {Nanni}, {van Loon}, \& {Zaggia}}]{pastorelli2020}
---. 2020, \mnras, 498, 3283, \dodoi{10.1093/mnras/staa2565}

\bibitem[{{Pearson} {et~al.}(2022){Pearson}, {Price-Whelan}, {Hogg}, {Seth}, {Sand}, {Hunt}, \& {Crnojevi{\'c}}}]{pearson2022}
{Pearson}, S., {Price-Whelan}, A.~M., {Hogg}, D.~W., {et~al.} 2022, \apj, 941, 19, \dodoi{10.3847/1538-4357/ac9bfb}

\bibitem[{{Pearson} {et~al.}(2016){Pearson}, {Besla}, {Putman}, {Lutz}, {Fern{\'a}ndez}, {Stierwalt}, {Patton}, {Kim}, {Kallivayalil}, {Johnson}, \& {Sung}}]{pearson2016}
{Pearson}, S., {Besla}, G., {Putman}, M.~E., {et~al.} 2016, \mnras, 459, 1827, \dodoi{10.1093/mnras/stw757}

\bibitem[{{Pearson} {et~al.}(2018){Pearson}, {Privon}, {Besla}, {Putman}, {Mart{\'\i}nez-Delgado}, {Johnston}, {Gabany}, {Patton}, \& {Kallivayalil}}]{pearson2018}
{Pearson}, S., {Privon}, G.~C., {Besla}, G., {et~al.} 2018, \mnras, 480, 3069, \dodoi{10.1093/mnras/sty2052}

\bibitem[{{Peng} {et~al.}(2009){Peng}, {Jord{\'a}n}, {Blakeslee}, {Mieske}, {C{\^o}t{\'e}}, {Ferrarese}, {Harris}, {Madrid}, \& {Meurer}}]{peng2009}
{Peng}, E.~W., {Jord{\'a}n}, A., {Blakeslee}, J.~P., {et~al.} 2009, \apj, 703, 42, \dodoi{10.1088/0004-637X/703/1/42}

\bibitem[{{Pucha} {et~al.}(2019){Pucha}, {Carlin}, {Willman}, {Strader}, {Sand}, {Bechtol}, {Brodie}, {Crnojevi{\'c}}, {Forbes}, {Garling}, {Hargis}, {Peter}, \& {Romanowsky}}]{pucha2019}
{Pucha}, R., {Carlin}, J.~L., {Willman}, B., {et~al.} 2019, \apj, 880, 104, \dodoi{10.3847/1538-4357/ab29fb}

\bibitem[{{Ramos} {et~al.}(2022){Ramos}, {Antoja}, {Yuan}, {Arentsen}, {Oria}, {Famaey}, {Ibata}, {Mateu}, \& {Carballo-Bello}}]{ramos2022}
{Ramos}, P., {Antoja}, T., {Yuan}, Z., {et~al.} 2022, \aap, 666, A64, \dodoi{10.1051/0004-6361/202142830}

\bibitem[{{Roussel} {et~al.}(2005){Roussel}, {Gil de Paz}, {Seibert}, {Helou}, {Madore}, \& {Martin}}]{roussel2005}
{Roussel}, H., {Gil de Paz}, A., {Seibert}, M., {et~al.} 2005, \apj, 632, 227, \dodoi{10.1086/432707}

\bibitem[{{Ry{\'s}} {et~al.}(2011){Ry{\'s}}, {Grocholski}, {van der Marel}, {Aloisi}, \& {Annibali}}]{rys2011}
{Ry{\'s}}, A., {Grocholski}, A.~J., {van der Marel}, R.~P., {Aloisi}, A., \& {Annibali}, F. 2011, \aap, 530, A23, \dodoi{10.1051/0004-6361/201015881}

\bibitem[{{Sand} {et~al.}(2024){Sand}, {Mutlu-Pakdil}, {Jones}, {Karunakaran}, {Andrews}, {Bennet}, {Crnojevic}, {Donatiello}, {Drlica-Wagner}, {Fielder}, {Martinez-Delgado}, {Martinez-Vazquez}, {Spekkens}, {Doliva-Dolinsky}, {Hunger}, {Carlin}, {Cerny}, {Hai}, {McQuinn}, {Pace}, \& {Smercina}}]{sand2024}
{Sand}, D.~J., {Mutlu-Pakdil}, B., {Jones}, M.~G., {et~al.} 2024, arXiv e-prints, arXiv:2409.16345, \dodoi{10.48550/arXiv.2409.16345}

\bibitem[{{Schlafly} \& {Finkbeiner}(2011)}]{schlafly2011}
{Schlafly}, E.~F., \& {Finkbeiner}, D.~P. 2011, \apj, 737, 103, \dodoi{10.1088/0004-637X/737/2/103}

\bibitem[{{Schlegel} {et~al.}(1998){Schlegel}, {Finkbeiner}, \& {Davis}}]{SFD1998}
{Schlegel}, D.~J., {Finkbeiner}, D.~P., \& {Davis}, M. 1998, \apj, 500, 525, \dodoi{10.1086/305772}

\bibitem[{{Sellwood}(2013)}]{sellwood2013}
{Sellwood}, J.~A. 2013, \apjl, 769, L24, \dodoi{10.1088/2041-8205/769/2/L24}

\bibitem[{{Sheldon}(2014)}]{sheldon2014}
{Sheldon}, E.~S. 2014, \mnras, 444, L25, \dodoi{10.1093/mnrasl/slu104}

\bibitem[{{Shipp} {et~al.}(2018){Shipp}, {Drlica-Wagner}, {Balbinot}, {Ferguson}, {Erkal}, {Li}, {Bechtol}, {Belokurov}, {Buncher}, {Carollo}, {Carrasco Kind}, {Kuehn}, {Marshall}, {Pace}, {Rykoff}, {Sevilla-Noarbe}, {Sheldon}, {Strigari}, {Vivas}, {Yanny}, {Zenteno}, {Abbott}, {Abdalla}, {Allam}, {Avila}, {Bertin}, {Brooks}, {Burke}, {Carretero}, {Castander}, {Cawthon}, {Crocce}, {Cunha}, {D'Andrea}, {da Costa}, {Davis}, {De Vicente}, {Desai}, {Diehl}, {Doel}, {Evrard}, {Flaugher}, {Fosalba}, {Frieman}, {Garc{\'\i}a-Bellido}, {Gaztanaga}, {Gerdes}, {Gruen}, {Gruendl}, {Gschwend}, {Gutierrez}, {Hartley}, {Honscheid}, {Hoyle}, {James}, {Johnson}, {Krause}, {Kuropatkin}, {Lahav}, {Lin}, {Maia}, {March}, {Martini}, {Menanteau}, {Miller}, {Miquel}, {Nichol}, {Plazas}, {Romer}, {Sako}, {Sanchez}, {Santiago}, {Scarpine}, {Schindler}, {Schubnell}, {Smith}, {Smith}, {Sobreira}, {Suchyta}, {Swanson}, {Tarle}, {Thomas}, {Tucker}, {Walker}, {Wechsler}, \& {DES Collaboration}}]{shipp2018}
{Shipp}, N., {Drlica-Wagner}, A., {Balbinot}, E., {et~al.} 2018, \apj, 862, 114, \dodoi{10.3847/1538-4357/aacdab}

\bibitem[{{Simon}(2019)}]{simon2019}
{Simon}, J.~D. 2019, \araa, 57, 375, \dodoi{10.1146/annurev-astro-091918-104453}

\bibitem[{{Smercina} {et~al.}(2023){Smercina}, {Bell}, {Price}, {Bailin}, {Dalcanton}, {de Jong}, {D'Souza}, {Gozman}, {Jang}, {Monachesi}, {Nidever}, \& {Slater}}]{smercina2023}
{Smercina}, A., {Bell}, E.~F., {Price}, P.~A., {et~al.} 2023, \apjl, 949, L37, \dodoi{10.3847/2041-8213/acd5d1}

\bibitem[{{Springel} {et~al.}(2006){Springel}, {Frenk}, \& {White}}]{springel2006}
{Springel}, V., {Frenk}, C.~S., \& {White}, S. D.~M. 2006, \nat, 440, 1137, \dodoi{10.1038/nature04805}

\bibitem[{{Starkenburg} {et~al.}(2016){Starkenburg}, {Helmi}, \& {Sales}}]{starkenburg2016}
{Starkenburg}, T.~K., {Helmi}, A., \& {Sales}, L.~V. 2016, \aap, 595, A56, \dodoi{10.1051/0004-6361/201528066}

\bibitem[{{Stinson} {et~al.}(2009){Stinson}, {Dalcanton}, {Quinn}, {Gogarten}, {Kaufmann}, \& {Wadsley}}]{stinson2009}
{Stinson}, G.~S., {Dalcanton}, J.~J., {Quinn}, T., {et~al.} 2009, \mnras, 395, 1455, \dodoi{10.1111/j.1365-2966.2009.14555.x}

\bibitem[{{Strader} {et~al.}(2012){Strader}, {Seth}, \& {Caldwell}}]{strader2012}
{Strader}, J., {Seth}, A.~C., \& {Caldwell}, N. 2012, \aj, 143, 52, \dodoi{10.1088/0004-6256/143/2/52}

\bibitem[{{Taibi} {et~al.}(2020){Taibi}, {Battaglia}, {Rejkuba}, {Leaman}, {Kacharov}, {Iorio}, {Jablonka}, \& {Zoccali}}]{taibi2020}
{Taibi}, S., {Battaglia}, G., {Rejkuba}, M., {et~al.} 2020, \aap, 635, A152, \dodoi{10.1051/0004-6361/201937240}

\bibitem[{{Taibi} {et~al.}(2018){Taibi}, {Battaglia}, {Kacharov}, {Rejkuba}, {Irwin}, {Leaman}, {Zoccali}, {Tolstoy}, \& {Jablonka}}]{taibi2018}
{Taibi}, S., {Battaglia}, G., {Kacharov}, N., {et~al.} 2018, \aap, 618, A122, \dodoi{10.1051/0004-6361/201833414}

\bibitem[{{Tan} {et~al.}(2024){Tan}, {Cerny}, {Drlica-Wagner}, {Pace}, {Geha}, {Ji}, {Li}, {Adam{\'o}w}, {Anbajagane}, {Bom}, {Carballo-Bello}, {Carlin}, {Chang}, {Choi}, {Collins}, {Doliva-Dolinsky}, {Ferguson}, {Gruendl}, {James}, {Limberg}, {Navabi}, {Mart{\'\i}nez-Delgado}, {Mart{\'\i}nez-V{\'a}zquez}, {Medina}, {Mutlu-Pakdil}, {Nidever}, {No{\"e}l}, {Riley}, {Sakowska}, {Sand}, {Sharp}, {Stringfellow}, {Tolley}, \& {Vivas}}]{tan2024}
{Tan}, C.~Y., {Cerny}, W., {Drlica-Wagner}, A., {et~al.} 2024, arXiv e-prints, arXiv:2408.00865, \dodoi{10.48550/arXiv.2408.00865}

\bibitem[{{Tang} {et~al.}(2014){Tang}, {Bressan}, {Rosenfield}, {Slemer}, {Marigo}, {Girardi}, \& {Bianchi}}]{tang2014}
{Tang}, J., {Bressan}, A., {Rosenfield}, P., {et~al.} 2014, \mnras, 445, 4287, \dodoi{10.1093/mnras/stu2029}

\bibitem[{{Taylor} {et~al.}(2011){Taylor}, {Hopkins}, {Baldry}, {Brown}, {Driver}, {Kelvin}, {Hill}, {Robotham}, {Bland-Hawthorn}, {Jones}, {Sharp}, {Thomas}, {Liske}, {Loveday}, {Norberg}, {Peacock}, {Bamford}, {Brough}, {Colless}, {Cameron}, {Conselice}, {Croom}, {Frenk}, {Gunawardhana}, {Kuijken}, {Nichol}, {Parkinson}, {Phillipps}, {Pimbblet}, {Popescu}, {Prescott}, {Sutherland}, {Tuffs}, {van Kampen}, \& {Wijesinghe}}]{taylor2011}
{Taylor}, E.~N., {Hopkins}, A.~M., {Baldry}, I.~K., {et~al.} 2011, \mnras, 418, 1587, \dodoi{10.1111/j.1365-2966.2011.19536.x}

\bibitem[{{Thilker} {et~al.}(2010){Thilker}, {Bianchi}, {Schiminovich}, {Gil de Paz}, {Seibert}, {Madore}, {Wyder}, {Rich}, {Yi}, {Barlow}, {Conrow}, {Forster}, {Friedman}, {Martin}, {Morrissey}, {Neff}, \& {Small}}]{thilker2010}
{Thilker}, D.~A., {Bianchi}, L., {Schiminovich}, D., {et~al.} 2010, \apjl, 714, L171, \dodoi{10.1088/2041-8205/714/1/L171}

\bibitem[{{Tikhonov} {et~al.}(2005){Tikhonov}, {Galazutdinova}, \& {Drozdovsky}}]{tikhonov2005}
{Tikhonov}, N.~A., {Galazutdinova}, O.~A., \& {Drozdovsky}, I.~O. 2005, \aap, 431, 127, \dodoi{10.1051/0004-6361:20047042}

\bibitem[{{Tody}(1986{\natexlab{a}})}]{Tody1986}
{Tody}, D. 1986{\natexlab{a}}, in Society of Photo-Optical Instrumentation Engineers (SPIE) Conference Series, Vol. 627, Instrumentation in astronomy VI, ed. D.~L. {Crawford}, 733, \dodoi{10.1117/12.968154}

\bibitem[{{Tody}(1986{\natexlab{b}})}]{iraf}
{Tody}, D. 1986{\natexlab{b}}, in Society of Photo-Optical Instrumentation Engineers (SPIE) Conference Series, Vol. 627, Instrumentation in astronomy VI, ed. D.~L. {Crawford}, 733, \dodoi{10.1117/12.968154}

\bibitem[{{Toloba} {et~al.}(2016{\natexlab{a}}){Toloba}, {Guhathakurta}, {Romanowsky}, {Brodie}, {Mart{\'\i}nez-Delgado}, {Arnold}, {Ramachandran}, \& {Theakanath}}]{toloba2016}
{Toloba}, E., {Guhathakurta}, P., {Romanowsky}, A.~J., {et~al.} 2016{\natexlab{a}}, \apj, 824, 35, \dodoi{10.3847/0004-637X/824/1/35}

\bibitem[{{Toloba} {et~al.}(2016{\natexlab{b}}){Toloba}, {Sand}, {Spekkens}, {Crnojevi{\'c}}, {Simon}, {Guhathakurta}, {Strader}, {Caldwell}, {McLeod}, \& {Seth}}]{toloba2016a}
{Toloba}, E., {Sand}, D.~J., {Spekkens}, K., {et~al.} 2016{\natexlab{b}}, \apjl, 816, L5, \dodoi{10.3847/2041-8205/816/1/L5}

\bibitem[{{VandenBerg} {et~al.}(2006){VandenBerg}, {Bergbusch}, \& {Dowler}}]{vandenberg2006}
{VandenBerg}, D.~A., {Bergbusch}, P.~A., \& {Dowler}, P.~D. 2006, \apjs, 162, 375, \dodoi{10.1086/498451}

\bibitem[{{Vansevi{\v{c}}ius} {et~al.}(2004){Vansevi{\v{c}}ius}, {Arimoto}, {Hasegawa}, {Ikuta}, {Jablonka}, {Narbutis}, {Ohta}, {Stonkut{\.{e}}}, {Tamura}, {Vansevi{\v{c}}ius}, \& {Yamada}}]{vansevivcius2004}
{Vansevi{\v{c}}ius}, V., {Arimoto}, N., {Hasegawa}, T., {et~al.} 2004, \apjl, 611, L93, \dodoi{10.1086/423802}

\bibitem[{{Vasiliev} {et~al.}(2022){Vasiliev}, {Belokurov}, \& {Evans}}]{vasiliev2022}
{Vasiliev}, E., {Belokurov}, V., \& {Evans}, N.~W. 2022, \apj, 926, 203, \dodoi{10.3847/1538-4357/ac4fbc}

\bibitem[{{Veronese} {et~al.}(2023){Veronese}, {de Blok}, \& {Walter}}]{veronese2023}
{Veronese}, S., {de Blok}, W.~J.~G., \& {Walter}, F. 2023, \aap, 672, A55, \dodoi{10.1051/0004-6361/202245423}

\bibitem[{Virtanen {et~al.}(2020)Virtanen, Gommers, Oliphant, Haberland, Reddy, Cournapeau, Burovski, Peterson, Weckesser, Bright, {van der Walt}, Brett, Wilson, Millman, Mayorov, Nelson, Jones, Kern, Larson, Carey, Polat, Feng, Moore, {VanderPlas}, Laxalde, Perktold, Cimrman, Henriksen, Quintero, Harris, Archibald, Ribeiro, Pedregosa, {van Mulbregt}, \& {SciPy 1.0 Contributors}}]{2020SciPy-NMeth}
Virtanen, P., Gommers, R., Oliphant, T.~E., {et~al.} 2020, Nature Methods, 17, 261, \dodoi{10.1038/s41592-019-0686-2}

\bibitem[{{Vlaji{\'c}} {et~al.}(2009){Vlaji{\'c}}, {Bland-Hawthorn}, \& {Freeman}}]{vlajic2009}
{Vlaji{\'c}}, M., {Bland-Hawthorn}, J., \& {Freeman}, K.~C. 2009, \apj, 697, 361, \dodoi{10.1088/0004-637X/697/1/361}

\bibitem[{{Westmeier} {et~al.}(2011){Westmeier}, {Braun}, \& {Koribalski}}]{westmeier2011}
{Westmeier}, T., {Braun}, R., \& {Koribalski}, B.~S. 2011, \mnras, 410, 2217, \dodoi{10.1111/j.1365-2966.2010.17596.x}

\bibitem[{{White} \& {Frenk}(1991)}]{white1991}
{White}, S. D.~M., \& {Frenk}, C.~S. 1991, \apj, 379, 52, \dodoi{10.1086/170483}

\bibitem[{{Williams} {et~al.}(2013){Williams}, {Dalcanton}, {Stilp}, {Dolphin}, {Skillman}, \& {Radburn-Smith}}]{williams2013}
{Williams}, B.~F., {Dalcanton}, J.~J., {Stilp}, A., {et~al.} 2013, \apj, 765, 120, \dodoi{10.1088/0004-637X/765/2/120}

\bibitem[{{Willmer}(2018)}]{willmer2018}
{Willmer}, C. N.~A. 2018, \apjs, 236, 47, \dodoi{10.3847/1538-4365/aabfdf}

\bibitem[{{Wojno} {et~al.}(2020){Wojno}, {Gilbert}, {Kirby}, {Escala}, {Beaton}, {Tollerud}, {Majewski}, \& {Guhathakurta}}]{wojno2020}
{Wojno}, J., {Gilbert}, K.~M., {Kirby}, E.~N., {et~al.} 2020, \apj, 895, 78, \dodoi{10.3847/1538-4357/ab8ccb}

\bibitem[{{Younger} {et~al.}(2007){Younger}, {Cox}, {Seth}, \& {Hernquist}}]{younger2007}
{Younger}, J.~D., {Cox}, T.~J., {Seth}, A.~C., \& {Hernquist}, L. 2007, \apj, 670, 269, \dodoi{10.1086/521976}

\bibitem[{{Yuan} {et~al.}(2022){Yuan}, {Malhan}, {Sestito}, {Ibata}, {Martin}, {Chang}, {Li}, {Caffau}, {Bonifacio}, {Bellazzini}, {Huang}, {Voggel}, {Longeard}, {Arentsen}, {Doliva-Dolinsky}, {Navarro}, {Famaey}, {Starkenburg}, \& {Aguado}}]{yuan2022}
{Yuan}, Z., {Malhan}, K., {Sestito}, F., {et~al.} 2022, \apj, 930, 103, \dodoi{10.3847/1538-4357/ac616f}

\bibitem[{{Zibetti} {et~al.}(2009){Zibetti}, {Charlot}, \& {Rix}}]{zibetti2009}
{Zibetti}, S., {Charlot}, S., \& {Rix}, H.-W. 2009, \mnras, 400, 1181, \dodoi{10.1111/j.1365-2966.2009.15528.x}

\bibitem[{{Zuntz} {et~al.}(2018){Zuntz}, {Sheldon}, {Samuroff}, {Troxel}, {Jarvis}, {MacCrann}, {Gruen}, {Prat}, {S{\'a}nchez}, {Choi}, {Bridle}, {Bernstein}, {Dodelson}, {Drlica-Wagner}, {Fang}, {Gruendl}, {Hoyle}, {Huff}, {Jain}, {Kirk}, {Kacprzak}, {Krawiec}, {Plazas}, {Rollins}, {Rykoff}, {Sevilla-Noarbe}, {Soergel}, {Varga}, {Abbott}, {Abdalla}, {Allam}, {Annis}, {Bechtol}, {Benoit-L{\'e}vy}, {Bertin}, {Buckley-Geer}, {Burke}, {Carnero Rosell}, {Carrasco Kind}, {Carretero}, {Castander}, {Crocce}, {Cunha}, {D'Andrea}, {da Costa}, {Davis}, {Desai}, {Diehl}, {Dietrich}, {Doel}, {Eifler}, {Estrada}, {Evrard}, {Fausti Neto}, {Fernandez}, {Flaugher}, {Fosalba}, {Frieman}, {Garc{\'\i}a-Bellido}, {Gaztanaga}, {Gerdes}, {Giannantonio}, {Gschwend}, {Gutierrez}, {Hartley}, {Honscheid}, {James}, {Jeltema}, {Johnson}, {Johnson}, {Kuehn}, {Kuhlmann}, {Kuropatkin}, {Lahav}, {Li}, {Lima}, {Maia}, {March}, {Martini}, {Melchior}, {Menanteau}, {Miller}, {Miquel}, {Mohr}, {Neilsen}, {Nichol}, {Ogando}, {Roe}, {Romer},
  {Roodman}, {Sanchez}, {Scarpine}, {Schindler}, {Schubnell}, {Smith}, {Smith}, {Soares-Santos}, {Sobreira}, {Suchyta}, {Swanson}, {Tarle}, {Thomas}, {Tucker}, {Vikram}, {Walker}, {Wechsler}, {Zhang}, \& {DES Collaboration}}]{zuntz2018}
{Zuntz}, J., {Sheldon}, E., {Samuroff}, S., {et~al.} 2018, \mnras, 481, 1149, \dodoi{10.1093/mnras/sty2219}

\end{thebibliography}
\bibliographystyle{aasjournal}



\end{document}